\documentclass[a4paper, 11pt]{article}
\usepackage[utf8]{inputenc}
\usepackage[T1]{fontenc}

\usepackage{geometry}
\usepackage{graphicx}
\usepackage{setspace}

\usepackage{amsmath}
\usepackage{bm}

\usepackage{booktabs}
\usepackage{longtable}
\usepackage{threeparttable}

\usepackage{charter}
\usepackage[bitstream-charter]{mathdesign}

\usepackage[bookmarks=true, bookmarksopen=true, colorlinks, citecolor=blue, urlcolor=blue, linkcolor=blue]{hyperref}
\usepackage[nameinlink,capitalize]{cleveref}
\crefname{chapter}{Chapter}{Chapters}
\crefname{section}{Section}{Sections}
\crefname{subsection}{Section}{Sections}
\crefname{subsubsection}{Section}{Sections}
\crefname{figure}{Figure}{Fig.}
\crefname{table}{Table}{Tab.}
\crefname{equation}{Eq.}{Eqs.}
\crefname{appendix}{Appendix}{Appendices}
\crefname{appsec}{Appendix}{Appendices}
\usepackage{bookmark}
\usepackage{enumitem}

\usepackage[hang,flushmargin]{footmisc} 
\usepackage[yyyymmdd]{datetime}
\usepackage[british]{babel}

\usepackage{float}
\usepackage{bigstrut}
\restylefloat{table}

\usepackage[title,titletoc]{appendix}
\makeatletter

\renewenvironment{appendices}{%
    \begin{oldappendices}%
    \renewcommand{\thefigure}{\ifnum \c@section>\z@ \thesection.\fi\@arabic\c@figure}%
    \@addtoreset{figure}{section}%
    \renewcommand{\thetable}{\ifnum \c@section>\z@ \thesection.\fi\@arabic\c@table}%
    \@addtoreset{table}{section}}{%
    \end{oldappendices}%
}\makeatother

\usepackage{natbib}
\let\natbibcitet\citet
\renewcommand\citet{\bibpunct{(}{)}{,}{a}{,}{,}\natbibcitet}
\let\natbibcitep\citep
\renewcommand\citep{\bibpunct{(}{)}{;}{a}{,}{;}\natbibcitep}
\newcommand{\bi}{\begin{itemize}}
\newcommand{\ei}{\end{itemize}}
\newcommand{\be}{\begin{equation}}
\newcommand{\ee}{\end{equation}}
\defcitealias{ieo14}{IEO, 2014}

\long\def\symbolfootnote[#1]#2{\begingroup%
\def\thefootnote{\fnsymbol{footnote}}\footnote[#1]{#2}\endgroup}

\makeatletter
\def\ubar#1{\underline{\sbox\tw@{$#1$}\dp\tw@\z@\box\tw@}}
\def\obar#1{\overline{\sbox\tw@{$#1$}\dp\tw@\z@\box\tw@}}
\makeatother

\usepackage{caption}

\makeatletter\let\p@subfigure\thefigure\makeatother

\usepackage[labelformat=simple]{subcaption}

\makeatletter

\def\Autoref#1{%
  \begingroup
  \edef\reserved@a{\cpttrimspaces{#1}}%
  \ifcsndefTF{r@#1}{%
    \xaftercsname{\expandafter\testreftype\@fourthoffive}
      {r@\reserved@a}.\\{#1}%
  }{%
    \ref{#1}%
  }%
  \endgroup
}
\def\testreftype#1.#2\\#3{%
  \ifcsndefTF{#1autorefname}{%
    \def\reserved@a##1##2\@nil{%
      \uppercase{\def\ref@name{##1}}%
      \csn@edef{#1autorefname}{\ref@name##2}%
      \autoref{#3}%
    }%
    \reserved@a#1\@nil
  }{%
    \autoref{#3}%
  }%
}
\makeatother

\title{The transmission of uncertainty shocks on income inequality: \\State-level evidence from the United States}
\date{}
\usepackage{authblk}

\author{Manfred M. Fischer, Florian Huber\footnote{Corresponding author: Florian Huber, Vienna University of Economics and Business, Welthandelsplatz 1, A-1020, Vienna. E-mail: \href{mailto:fhuber@wu.ac.at}{fhuber@wu.ac.at}. The authors gratefully acknowledge financial support from the Austrian National Bank, Jubilaeumsfond grant no. 17650.} ~and Michael Pfarrhofer\\ \large Vienna University of Economics and Business (WU)}


\def\equationautorefname~#1\null{%
  Eq.~(#1)\null
}
\def\equationautorefname~#1\null{
Eq.~(#1)\null
}
\begin{document}
\clearpage\maketitle
\thispagestyle{empty}
%\graphicspath{{results/}}

\onehalfspacing

\begin{abstract}
\noindent In this paper, we explore the relationship between state-level household income inequality and macroeconomic uncertainty in the United States. Using a novel large-scale macroeconometric model, we shed light on regional disparities of inequality responses to a national uncertainty shock. The results suggest that income inequality decreases in most states, with a pronounced degree of heterogeneity in terms of shapes and magnitudes of the dynamic responses. By contrast, some few states, mostly located in the West and South census region, display increasing levels of income inequality over time. We find that this directional pattern in responses is mainly driven by the income composition and labor market fundamentals. In addition, forecast error variance decompositions allow for a quantitative assessment of the importance of uncertainty shocks in explaining income inequality. The findings highlight that volatility shocks account for a considerable fraction of forecast error variance for most states considered. Finally, a regression-based analysis sheds light on the driving forces behind differences in state-specific inequality responses.
\end{abstract}

\bigskip
\begin{tabular}{p{0.2\hsize}p{0.65\hsize}} %0.15
\textbf{Keywords:}  &income distribution, inequality, uncertainty shocks, US states,  global vector autoregressive model\\
\end{tabular}

\smallskip
\begin{tabular}{p{0.2\hsize}p{0.4\hsize}}
\textbf{JEL Codes:} &C11, C30, E3, D31 \\
\end{tabular}
\vspace{0.6cm}

\bigskip
\newpage
\section{Introduction}\label{sec:intro}
This paper explores the nexus between uncertainty shocks and income inequality at the US state level. The literature on uncertainty shocks \citep{bloom2009, caggiano2014uncertainty,jurado2015measuring, caldara2016macroeconomic, carriero2016measuring, basu2017uncertainty, mumtaz2017changing} identifies a range of channels through which volatility impacts the wider macroeconomy. Movements in  quantities related to these channels are typically perceived as important determinants of income inequality \citep{doi:10.1162/00335530360535135,ROINE2009974,COIBION201770}. In this contribution, we aim to link these strands of the literature by proposing a large-scale dynamic macroeconometric model. This allows for capturing dynamics between national US quantities and a set of state-specific variables related to the distribution of income across space and time.

The recent literature on uncertainty shocks \citep[see, among many others,][]{caldara2016macroeconomic, doi:10.1093/qje/qjw024} increasingly discriminates between different types of uncertainty. In his seminal contribution, \citet{bloom2009}, for instance, uses the volatility index (VIX) of the Chicago Board Options Exchange as an observed measure of uncertainty that is closely related to financial market uncertainty. As opposed to uncertainty arising from financial markets, real macroeconomic uncertainty is associated with unexpected fluctuations in output or prices. Other studies highlight that uncertainty might also be linked to unexpected actions of policy makers in central banks and the government \citep{doi:10.1093/qje/qjw024}. All types of uncertainty have in common, however, that they are generally perceived to be detrimental for economic performance, at least in the short-run. For instance, the latest global financial crisis can also be viewed as a US-based uncertainty shock that ultimately engulfed the world economy and led to a sharp decline in economic activity.

During economic downturns, income inequality has been found to decrease in multiple contributions to the literature \citep[see, for instance,][]{heathcote2010unequal, petev2011consumption, meyer2013consumption, mumtaz2017impact}. This finding, however, strongly depends on the composition of income within a given country or region. A potential causal mechanism behind this finding might that capital owners are comparatively more exposed to adverse business cycle movements, which are often accompanied by sharp declines in corporate profits and stock prices.
By contrast, if the income share of capital is comparatively low in a given economy, inequality could also increase during recessions. This is due to the notion that less skilled workers are typically more vulnerable to labor market shifts, and may be forced to accept wage cuts during recessions with unemployment being the alternative. Thus, the impact of uncertainty shocks on the income distribution in recessionary episodes is unclear a priori. Understanding the causal mechanisms that give rise to changes in income inequality proves to be important for policy makers in governmental institutions and central banks.\footnote{Several studies highlight the relation between household income inequality and the emergence of crises \citep[see, for instance, ][and the references therein]{doi:10.1080/19452829.2011.643098,doi:10.1111/joes.12028}. Based on findings that inequality increased in the build-up to both the Great Depression and the Great Recession, \citet{10.1257/aer.20110683} identify a causal relationship between inequality, household debt and economic depressions in a DSGE model.} 

The empirical literature dealing with the dynamic relationship between uncertainty and income inequality is, however, relatively sparse. This contribution attempts to fill the gap by considering data on unemployment, real income, employment, and a survey-based measure on income inequality for all US states and the District of Columbia. State-specific information is complemented by a set of US macroeconomic aggregates that serve as common driving factors of regional business cycle movements. Taking such a state-level perspective enables a detailed investigation on whether national uncertainty shocks yield asymmetric responses across states while the inclusion of additional covariates at the country-level provides the possibility to inspect the transmission mechanisms of uncertainty shocks on state-level income inequality in more detail.
 
Since the data set considered is large, we suggest a parsimonious multi-state framework closely related to the global vector autoregressive (GVAR) model proposed in \citet{Pesaran2004}. The model differs from a standard GVAR model along several important dimensions. First, inspired by the panel data literature, state-specific regression coefficients are assumed to arise from an underlying common distribution. This improves estimation accuracy while maintaining sufficient flexibility for state-specific idiosyncrasies. Second, one key assumption of the model is that contemporaneous relations among states and variables are driven by a small number of latent factors. This reduces the amount of free parameters to be estimated significantly. Third, we assume that all shocks to the system are heteroscedastic and follow a flexible stochastic volatility specification. Finally, structural identification is achieved by using the measure proposed in \citet{doi:10.1093/qje/qjw024} that approximates general economic policy uncertainty.

The empirical findings reveal that uncertainty shocks lead to heterogeneous responses across states. Some display a significant positive reaction of household income inequality, others show a significant decrease in income dispersion. Forecast error variance decompositions (FEVDs) identify uncertainty to be an important driver of variation in income inequality, especially for certain states. Conducting an exploratory regression analysis suggests that the specific composition of income is crucial in determining the reaction of income inequality. In order to shed light on the specific transmission mechanisms at the state-level, we further investigate the reactions of additional macroeconomic quantities. Pronounced shifts in unemployment, employment as well as  total private income point towards a prominent role of the income-wealth channel in explaining the dynamic responses of inequality.

The remainder of the paper is structured as follows. Section \ref{sec:econometrics} presents the econometric framework adopted while Section \ref{sec:data} provides a brief summary of the dataset used. Section \ref{sec:results}  describes the empirical findings based on structural impulse responses and forecast error variance decompositions. Section \ref{sec:transmission} sheds further light on the transmission mechanisms of uncertainty shocks on income inequality and attempts to explain state-specific differences in the responses. Finally, the last section summarizes and concludes the paper.

\section{Econometric framework}\label{sec:econometrics}
In order to measure the impact of uncertainty on income inequality across regions and variable types, a suitable econometric framework is necessary. The large number of US states alongside a moderate number of region-specific endogenous variables calls for a modeling approach that adequately captures dynamic relations in the data. Here, we follow \cite{Pesaran2004} and propose a variant of the GVAR involving $N$ small-scale region-specific models. These models feature domestic variables of regional economies collected in the $k$-dimensional vector $\bm{y}_{it}$ besides region-specific cross-sectional averages of foreign variables, collected in the $k$-dimensional vector  
\begin{equation}
\bm{y}^{\ast}_{it} = \sum_{j=1}^{N}w_{ij} \bm{y}_{jt}, \label{eq: WEX}
\end{equation}
where the weights $w_{ij}~(i,j=1,\hdots,N)$ are elements of a conventional $N\times N$ row-stochastic matrix that represents the connectivity relationships between the $N$ regions. By convention, $w_{ii} = 0$ for all $i$. Note that the higher the connectedness is between region $i$ and region $j$ (that is, the larger $w_{ij}$ is), the more region $i$ is exposed to externalities arising in region $j$.\footnote{For the purpose of this paper, we employ a weighting scheme based on the inverse distance between centroids of the regions. Alternative specifications do not alter the results significantly, adding robustness to our findings.}

The regional economies may then be modeled as a VAR augmented by a vector of lagged foreign variables, and a set of national macroeconomic aggregates that are assumed to be important determinants of regional business cycle dynamics, 
\begin{equation}
\bm{y}_{it} = \bm{\theta}_i + \sum_{p=1}^{P} \bm{A}_{ip} \bm{y}_{it-p} + \sum_{q=1}^{Q} \bm{B}_{iq} \bm{y}^{\ast}_{it-q} + \bm{C}_{i} \bm{z}_{t-1} + \bm{\epsilon}_{it}, \quad \bm{\epsilon}_{it}\sim \mathcal{N}(\bm{0},\bm{\Sigma}_{it}).\label{eq:varx}
\end{equation}
Hereby, $\{\bm{y}_{it}\}_{t=1}^{T}$ is a $k$-dimensional vector of macroeconomic time series specific to region $i = 1, \hdots, N$. $\bm{\theta}_i$ is a $k$-dimensional intercept vector, while $\bm{A}_{ip}~(p = 1, \hdots, P)$ and $\bm{B}_{iq}~(q = 1, \hdots, Q)$ are $k\times k$  matrices of unknown parameters, respectively. $\bm{C}_i$ is a $k \times \ell$ matrix of regression coefficients associated with $\ell$ national macroeconomic aggregates collected in $\bm{z}_t$. The error term $\bm{\epsilon}_{it}$ follows a zero mean Gaussian distribution with a time-varying variance-covariance matrix $\bm{\Sigma}_{it}$.

The national aggregates in $\bm{z}_{t}$ follow a VAR process,
\begin{equation}
\bm{z}_{t} = \sum_{p=1}^{P} \bm{D}_{p} \bm{z}_{t-p} + \sum_{q=1}^{Q} \bm{S}_{q} \bm{z}^{\ast}_{t-q} + \bm{u}_{t}, \quad \bm{u}_{t} \sim \mathcal{N}(\bm{0}, \bm{\Xi}_t)
\label{eq:nationalVAR}
\end{equation}
with $\bm{D}_p~(p = 1,\hdots,P)$ and $\bm{S}_q~(q = 1,\hdots,Q)$ denoting $\ell \times \ell$ and $\ell \times k$ coefficient matrices. To establish dependencies between the national aggregates and the regions, we include simple arithmetic averages of the $k$ region-level quantities over $N$ regions denoted by $\bm{z}^{\ast}_{t} = ( z_{1t}^{\ast}, \dots, z_{kt}^{\ast})'$. Again, we assume the error term $\bm{u}_{t}$ to follow a Gaussian distribution centered on zero with time-varying variance covariance matrix $\bm{\Xi}_t$.

To capture contemporaneous relations among the elements in $\bm{y}_t = (\bm{y}_{1t}', \hdots, \bm{y}_{Nt}')'$ and $\bm{z}_t$, we assume that the shock vector $\bm{\varepsilon}_{t} = (\bm{u}_{t}',\bm{\epsilon}_{1t}',\hdots,\bm{\epsilon}_{Nt}')'$ of size $L=kN+\ell$ features a factor stochastic volatility structure \citep{aguilar2000bayesian}, that is,
\begin{equation}
\bm{\varepsilon}_{t} = \bm{\Lambda} \bm{f}_t + \bm{\eta}_t.
\end{equation}
$\bm{f}_t\sim\mathcal{N}(\bm{0},\bm{H}_t)$ represents a set of $F (\ll L)$ common static factors, $\bm{\Lambda}$ is an $L\times F$ matrix of factor loadings, and $\bm{\eta}_t\sim\mathcal{N}(\bm{0},\bm{\Omega}_t)$ is an $L$-dimensional idiosyncratic noise vector. The variance-covariance matrices $\bm{H}_t = \text{diag}[\exp(h_{1t}),\hdots,\exp(h_{Ft})]$ and $\bm{\Omega}_t = \text{diag}[\exp(\omega_{1t}),\hdots,\exp(\omega_{Lt})]$ are diagonal matrices, implying that any comovement across the elements in $\bm{\varepsilon}_t$ stems from the common factors. We control for heteroscedasticity of the shocks by assuming that the logarithm of the main diagonal elements follows an autoregressive process of order one. This setup implies that $\text{Var}(\bm{\varepsilon}_t) = \bm{\Lambda} \bm{H}_t \bm{\Lambda}' + \bm{\Omega}_t := \bm{\Theta}_t$.

Since unrestricted estimation of the  model typically translates into issues associated with overfitting, we introduce additional structure on the coefficients of the model in \autoref{eq:varx}. In what follows, we assume that the  $M=k (1 + Pk + Qk + \ell)$ vectorized regression coefficients $\bm{\beta}_i = \text{vec}[(\bm{\theta}_{i},\bm{A}_{i1},\hdots,\bm{A}_{iP},\bm{B}_{i1},\hdots,\bm{B}_{iQ}, \bm{C}_i)]$ for region $i = 1, \hdots, N$ arise from a common distribution,
\begin{equation}
\bm{\beta}_i\sim\mathcal{N}(\bm{\mu},\bm{V}),
\end{equation}
where $\bm{\mu}$ denotes a common mean and $\bm{V}=\text{diag}(v_1, \dots, v_M)$ a variance-covariance matrix. Notice that $v_j~(j = 1,\hdots,M)$ provides a natural measure of similarity between the $j$th and the $r$th element in $\bm{\beta}_i$ across regions and controls the magnitude of potential deviations from $\mu_j$, the $j$th element of $\bm{\mu}$. The presence of the common distribution implies that our framework is a hierarchical model that is related to random coefficient models in microeconometrics \citep{verbeke1996linear, allenby1998heterogeneity} and the panel VAR specification outlined in \cite{jarocinski2010responses}.

Estimation and inference is carried out within a Bayesian framework. This implies that suitable priors need to be specified that are in turn combined with a likelihood function to yield proper posterior distributions for parameters. Appendices \ref{app:priors} and \ref{app:mcmc} provide details on prior specification and the employed Markov chain Monte Carlo (MCMC) algorithm.

\section{Data overview and model specification}\label{sec:data}
The objective of this paper is to analyze  state-level responses of household income inequality to a national uncertainty shock across all US states and the District of Columbia. Three of the four state-level quantities are taken from Federal Reserve Bank of St. Louis data base. In particular, we use quarterly time series for employment, unemployment and price adjusted total personal income per capita for the period 1985:Q1 to 2017:Q1 in $\bm{y}_{it}$. The inequality measure for all states is constructed using data from the Annual Social and Economic Supplement of the Current Population Survey \citep[CPS,][]{ipums-cps}. The income definition is based on equivalized household income employing the square root scale, where we set negative incomes equal to zero. This quantity is adjusted by the given survey weights to calculate the well-known Gini coefficient per year, while the quarterly time structure is obtained by applying spline interpolation. For the empirical specification, we thus have a set of $k = 4$ state-level quantities in $\bm{y}_{it}$ (household income inequality, total personal income, unemployment and employment) for the $N = 51$ US states.

National macroeconomic quantities included in $\bm{z}_{t}$ are the one-year treasury rate and the consumer price index, taken from the data set presented in \citet{doi:10.1080/07350015.2015.1086655} and consequently aggregated to quarterly frequency. Moreover, we use quarterly US gross domestic product (GDP) obtained from the National Income and Product Accounts provided by the Bureau of Economic Analysis.\footnote{Available for download at \url{https://www.bea.gov/iTable/index_nipa.cfm}.} To capture national economic uncertainty, we rely on the overall economic policy index provided by \citet{doi:10.1093/qje/qjw024}. This approach is based on newspaper coverage frequency and involves the search of a combination of selected uncertainty related terms in ten leading newspapers in the US from 1985 onwards on a monthly basis.

The national model in \autoref{eq:nationalVAR} thus includes $\ell = 4$ variables in the vector $\bm{z}_{t}$. We use a lag length of $P = Q = 1$. Based on classical information criteria, we opt for $F = 2$ of latent factors.\footnote{Using more factors leads to qualitatively similar results.} All time series are deseasonalized and aggregated to a quarterly structure if necessary. State-level data, except observations on unemployment, are transformed by the natural logarithm. On the national-level, US gross domestic product, the uncertainty index and the consumer price index are in logarithms, while the one-year treasury rate is first-differenced as suggested in \citet{doi:10.1080/07350015.2015.1086655}.

\section{The impact of uncertainty shocks on income inequality}\label{sec:results}
In the following section,  the main empirical findings of the paper are discussed. Section \ref{sec:macro_nat} displays the reactions of the national macroeconomic quantities with respect to movements in uncertainty while Section \ref{sec: IQ_resp} shows the dynamic responses of state-level inequality to national uncertainty shocks. To assess the quantitative importance of uncertainty shocks in shaping the income distribution, we conduct a forecast error variance decomposition in \cref{sec:fevd}.

\subsection{Dynamic responses of national macroeconomic quantities to uncertainty shocks}\label{sec:macro_nat}
\begin{figure}[ht]
\centering
\begin{subfigure}{.32\textwidth}
\subcaption{gross domestic product}\label{fig:gdp}
\includegraphics[width=\textwidth]{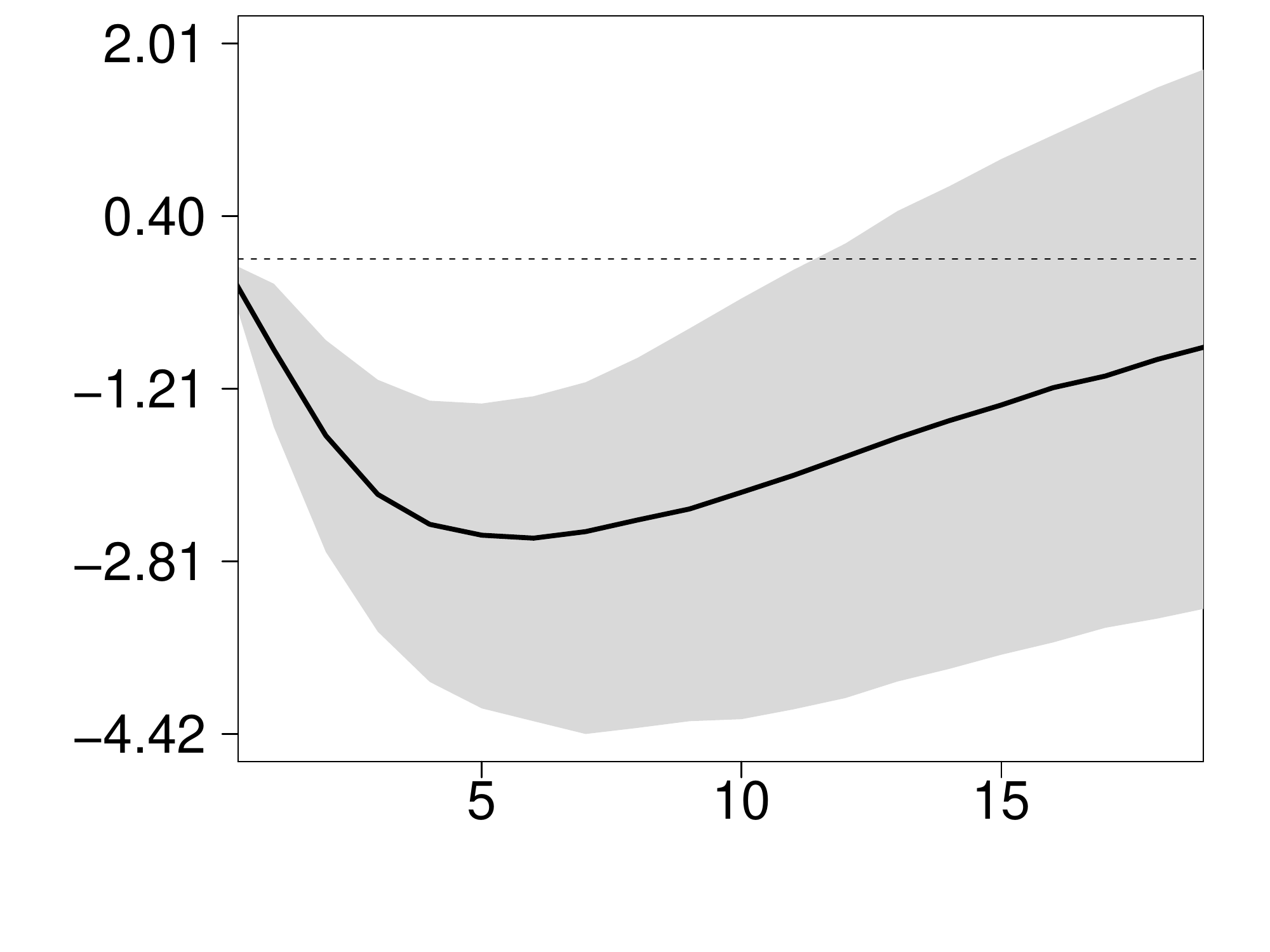}
\end{subfigure}
\begin{subfigure}{.32\textwidth}
\subcaption{consumer price index}\label{fig:cpi}
\includegraphics[width=\textwidth]{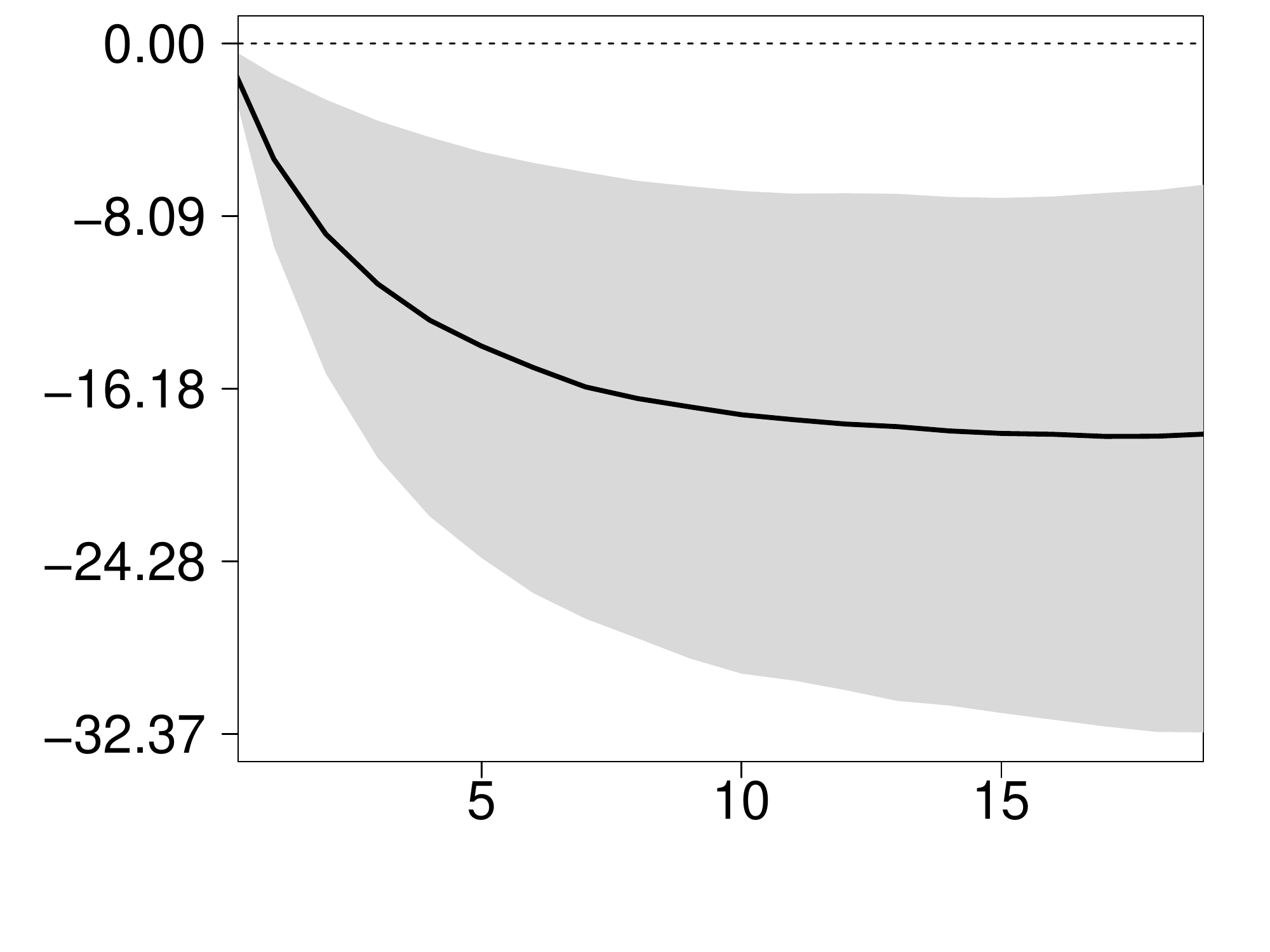}
\end{subfigure}
\begin{subfigure}{.32\textwidth}
\subcaption{one-year treasury rate}\label{fig:gs1}
\includegraphics[width=\textwidth]{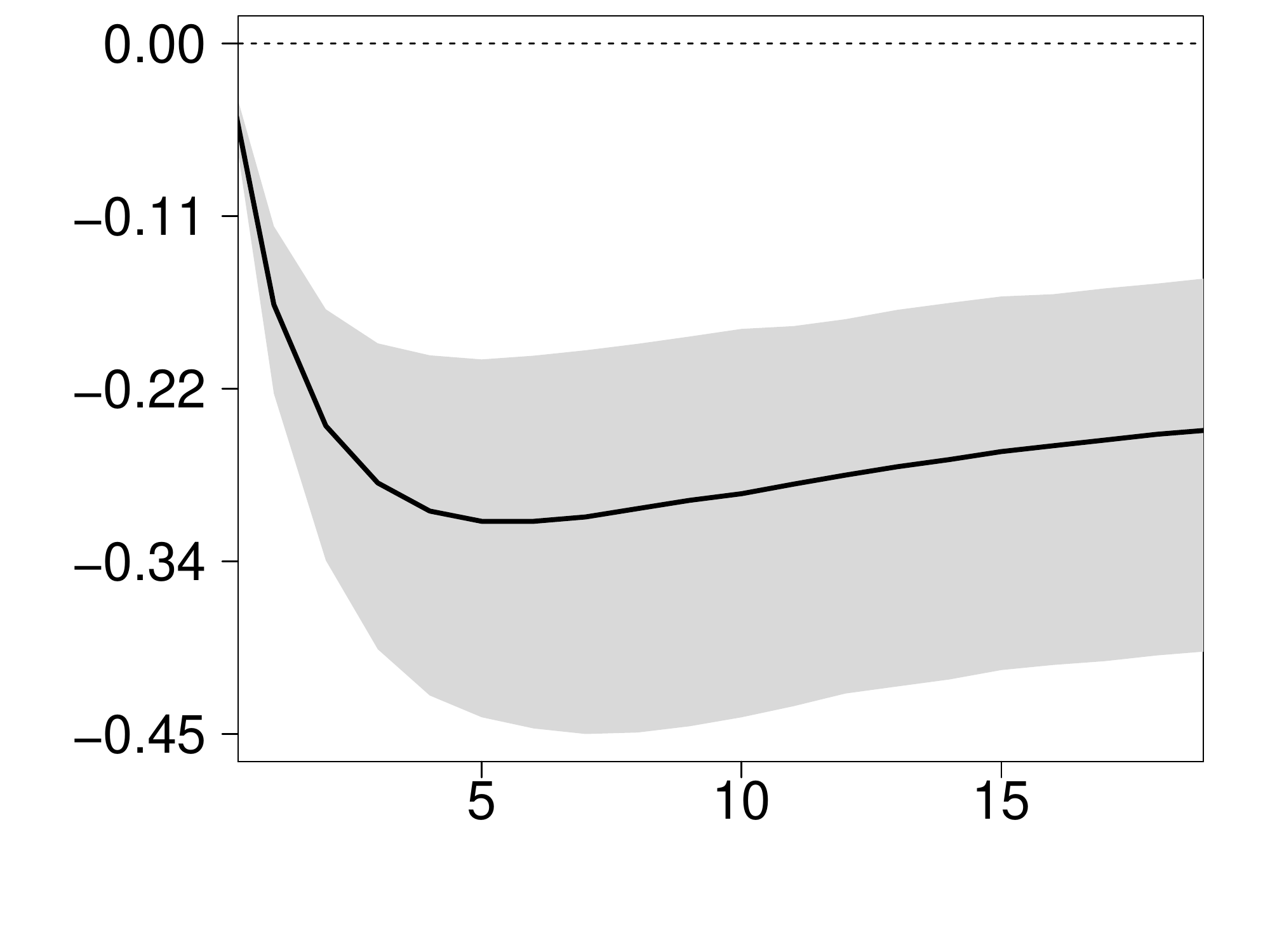}
\end{subfigure}\\
\caption{Impulse responses of national US quantities to an economic policy uncertainty shock.}
\caption*{\footnotesize{\textit{Notes}: The solid black line denotes the median response, the dashed line indicates zero, and the shaded bands (in light grey) the 68 percent posterior coverage interval. Sample period: 1985:Q1 -- 2017:Q1. Front axis: quarters after impact.}}
\label{fig:IRF_US}
\end{figure}

Most of the results regarding responses of national macroeconomic aggregates to uncertainty shocks mirror established findings in the literature. Before proceeding, a brief word on structural identification is in order. In the present paper, we follow \citet{doi:10.1093/qje/qjw024} and include the economic policy uncertainty (EPU) index as the first variable in $\bm{z}_t$. Identification of the model is then achieved by using a simple Cholesky decomposition of the variance-covariance matrix. Figure \ref{fig:IRF_US} presents the endogenous responses to a one standard error shock to the EPU index that amounts to an immediate reaction of the EPU of around 23 percent.

Starting with real activity, measured by gross domestic product, we find a pronounced and long-lasting negative effect of an uncertainty shock, as shown in \autoref{fig:gdp}. The peak response occurs around five quarters after impact, with effects turning insignificant after roughly three years. It is worth mentioning that there is no subsequent real activity overshoot as described by \citet{bloom2009}, which is consistent with more recent contributions in the literature \citep{jurado2015measuring, carriero2016measuring, mumtaz2017impact}. Uncertainty typically affects real economic activity via decreases in investment and hiring of firms, thereby depressing productivity growth \citep{bloom2009}. 

Considering the consumer price index, \autoref{fig:cpi} suggests that prices tend to decline in a rather persistent manner. This reaction can be traced back to depressed overall consumption which leads to decreases in the demand for goods and thus prices. This mechanism is commonly referred to as the aggregate demand channel discussed in \citet{10.1257/aer.101.6.2530}. By contrast, there is no evidence in favor of the upward pricing bias channel that states that firms increase prices to maximize profits in the aftermath of an uncertainty shock.

The final macroeconomic quantity on the US national level is shown in \autoref{fig:gs1}. Instead of using the Federal Funds rate as a measure of the monetary policy stance, we opt for the one-year treasury rate. This is due to the fact that a substantial period of our data set is characterized by short-term interest rates being close to the zero lower bound. Consistent with \citet{bloom2009}, we find that an economic policy uncertainty shock leads to a decline in interest rates. This implies that the Federal Reserve counteracts the negative impact of uncertainty shocks on real activity by conducting expansionary monetary policy. 

\subsection{Dynamic responses of state-level inequality}\label{sec: IQ_resp}
\begin{figure}[ht]
\centering
\includegraphics[width=\textwidth]{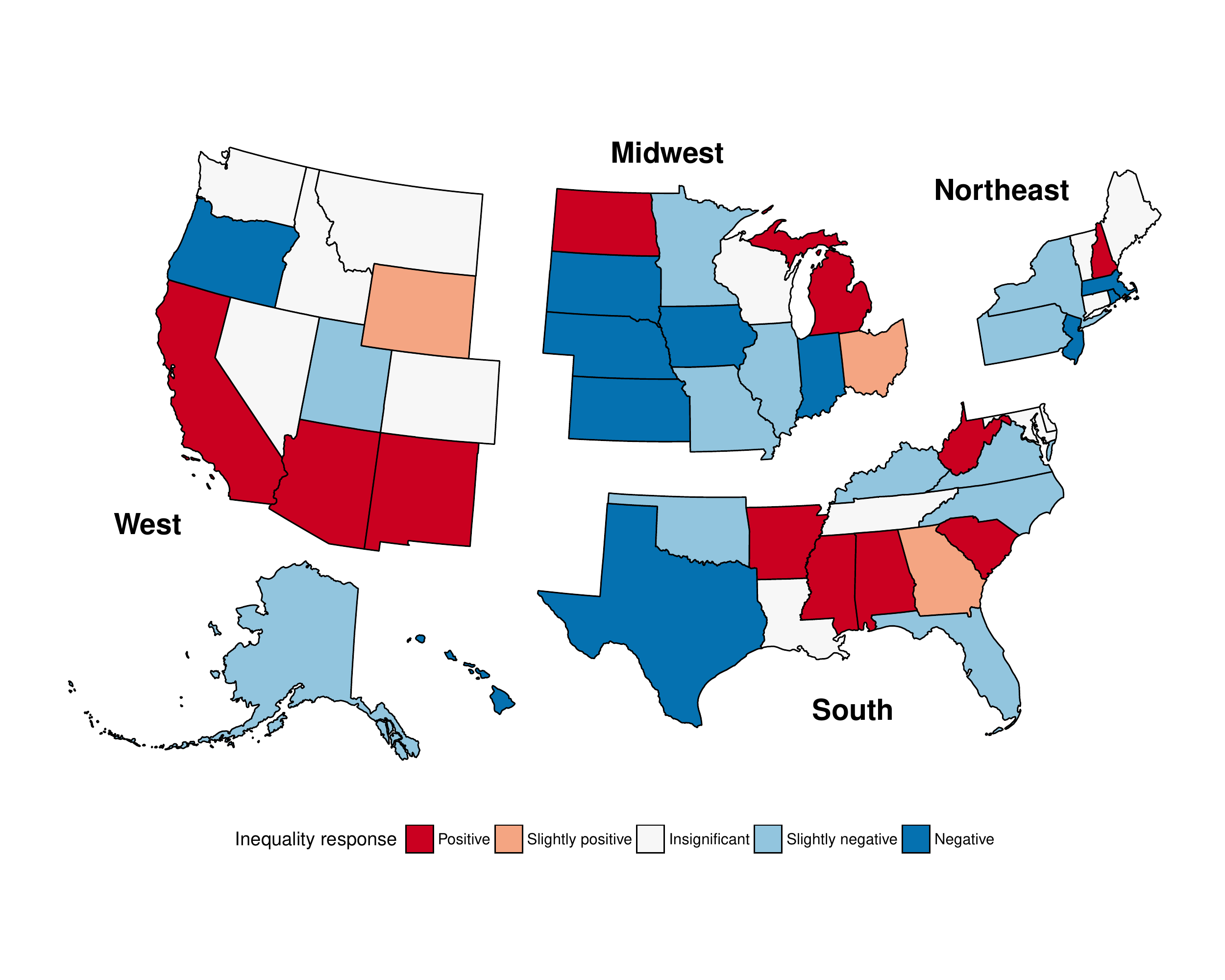}\vspace*{-1.5cm}
\caption*{\footnotesize{\noindent\textit{Notes}: US states divided into the four census regions, thin lines represent state borders. Classification: 'Positive' and 'Negative' refer to inequality responses exceeding thresholds based on the upper ($0.08$) and lower ($-0.10$) 20 percent of the states responses. 'Slightly positive' and 'Slightly negative' include all states with significant responses between zero and the upper and lower 20 percent of the responses. 'Insignificant' indicates non-significant responses based on the 68 percent posterior coverage interval.}}
\caption{Peak response of the median for equivalized household income inequality.}\label{fig:map_iq}
\end{figure}

An overview on regional state-level disparities with respect to the reaction of income inequality is given in \autoref{fig:map_iq}, which depicts the posterior median of the peak responses. Statistical significance is assessed by zeroing out peak responses of states where the 16th and 84th credible intervals contain zero. This figure serves as a means to illustrate the direction of the dynamic responses of inequality across space to see whether geographic patterns exist.  In order to provide a more detailed picture, we subsequently focus our analysis on selected states from the four census regions West, Midwest, South, and Northeast.
 
One key insight from the figure is that reactions of income inequality appear to be heterogenous across states. For some states, income inequality declines whereas other states exhibit increases in income inequality. While finding a geographical pattern behind the sign and magnitude of the responses is challenging, one commonality is that most states displaying negative inequality responses are located in the Midwest census region, with some exceptions. Positive reactions of income inequality to an uncertainty shock are observed mainly in the West and South census region. The finding that inequality decreases in response to shifts in economic uncertainty is consistent with the actual development of income inequality during the global financial crisis in 2008/2009 observed in the data from the CPS.
%The mixed reactions across states can be explained by considering the share of capital income. States that display declining inequality levels appear to be characterized by higher levels of capital income whereas states that show rising inequality levels are generally accompanied by lower levels of capital income and higher shares of labor income.

\begin{figure}[!ht]
\fbox{\begin{minipage}[t]{\textwidth}
\centering
\begin{minipage}[t]{.24\textwidth}
\centering
\large \textit{West}\vspace{0.1cm}\hrule\vspace{0.5cm}
\small California\\[0.1cm]
\includegraphics[width=\linewidth]{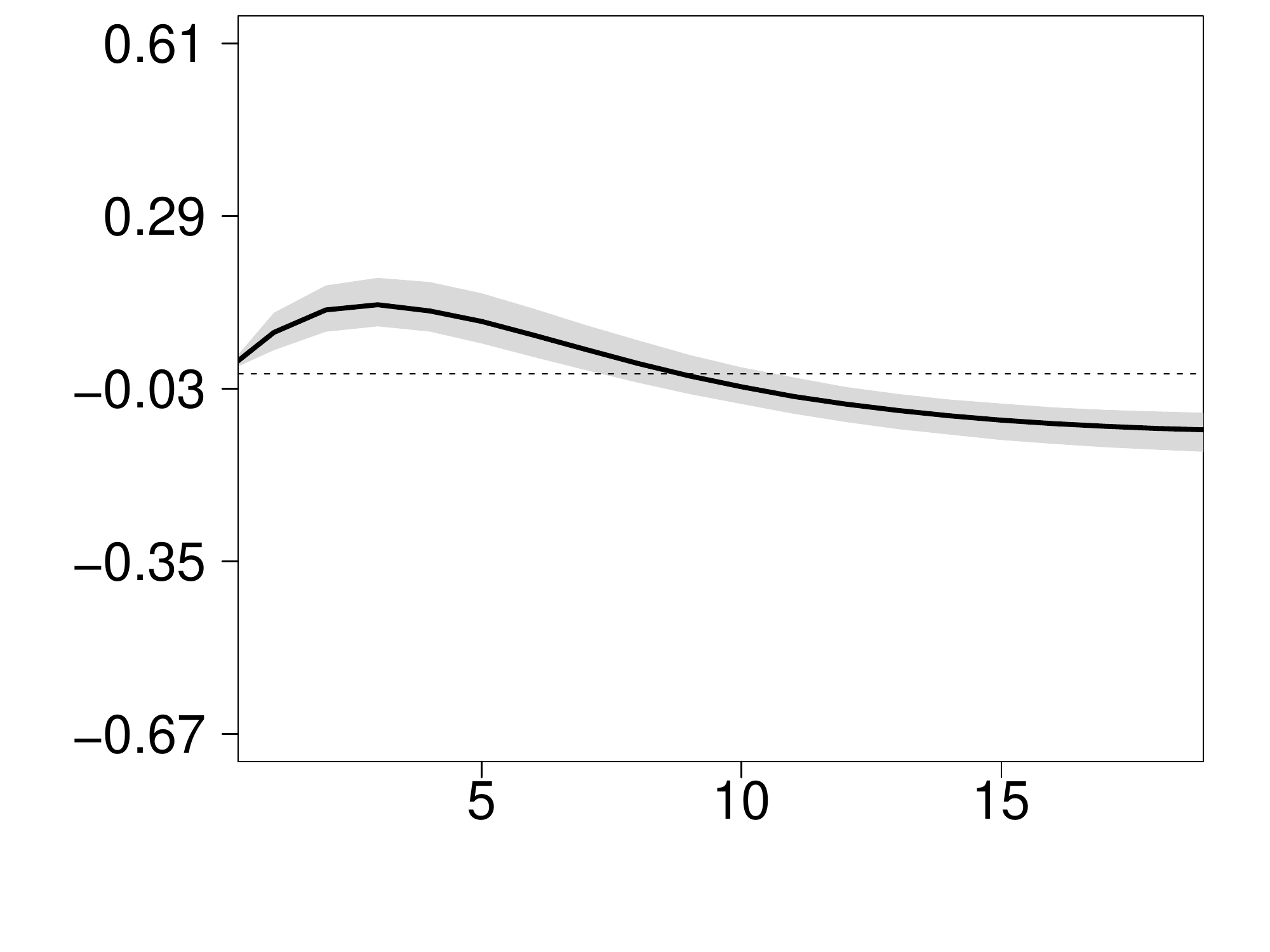}
\small New Mexico\\[0.1cm]
\includegraphics[width=\linewidth]{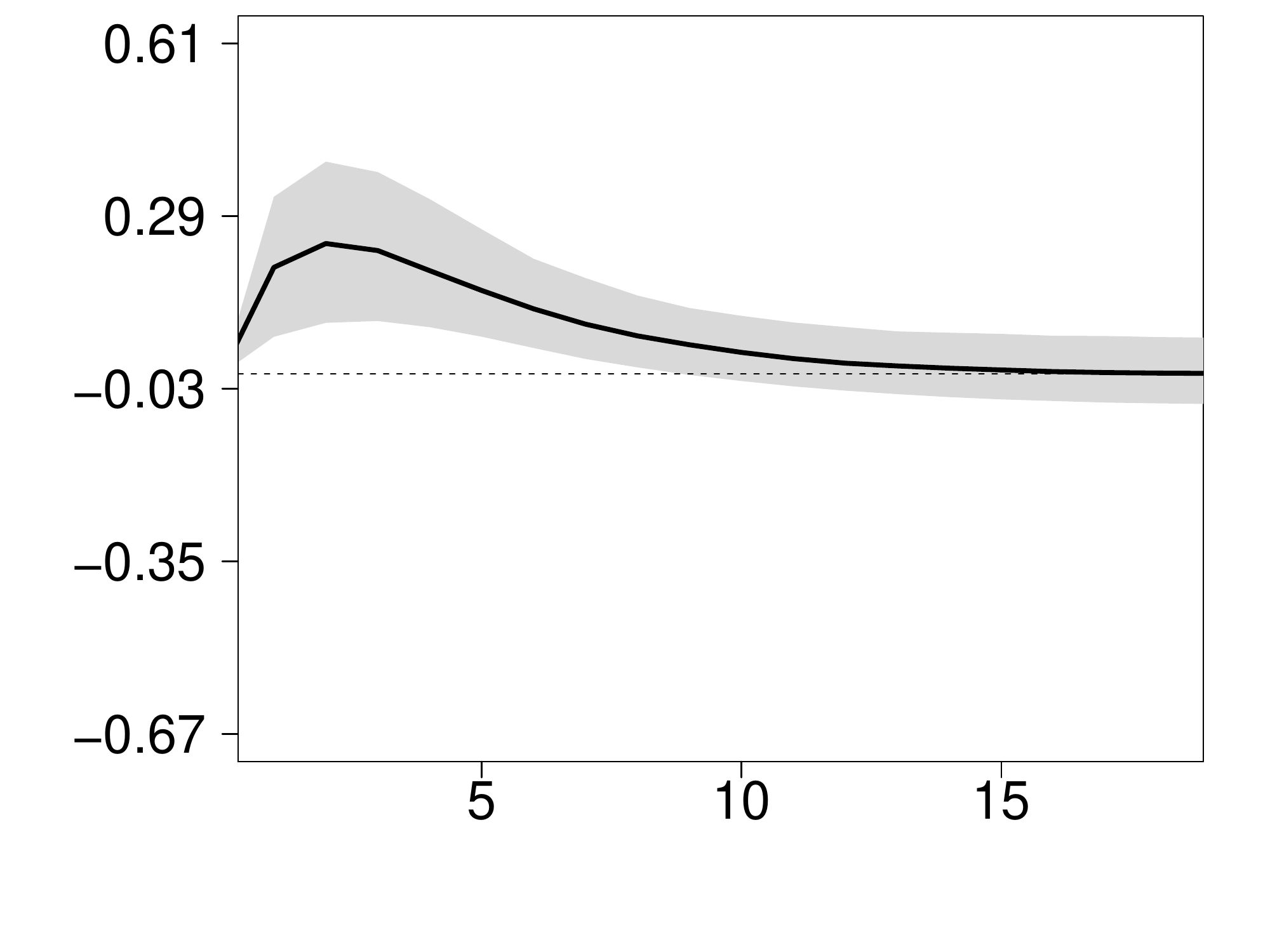}
\end{minipage}
\begin{minipage}[t]{0.24\textwidth}
\centering 
\large \textit{Midwest}\vspace{0.1cm}\hrule\vspace{0.5cm}
\small North Dakota\\[0.1cm]
\includegraphics[width=\linewidth]{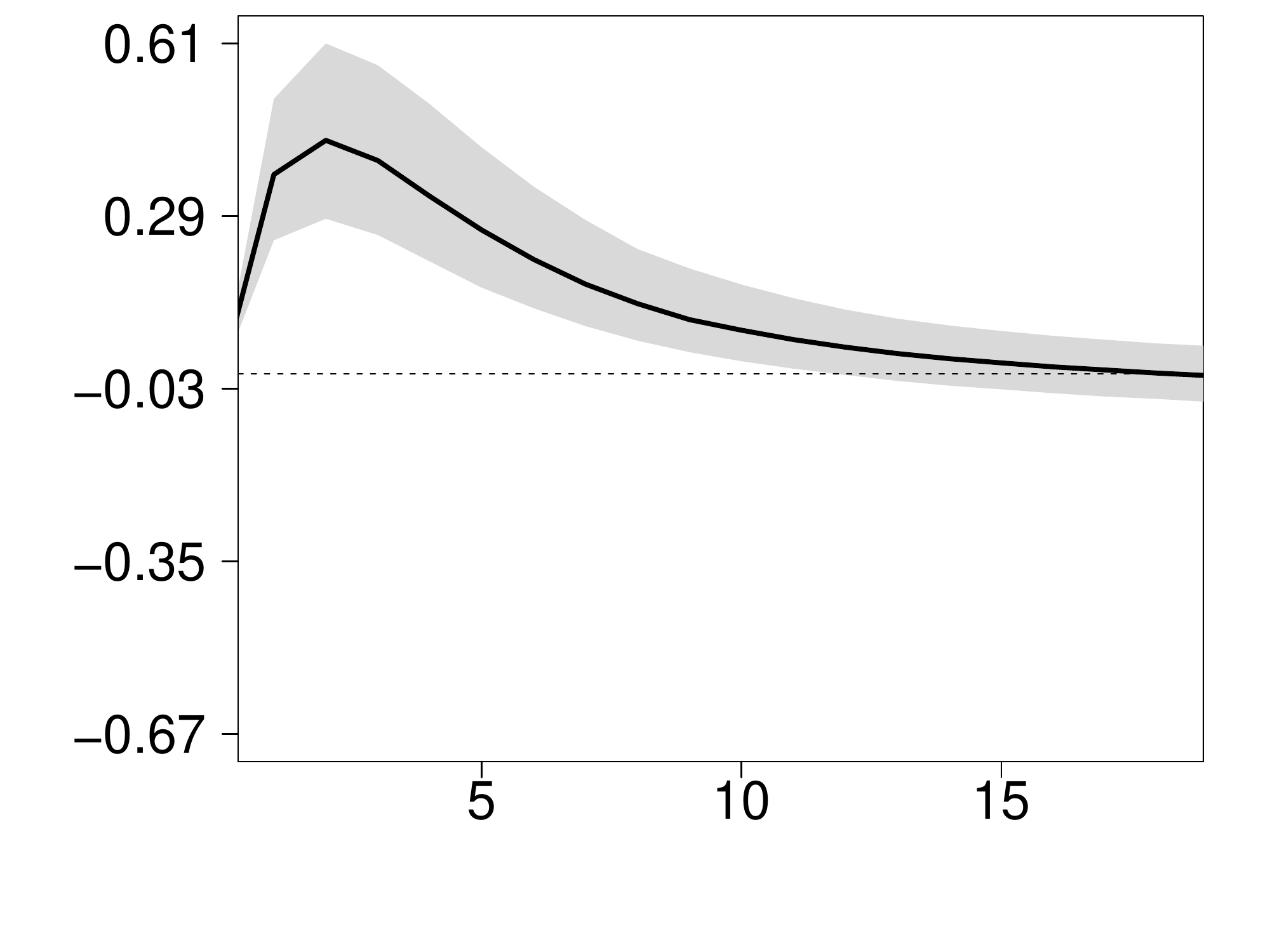}
\small Arkansas\\[0.1cm]
\includegraphics[width=\linewidth]{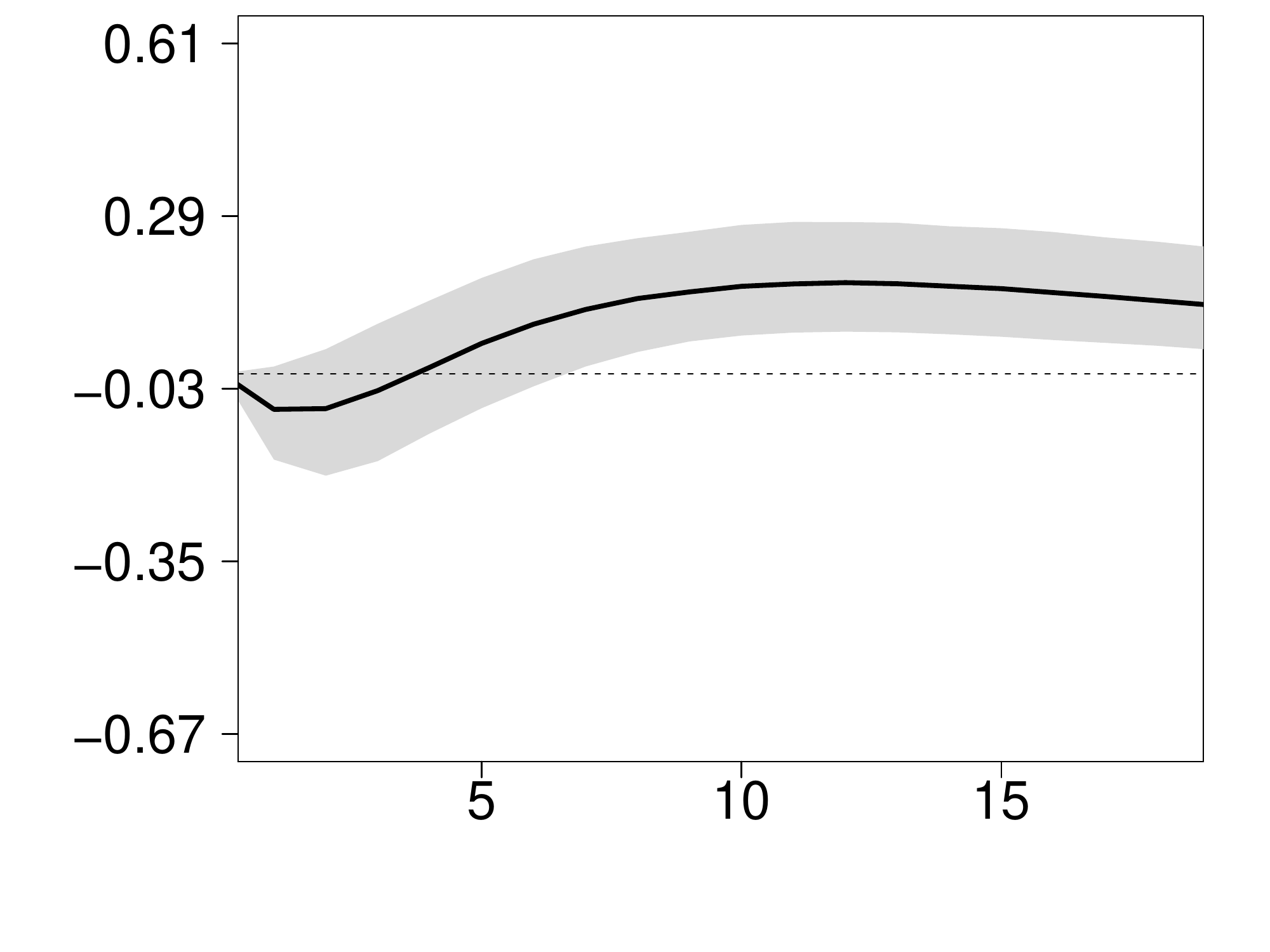}
\end{minipage}
\begin{minipage}[t]{0.24\textwidth}
\centering
\large \textit{South}\vspace{0.1cm}\hrule\vspace{0.5cm}
\small Texas\\[0.1cm]
\includegraphics[width=\linewidth]{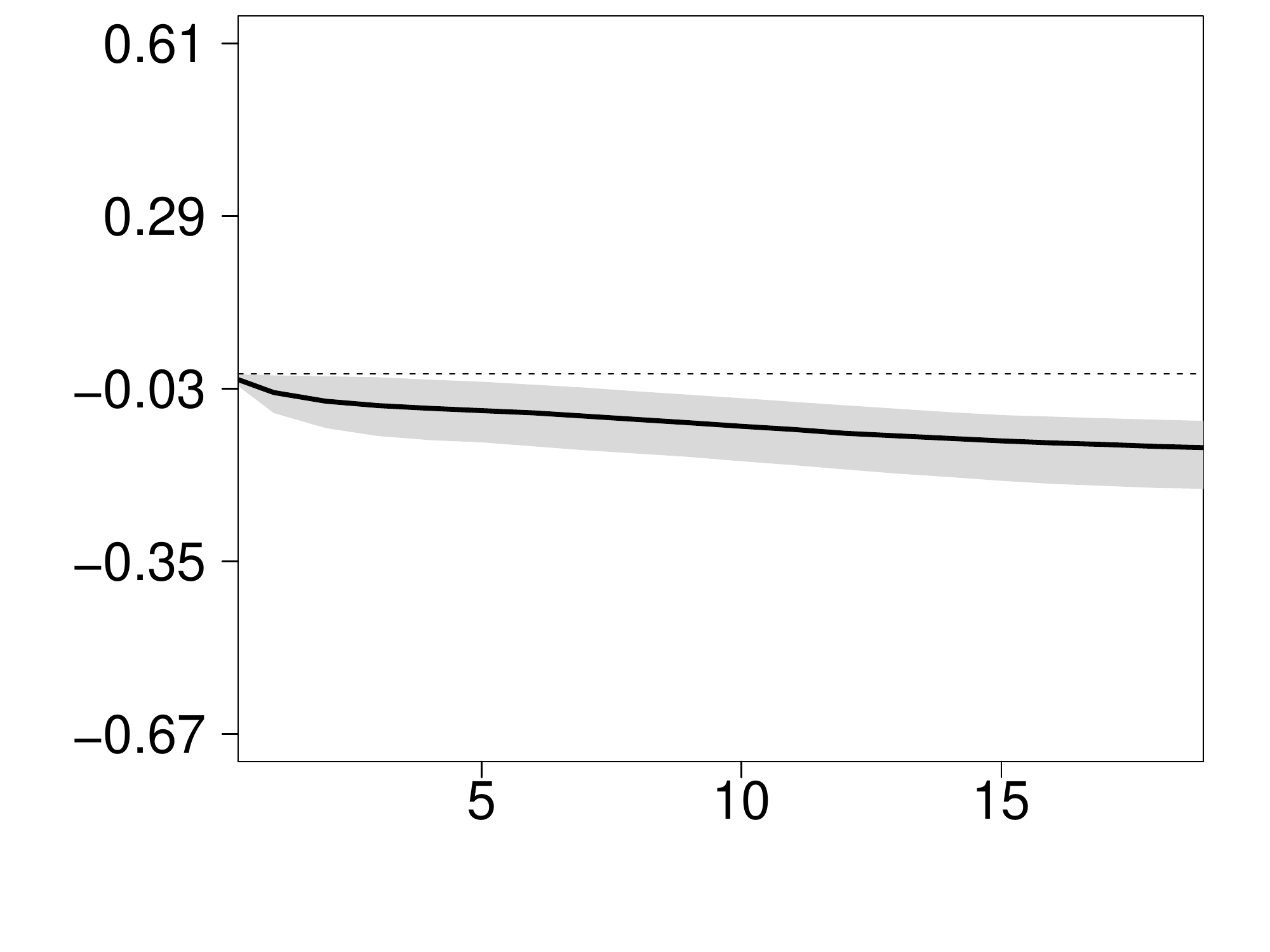}
\small Florida\\[0.1cm]
\includegraphics[width=\linewidth]{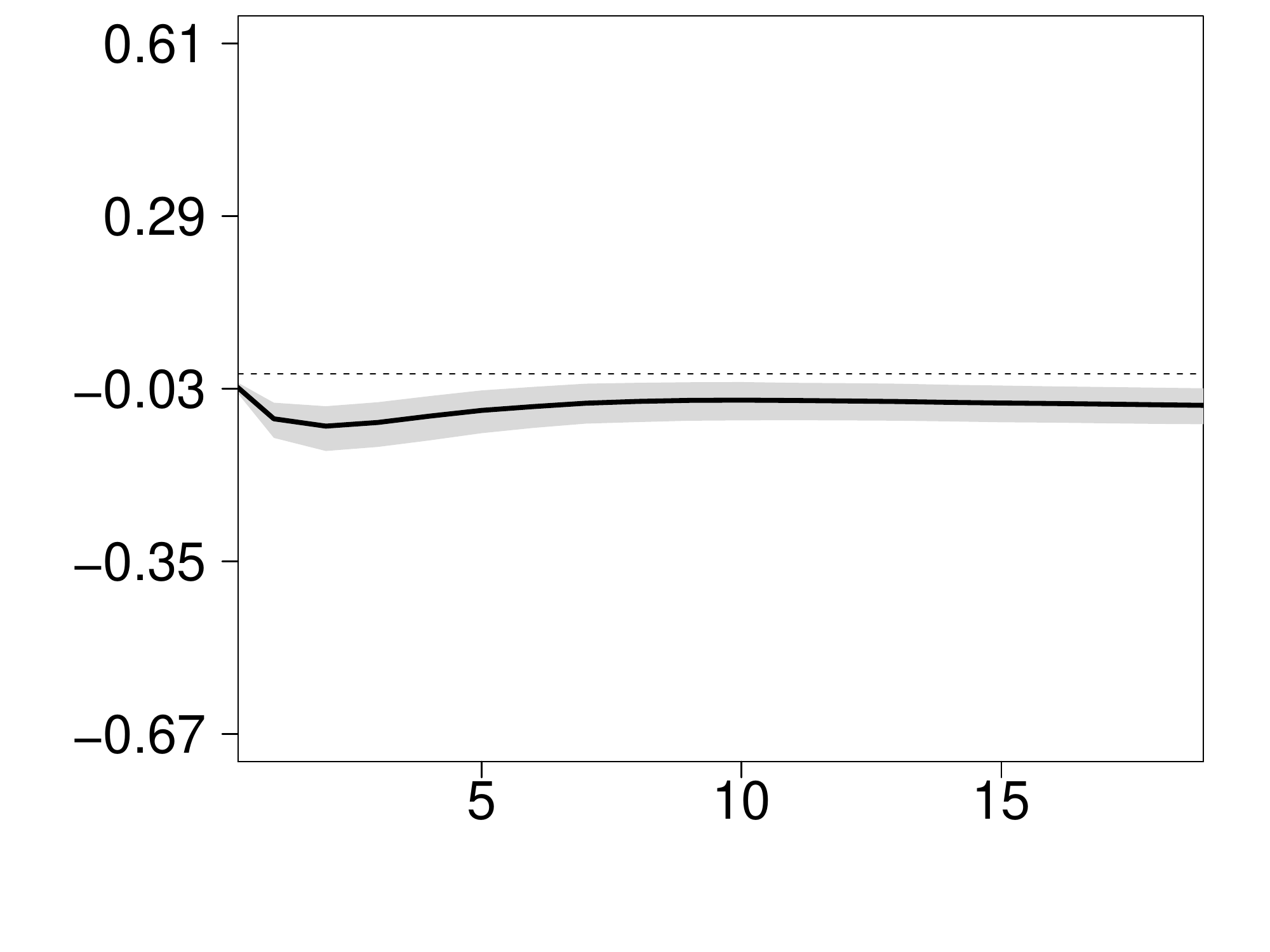}
\end{minipage}
\begin{minipage}[t]{0.24\textwidth}
\centering
\large \textit{Northeast}\vspace{0.1cm}\hrule\vspace{0.5cm}
\small New Jersey\\[0.1cm]
\includegraphics[width=\linewidth]{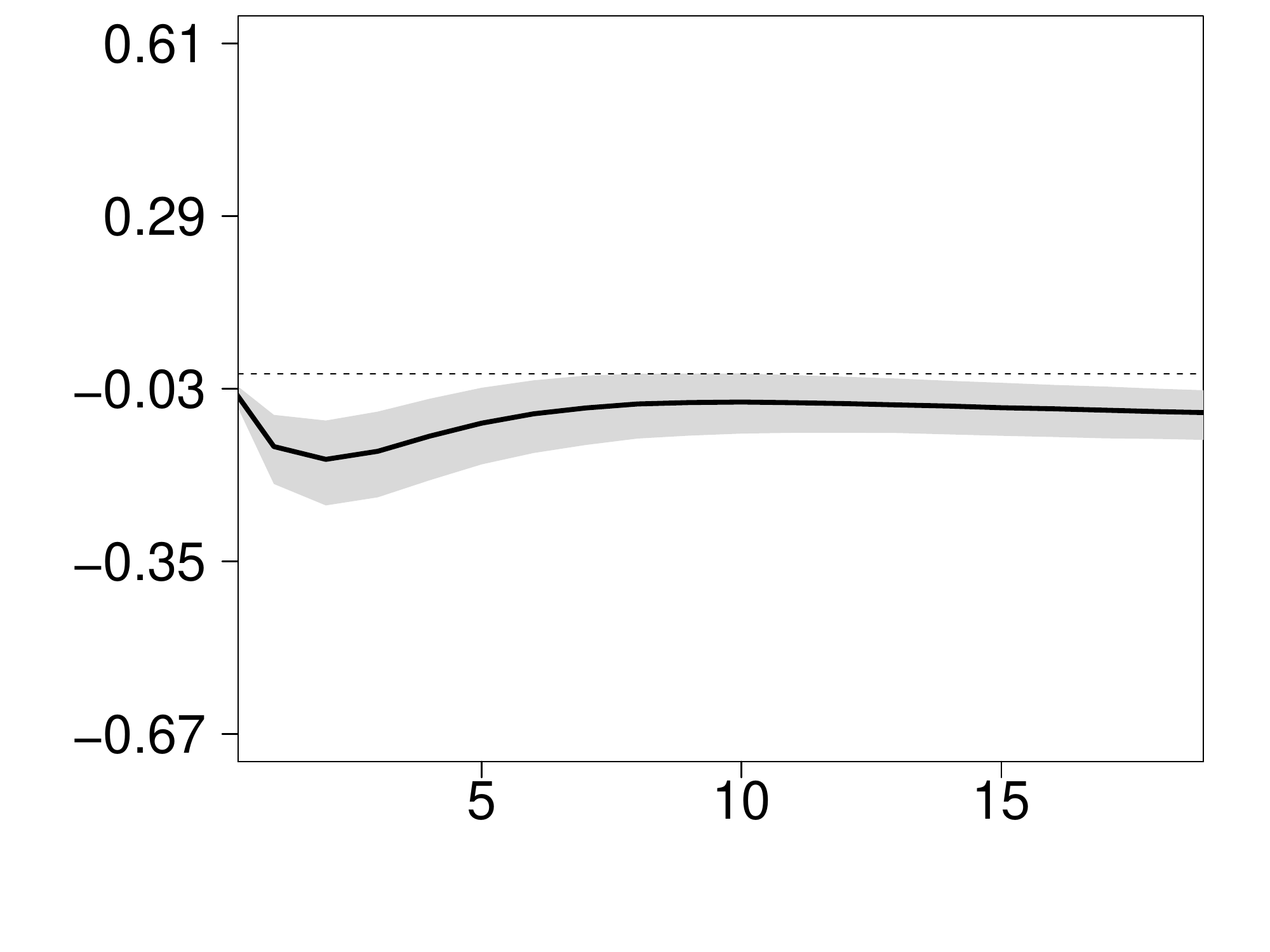}
\small Massachusetts\\[0.1cm]
\includegraphics[width=\linewidth]{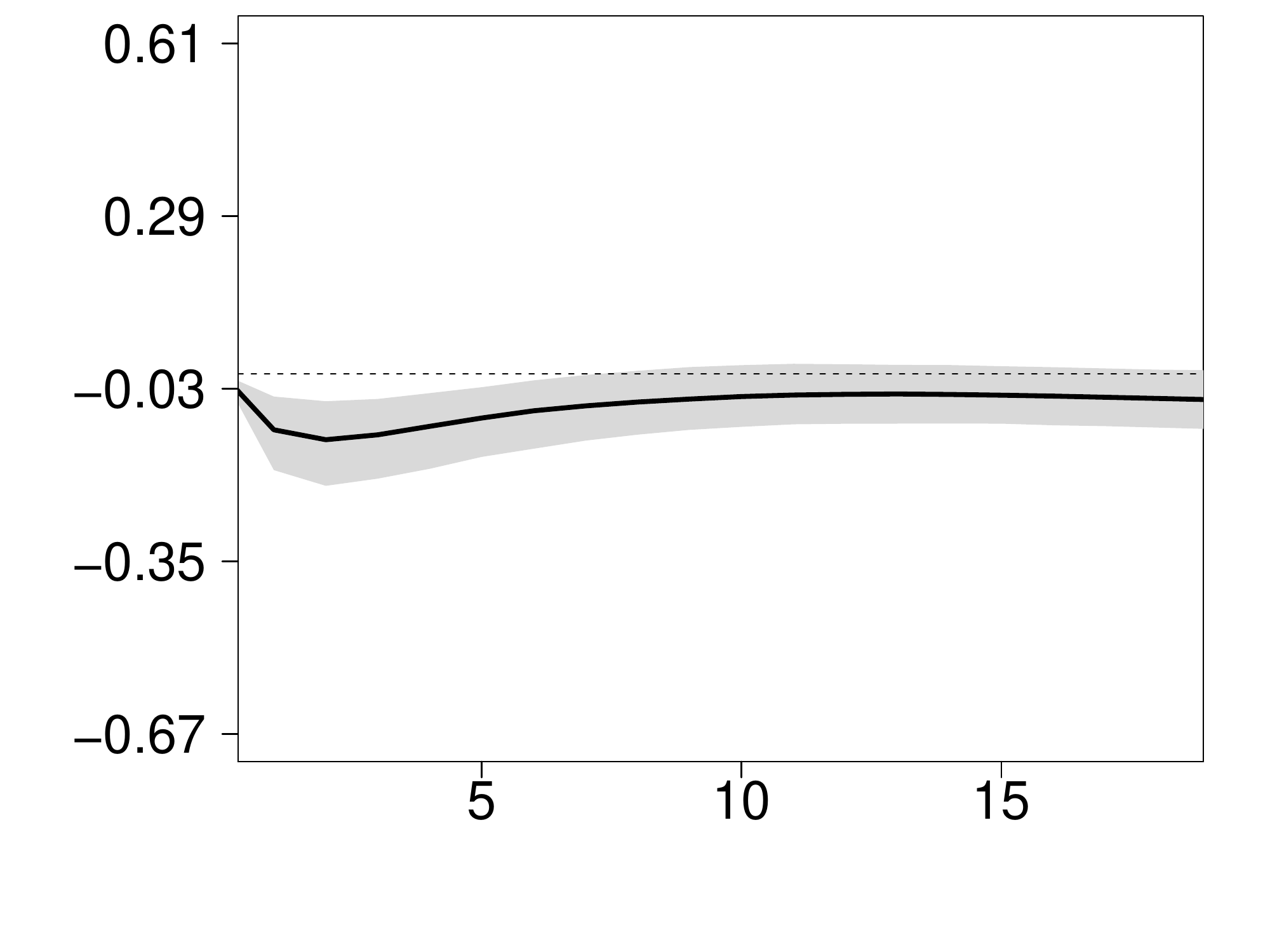}
\end{minipage}
\end{minipage}}
\begin{minipage}{16cm}\vspace{.3cm}
\footnotesize \textit{Notes}: The solid black line is the median response and the (gray) shaded area represents the 16th and 84th percentiles. The dotted line indicates the zero line. Results are based on 5,000 posterior draws. Sample period: 1985:Q1 -- 2017:Q1. Front axis: quarters after impact.
\end{minipage}%
\caption{Impulse response functions for household income inequality in US states.}\label{fig:iq_selected}
\end{figure}

To provide information on the shape as well as the statistical significance of the impulse responses, \autoref{fig:iq_selected} depicts the dynamic responses of income inequality for selected states across the US census regions West (California and New Mexico), Midwest (North Dakota and Arkansas), South (Texas and Florida), and Northeast (New Jersey and Massachusetts). Appendix \ref{app:add_results} provides information on the responses of the remaining states.

Figure \ref{fig:iq_selected} points towards pronounced heterogeneity with respect to the significance as well as the shape of the impulse response functions. Note that the majority of responses across states are statistically significant. Some of them provide evidence that inequality reacts strongly on impact with a tendency to turn insignificant after some few quarters, other responses imply that reactions of inequality on impact are insignificant but turn significant after around a year. 

Before discussing the responses for each state, it is worth emphasizing some findings that hold for a wide set of states considered and in particular the eight selected states shown in \autoref{fig:iq_selected}. In general, the results point towards marked heterogeneity with respect to the shape of the impulse response functions. While some states tend to react quickly, reaching peak responses within a few quarters, others display a somewhat slower response reaching their peak after around ten quarters. This group of states also displays a more persistent reaction that does not fade out within the five years time horizon considered. We conjecture that these differences in the time profile can be traced back to institutional factors and the income composition across states.

Turning to the state-specific reactions in \autoref{fig:iq_selected}, and starting with the West census region, increases in income inequality in the short-run are evident. In California, for instance, we observe that inequality tends to increase during the first six quarters before decreasing in the medium run (i.e. from around one year after impact). Considering the responses of income inequality in New Mexico yields a similar picture in the short run. After increasing on impact, responses reach a peak after around three quarters before slowly fading out. As compared to California, there is no evidence of a decrease in inequality in the medium run. 

Moving to the Midwest region provides different insights. For North Dakota, we find that inequality increases sharply during the first year, slowly fading out afterwards. By contrast, inequality in Arkansas displays a hump-shaped reaction, decreasing after around three to four quarters and increasing afterwards. Considering the cases of Texas and Florida, both located in the South census, suggests that inequality steadily decreases in response to uncertainty shocks. As compared to the remaining states considered, these two responses appear to be highly persistent and show no tendency to level out after around three years.

Finally, inequality in New Jersey and Massachusetts exhibits  a hump-shaped behavior that suggests decreasing levels of inequality that slowly die out after around two to three years. In sum, we find that for five out of eight states inequality shows a tendency to decline in response to uncertainty shocks. For the other three states considered -- two of them located in the West region -- we observe that inequality rises as a reaction to increases in uncertainty. As stated above, we conjecture that the mixed reactions across states can be explained by the specific composition of income. States that display declining inequality levels appear to be characterized by higher levels of capital income, whereas states that show rising inequality levels are generally accompanied by lower levels of capital and higher shares of labor income.

\subsection{The role of uncertainty shocks in explaining income inequality}\label{sec:fevd}
\begin{figure}[!ht]
\fbox{\begin{minipage}[t]{\textwidth}
\centering
\begin{minipage}[t]{.24\textwidth}
\centering
\large \textit{West}\vspace{0.1cm}\hrule\vspace{0.5cm}
\small California\\[0.1cm]
\includegraphics[width=\linewidth]{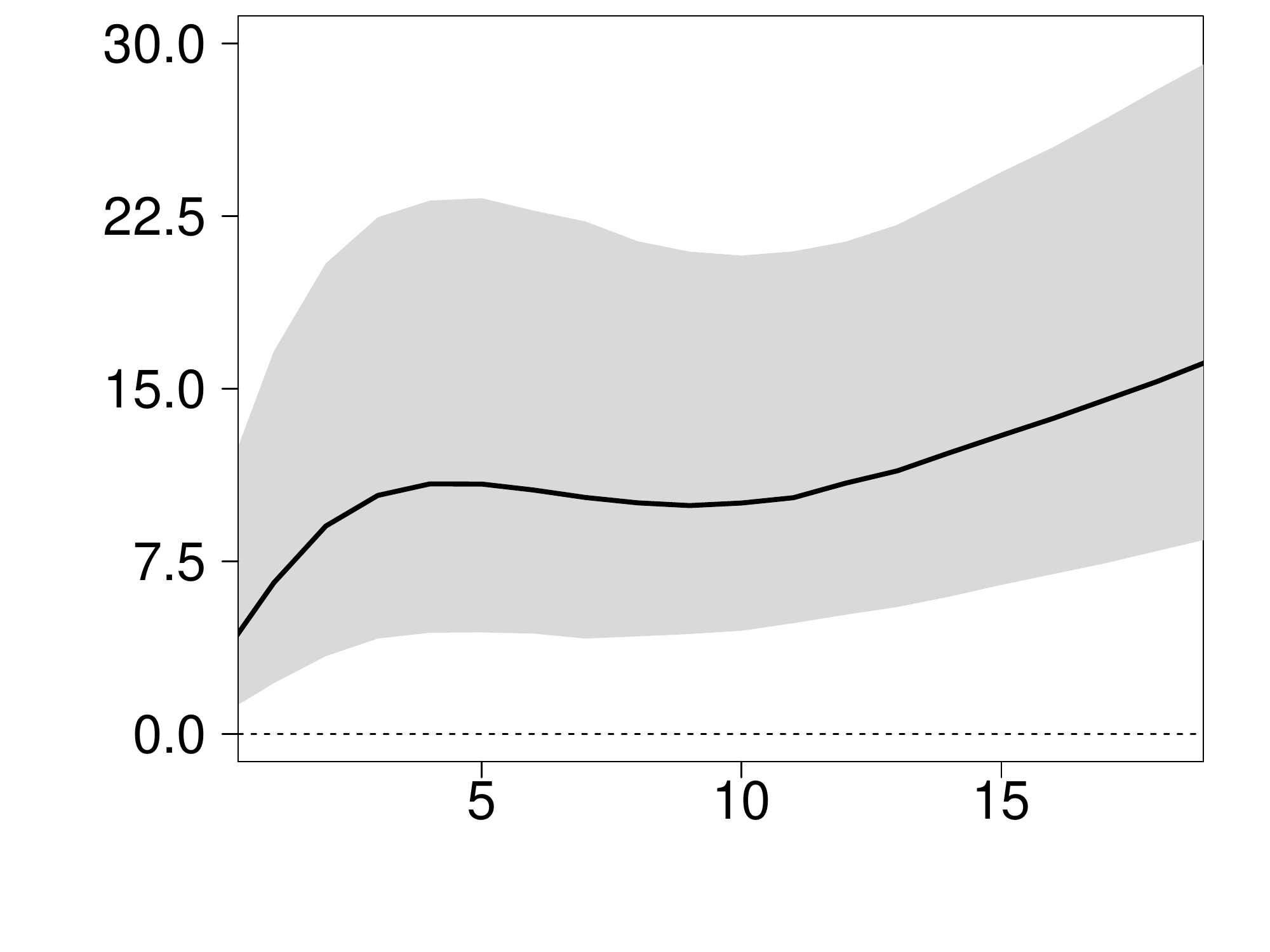}
\small New Mexico\\[0.1cm]
\includegraphics[width=\linewidth]{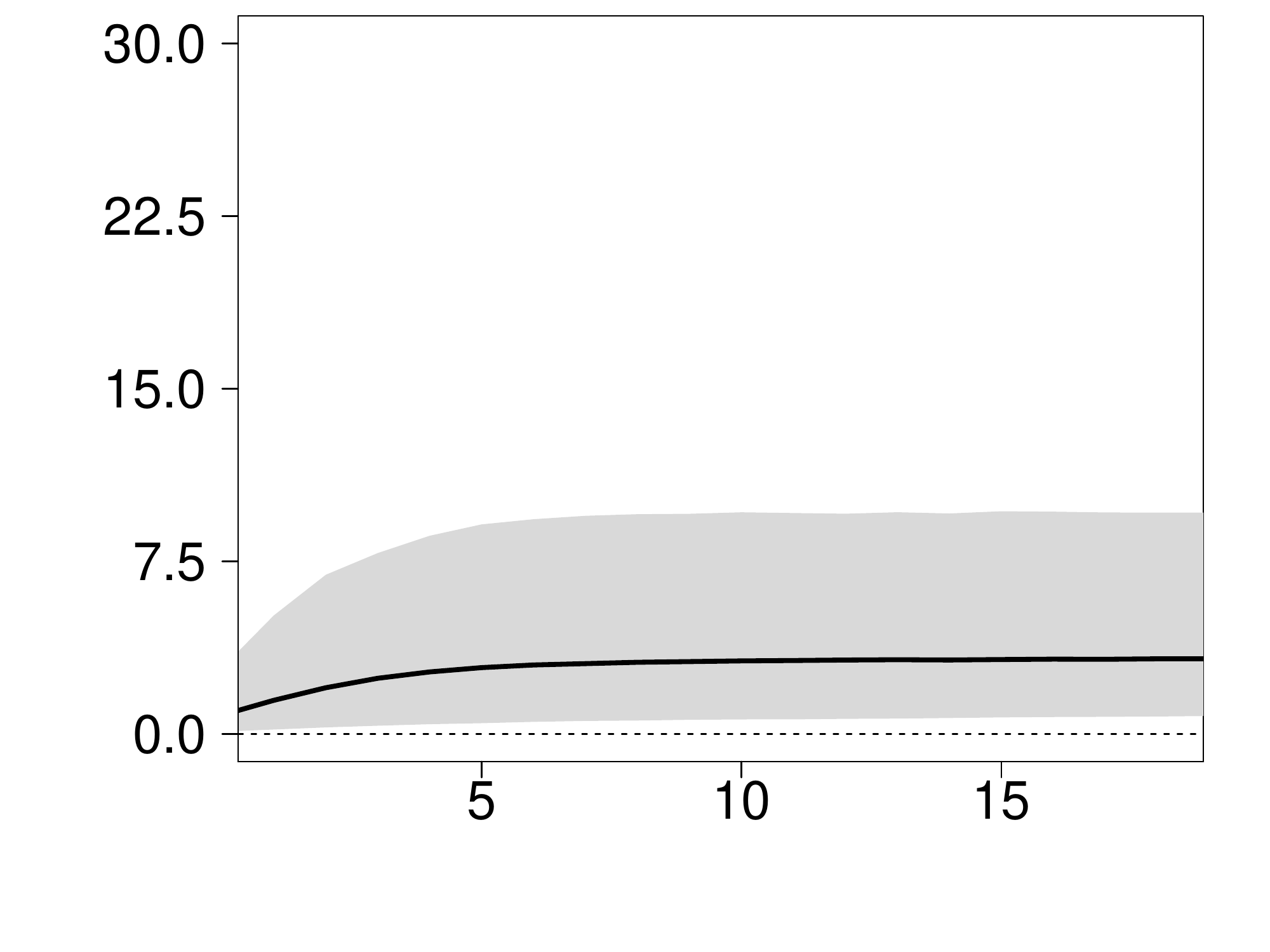}
\end{minipage}
\begin{minipage}[t]{0.24\textwidth}
\centering 
\large \textit{Midwest}\vspace{0.1cm}\hrule\vspace{0.5cm}
\small North Dakota\\[0.1cm]
\includegraphics[width=\linewidth]{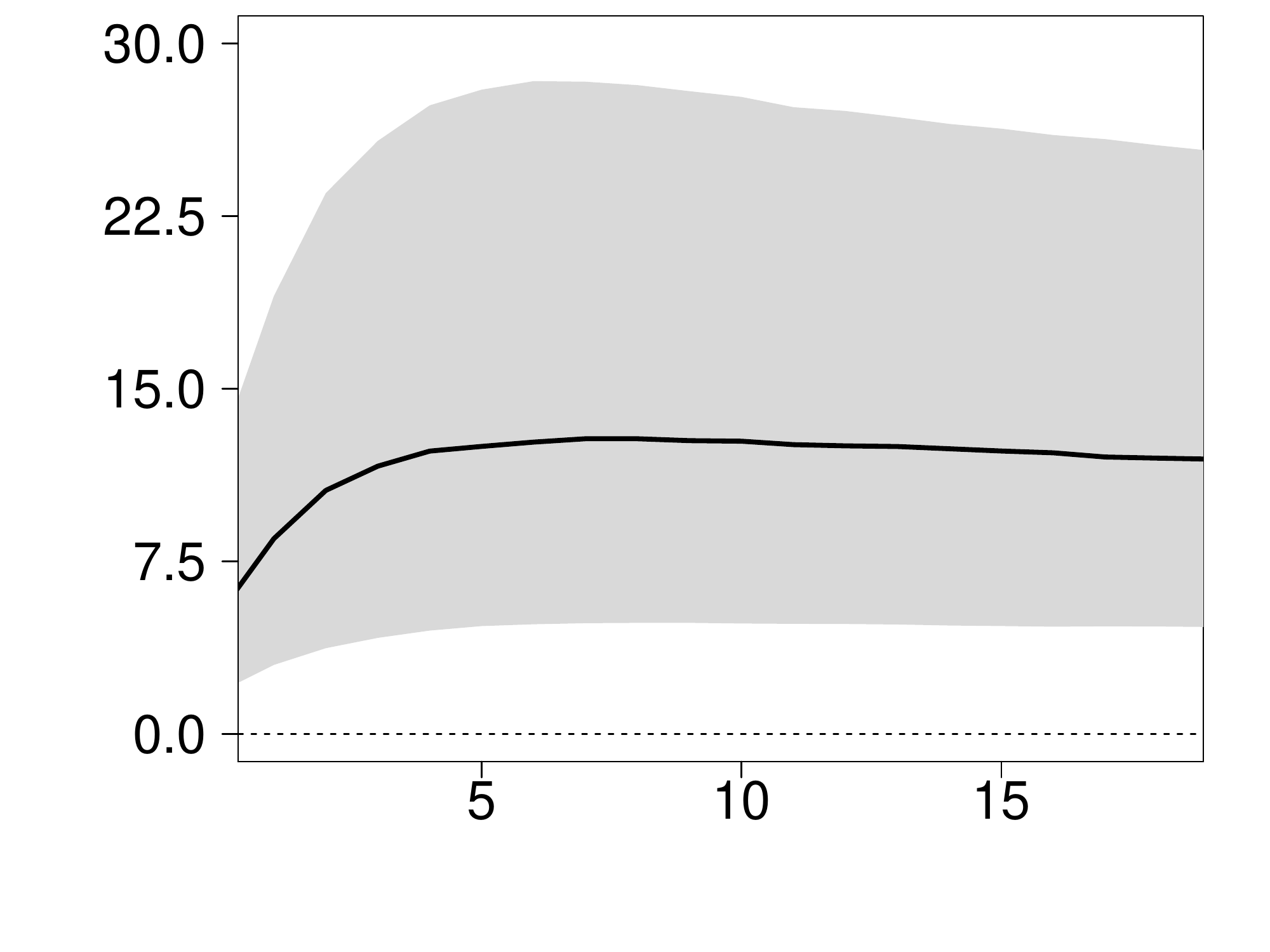}
\small Arkansas\\[0.1cm]
\includegraphics[width=\linewidth]{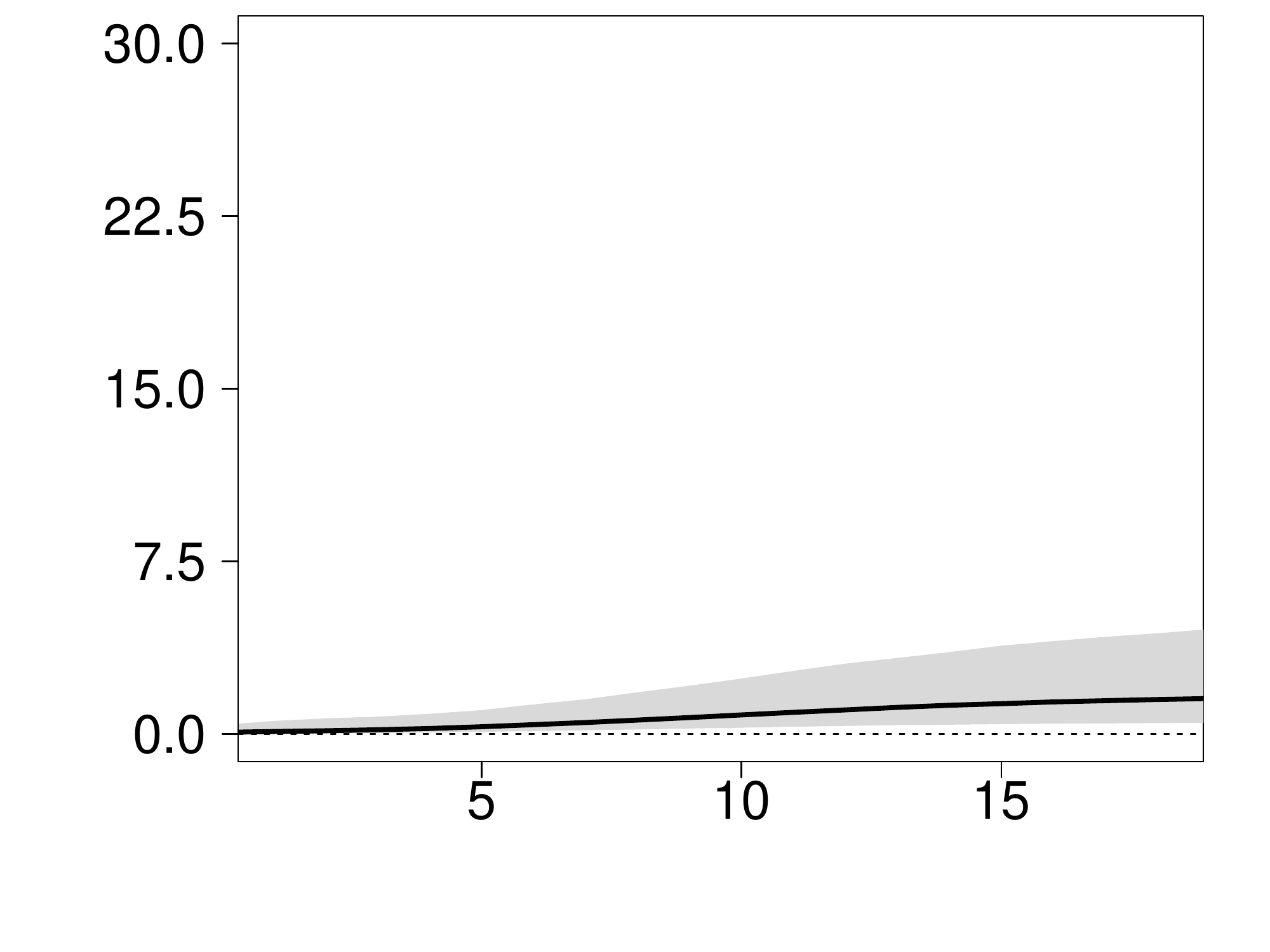}
\end{minipage}
\begin{minipage}[t]{0.24\textwidth}
\centering
\large \textit{South}\vspace{0.1cm}\hrule\vspace{0.5cm}
\small Texas\\[0.1cm]
\includegraphics[width=\linewidth]{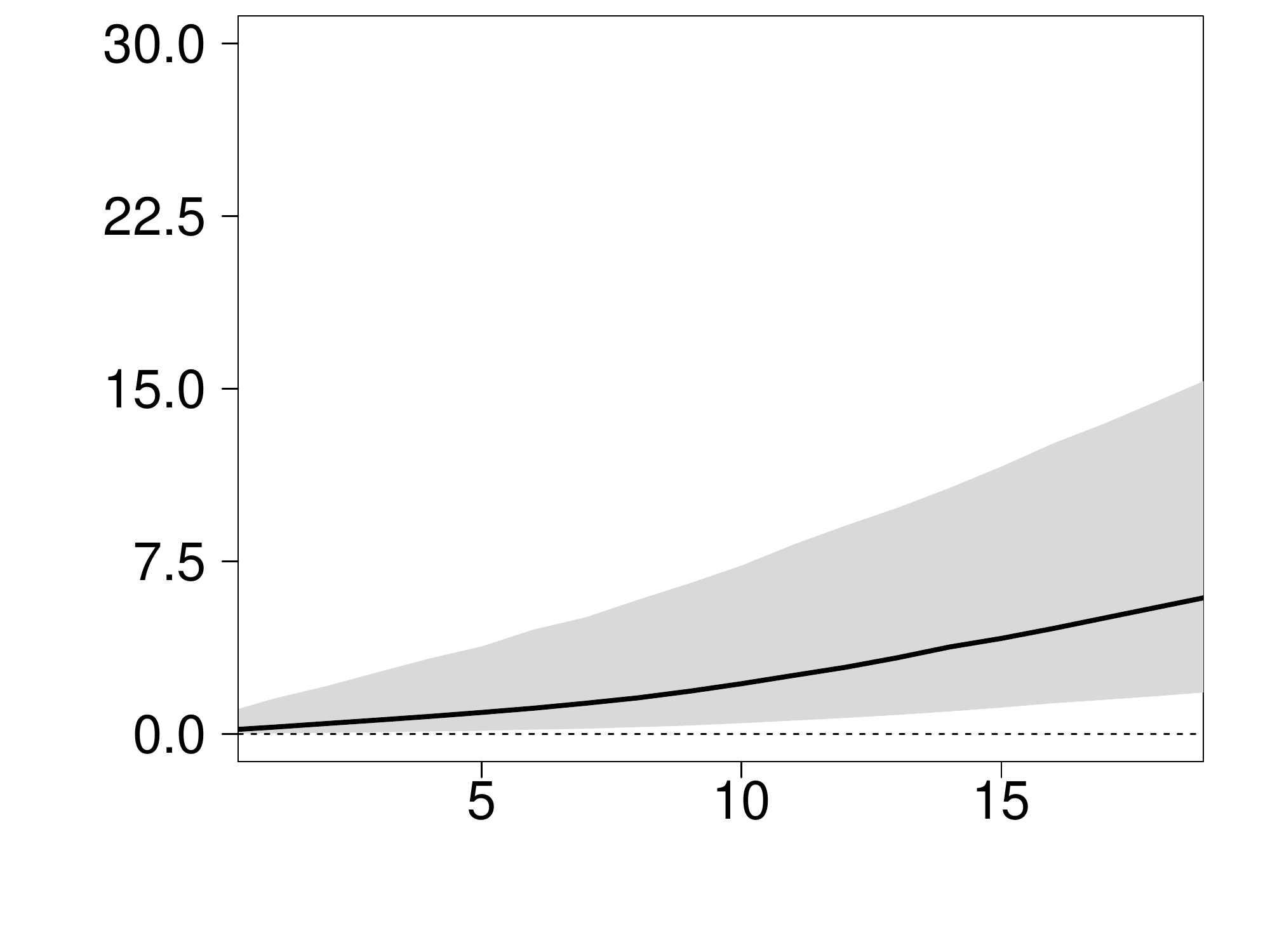}
\small Florida\\[0.1cm]
\includegraphics[width=\linewidth]{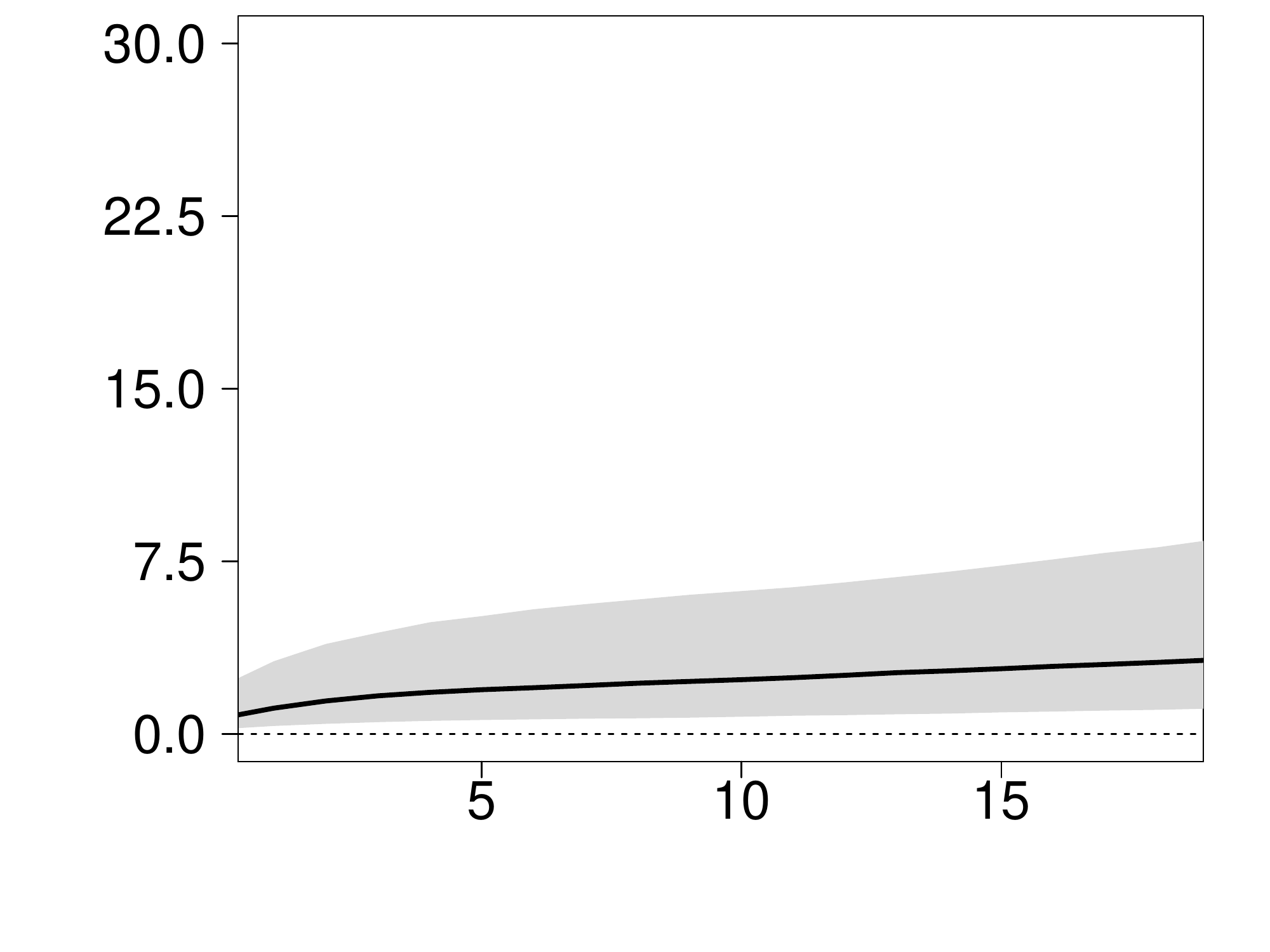}
\end{minipage}
\begin{minipage}[t]{0.24\textwidth}
\centering
\large \textit{Northeast}\vspace{0.1cm}\hrule\vspace{0.5cm}
\small New Jersey\\[0.1cm]
\includegraphics[width=\linewidth]{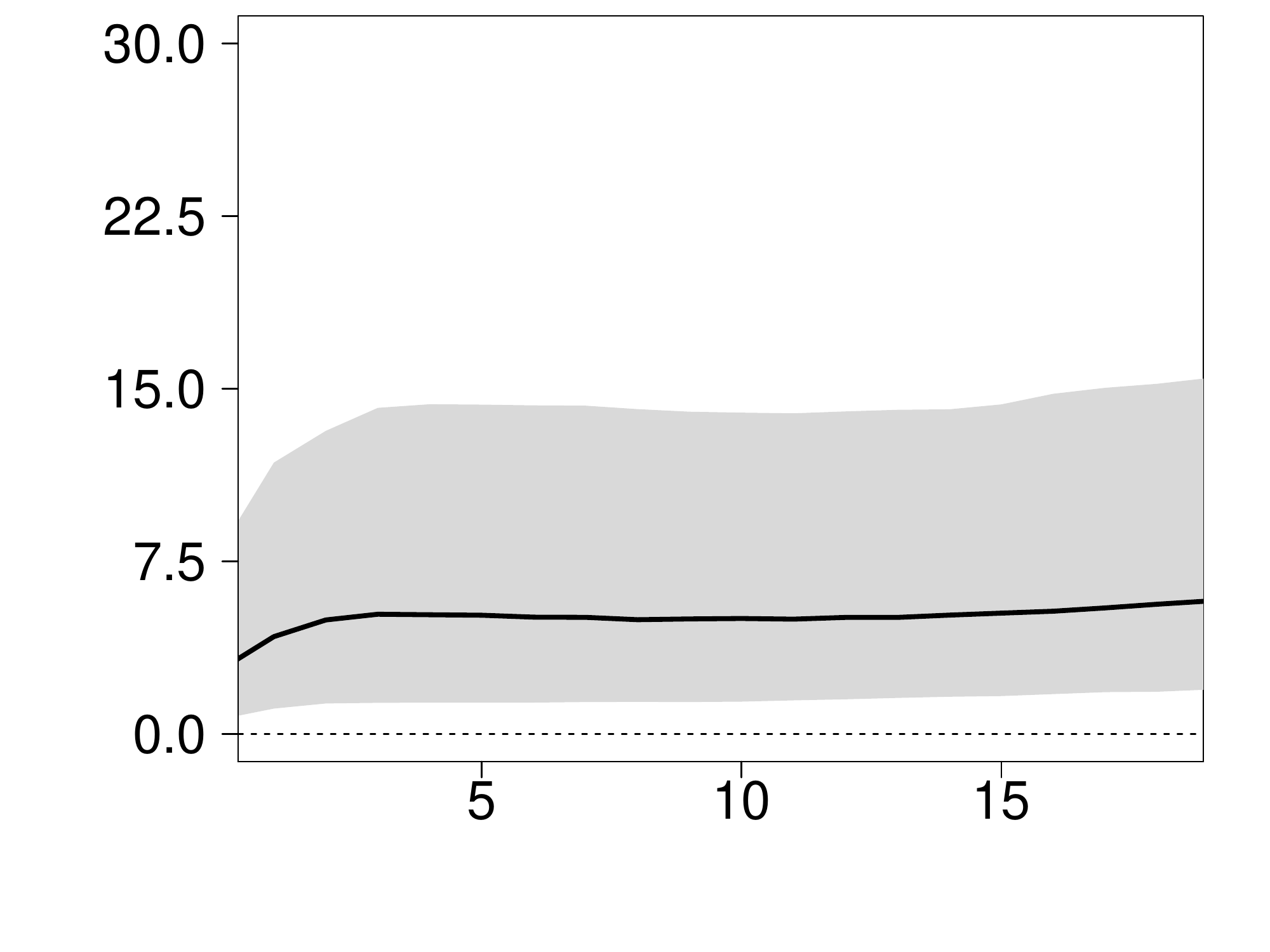}
\small Massachusetts\\[0.1cm]
\includegraphics[width=\linewidth]{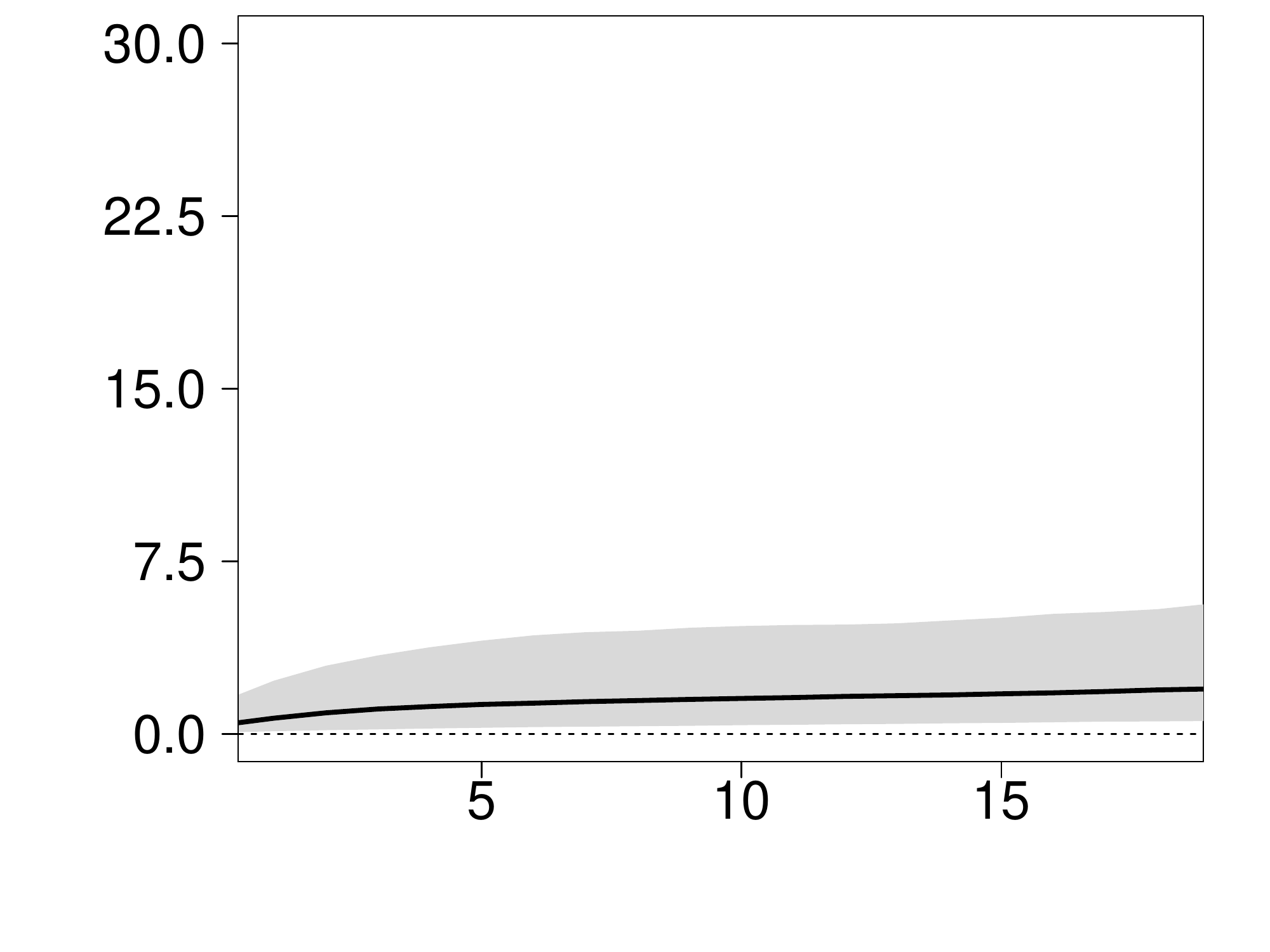}
\end{minipage}
\end{minipage}}
\begin{minipage}{16cm}\vspace{.3cm}
\footnotesize \textit{Notes}: The solid black line is the median and the (gray) shaded area represents the 16th and 84th percentiles. The dotted line indicates the zero line. Results are based on 5,000 posterior draws. Sample period: 1985:Q1 -- 2017:Q1. Front axis: quarters after impact.
\end{minipage}%
\caption{Forecast error variance decompositions for inequality in selected US states.}\label{fig:FEVD_selected}
\end{figure}

While the previous two sections aimed at establishing a dynamic relationship between income inequality and uncertainty, this section centers on assessing the quantitative contribution of uncertainty shocks to movements in the state-level income dispersion between households. For this purpose, we consider the share of forecast error variance of income inequality explained by the uncertainty shock in \cref{fig:FEVD_selected}.

Consistent with the findings for the impulse responses, pronounced differences across states are visible. For the eight selected states, we find that the quantitative contribution of the uncertainty shock in explaining the forecast error variance of income inequality ranges from around three percent in New Mexico and Arkansas at the one-step-ahead horizon to around 7.5 percent in New Jersey for the three-year-ahead horizon.  Notice that the share of variance explained increases for most states under consideration. For some states, the slope appears much steeper during the first few quarters (see, e.g., North Dakota) whereas for other states, the FEVDs appear to be increasing at a steady rate. 

\section{Inspecting the transmission mechanism}\label{sec:transmission}
In this section, we first discuss the reactions of the further state-level specific quantities (employment, unemployment and total personal income per capita) and assess whether there exist significant relations to the reactions of income inequality in Section \ref{sec:responsesdetail}. In a second step, we aim at explaining state-level differences in responses of income inequality using a set of additional state-level variables in Section \ref{sec:explanation}.

\subsection{Responses of state-level quantities}\label{sec:responsesdetail}
The discussion of what drives movements in inequality has mostly been based on theoretical economic reasoning up to this point. In the following, we investigate the driving forces of changes in income inequality within our model framework. Again, we focus attention on the set of eight states considered in the previous section. 

\begin{figure}[!t]
\fbox{\begin{minipage}[t]{\textwidth}
\centering
\begin{minipage}[t]{.24\textwidth}
\centering
\large \textit{West}\vspace{0.1cm}\hrule\vspace{0.5cm}
\small California\\[0.1cm]
\includegraphics[width=\linewidth]{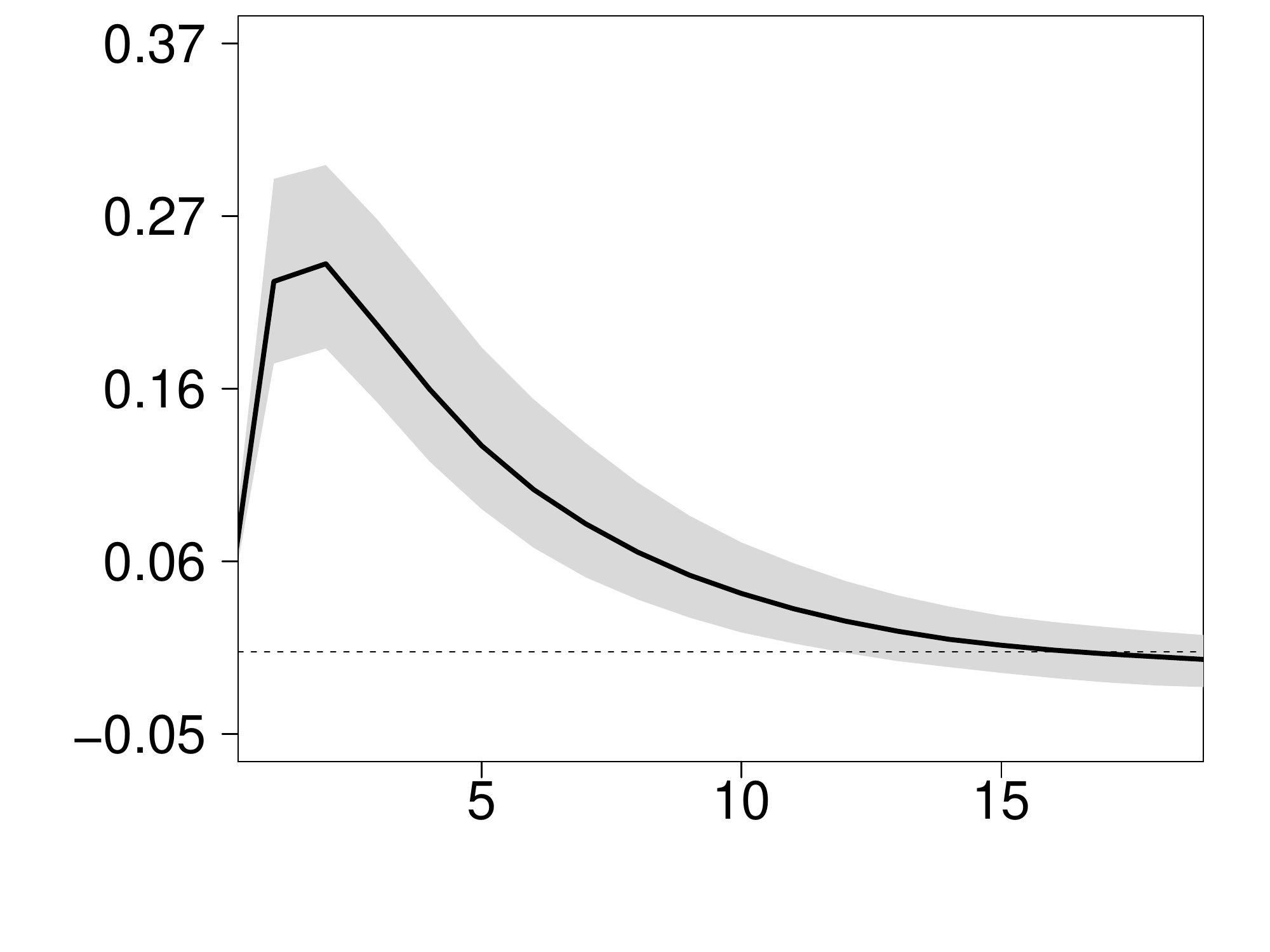}
\small New Mexico\\[0.1cm]
\includegraphics[width=\linewidth]{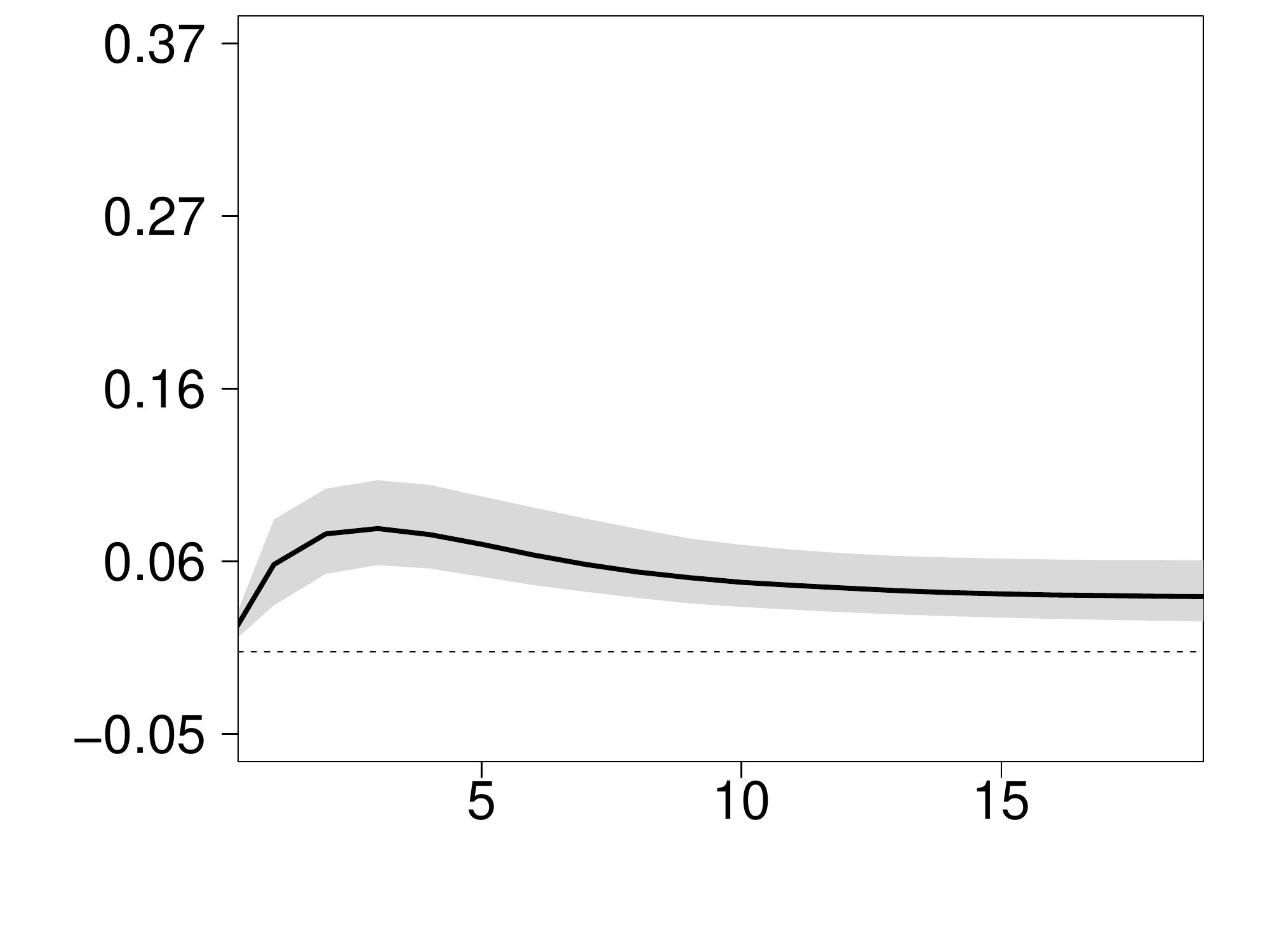}
\end{minipage}
\begin{minipage}[t]{0.24\textwidth}
\centering 
\large \textit{Midwest}\vspace{0.1cm}\hrule\vspace{0.5cm}
\small North Dakota\\[0.1cm]
\includegraphics[width=\linewidth]{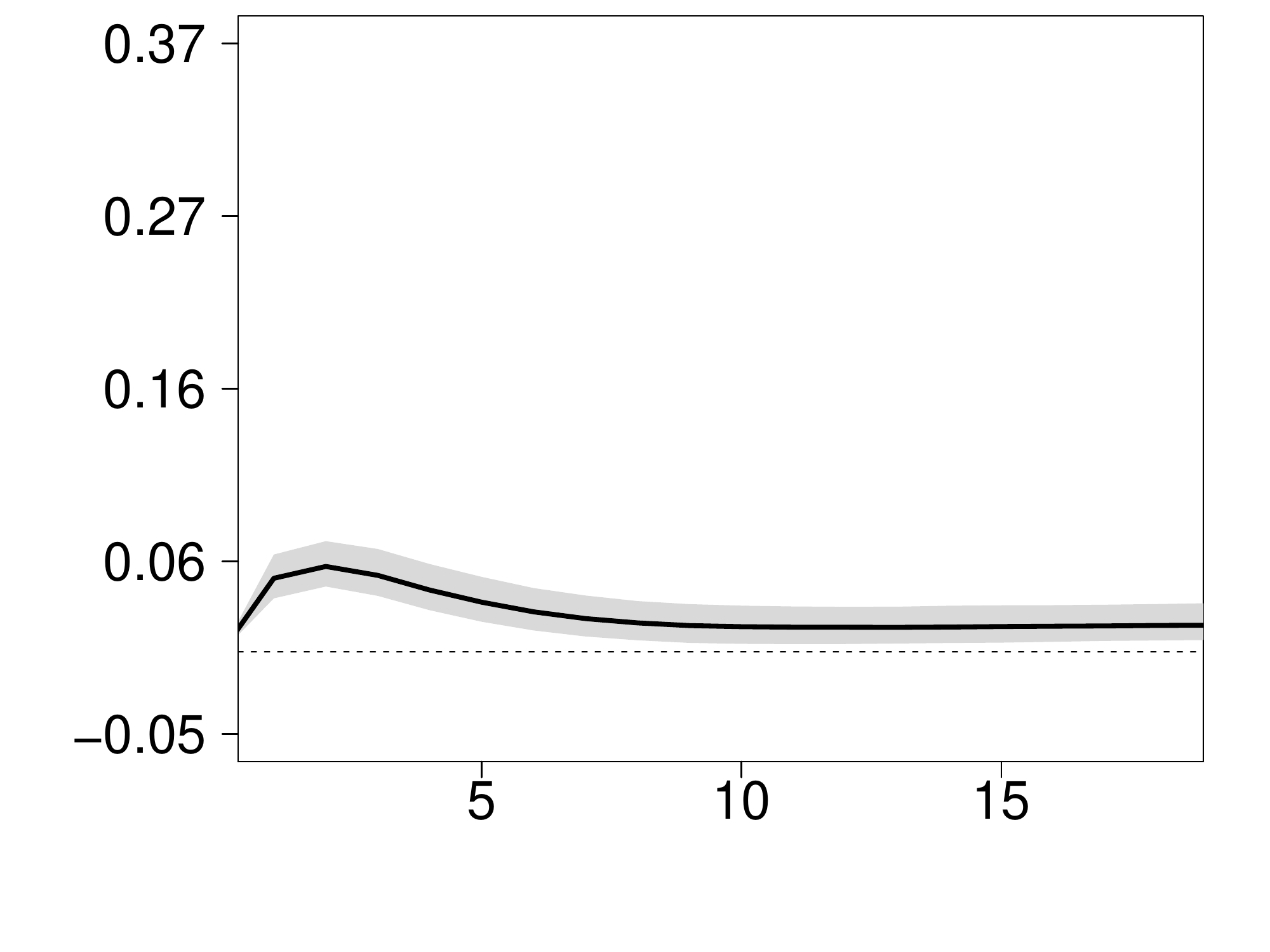}
\small Arkansas\\[0.1cm]
\includegraphics[width=\linewidth]{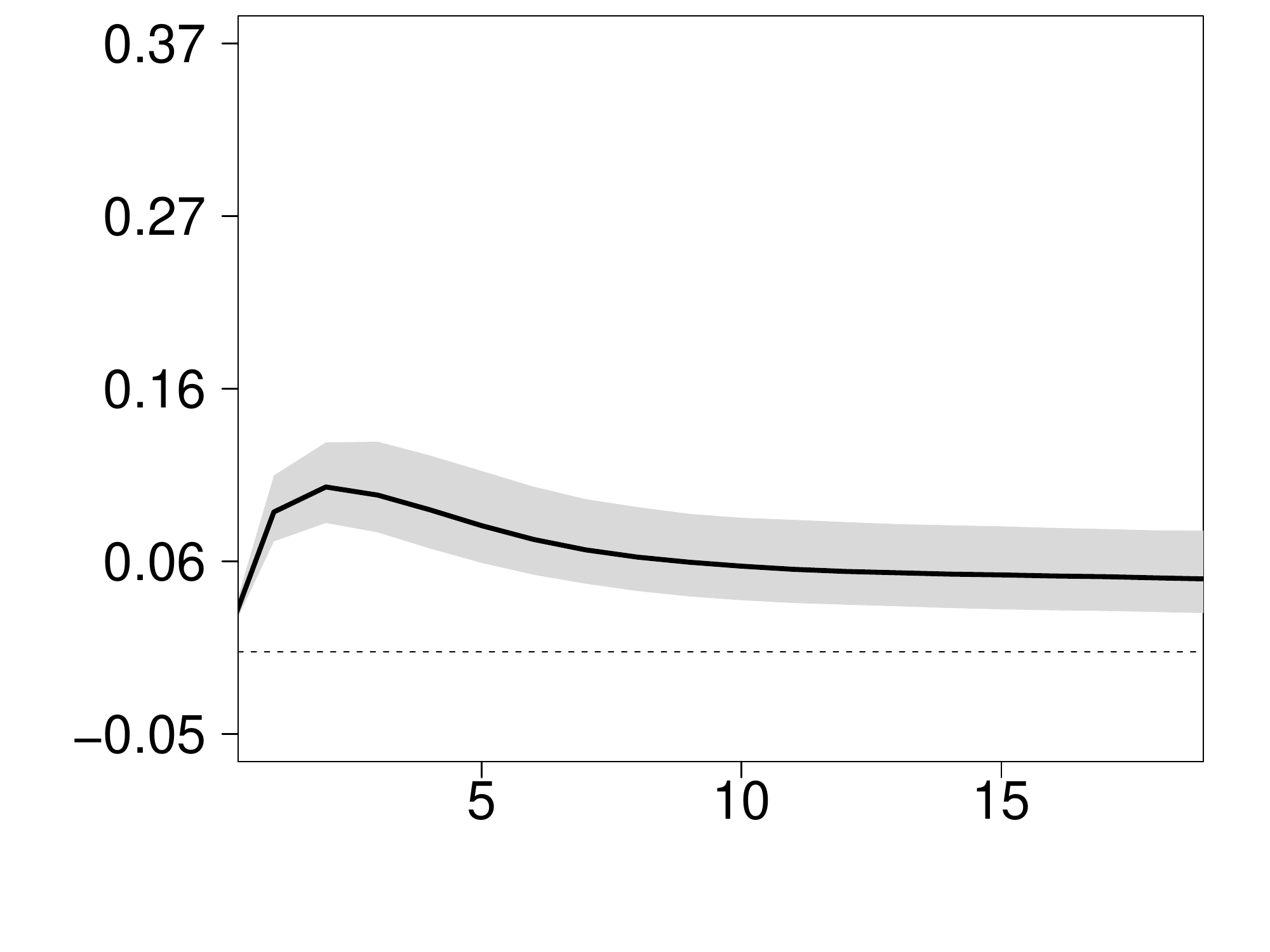}
\end{minipage}
\begin{minipage}[t]{0.24\textwidth}
\centering
\large \textit{South}\vspace{0.1cm}\hrule\vspace{0.5cm}
\small Texas\\[0.1cm]
\includegraphics[width=\linewidth]{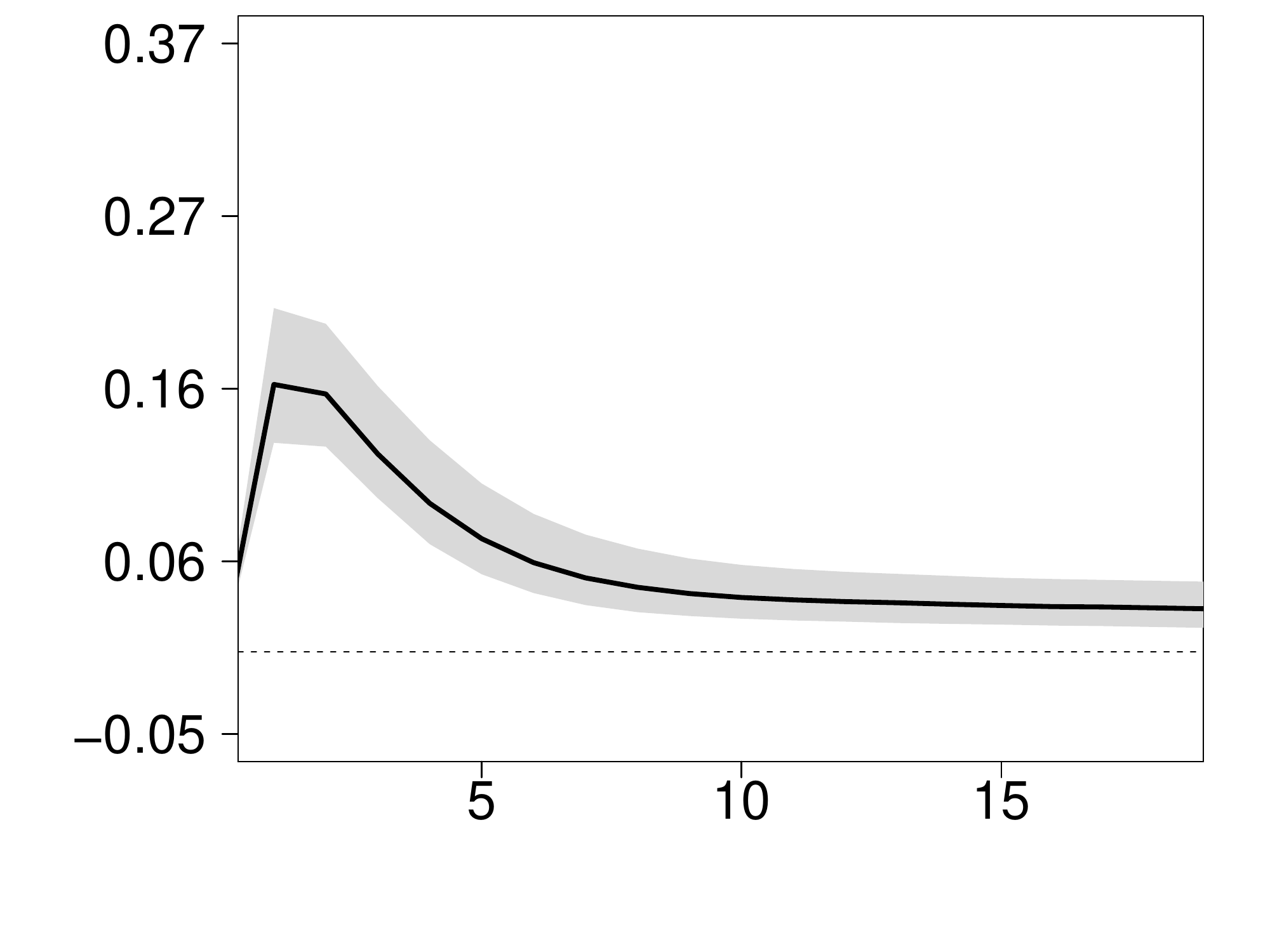}
\small Florida\\[0.1cm]
\includegraphics[width=\linewidth]{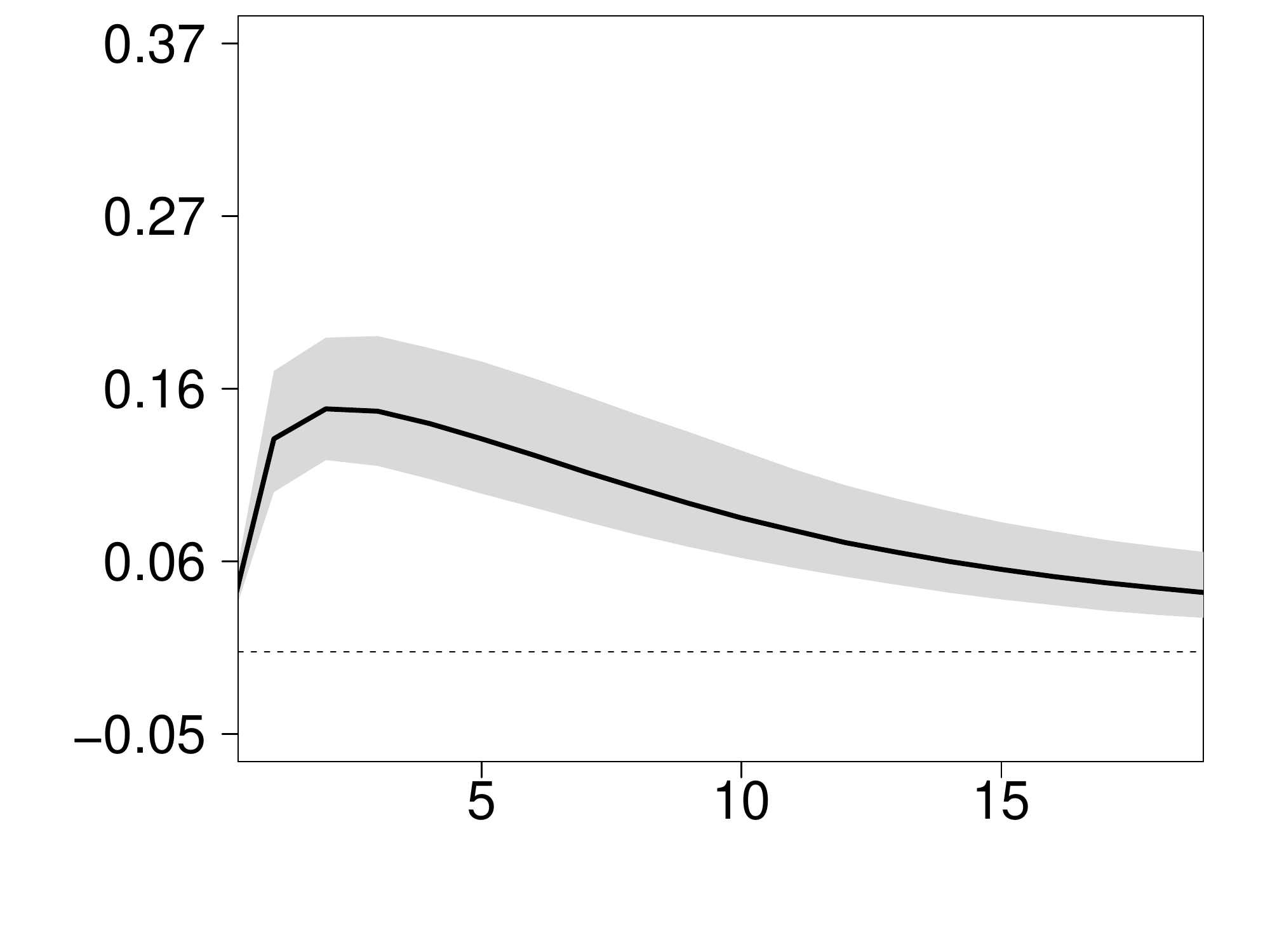}
\end{minipage}
\begin{minipage}[t]{0.24\textwidth}
\centering
\large \textit{Northeast}\vspace{0.1cm}\hrule\vspace{0.5cm}
\small New Jersey\\[0.1cm]
\includegraphics[width=\linewidth]{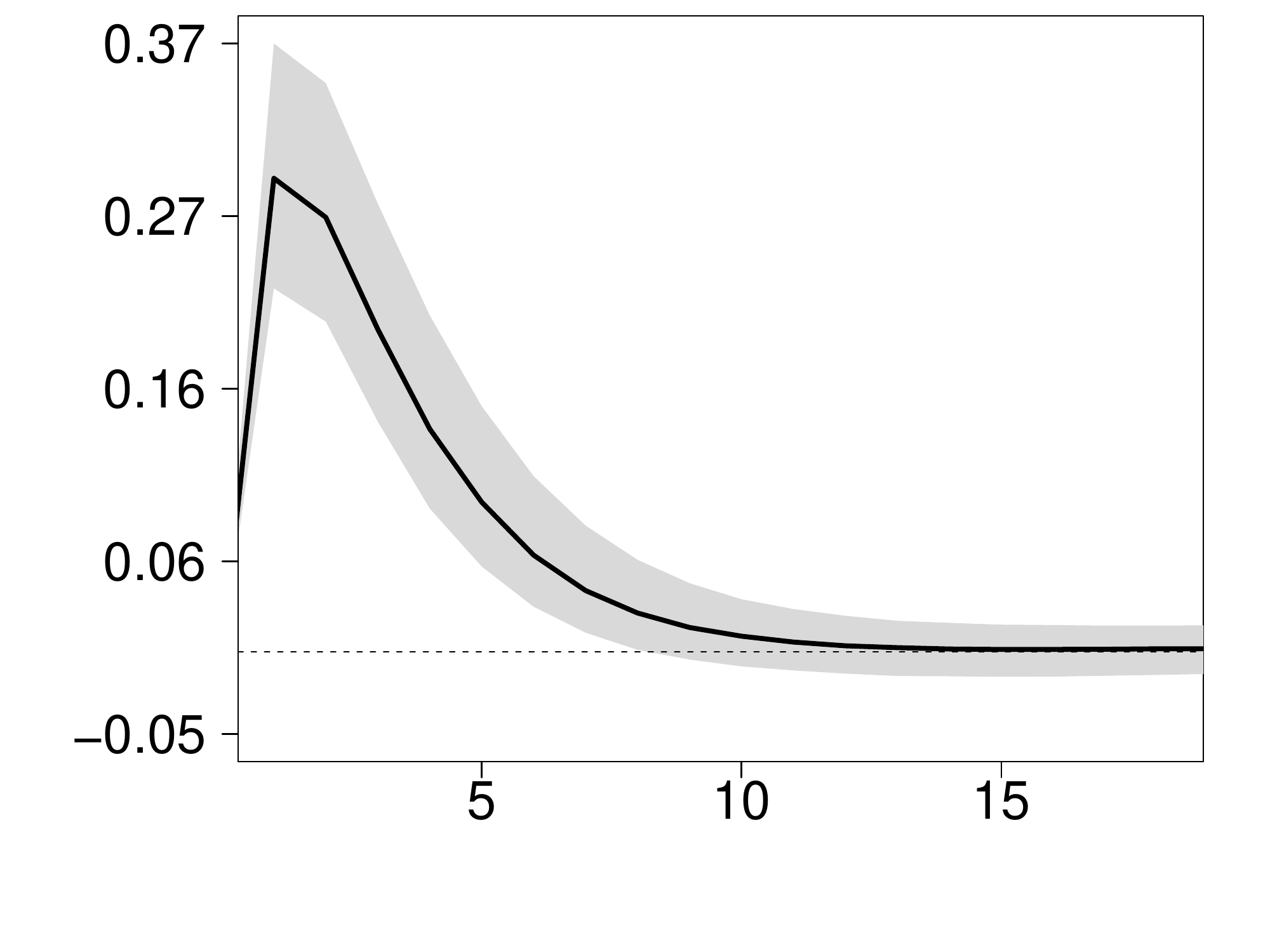}
\small Massachusetts\\[0.1cm]
\includegraphics[width=\linewidth]{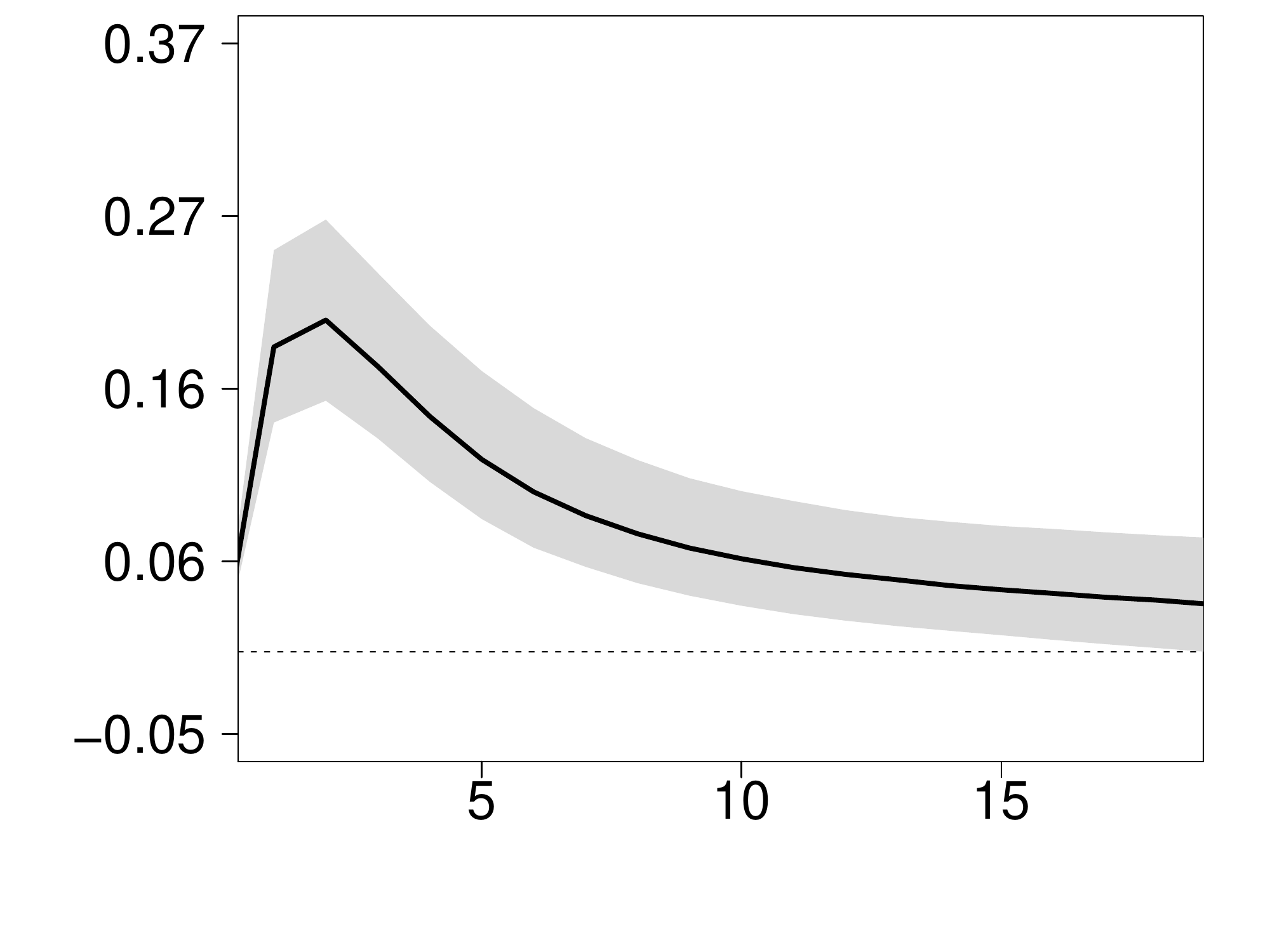}
\end{minipage}
\end{minipage}}
\begin{minipage}{16cm}\vspace{.3cm}
\footnotesize \textit{Notes}: The solid black line is the median response and the (gray) shaded area represents the 16th and 84th percentiles. The dotted line indicates the zero line. Results are based on 5,000 posterior draws. Sample period: 1985:Q1 -- 2017:Q1. Front axis: quarters after impact.
\end{minipage}%
\caption{Impulse response functions for unemployment in selected US states.}\label{fig:UN_selected}
\end{figure}

Starting with the responses of unemployment across states (see \autoref{fig:UN_selected}), we find that for all eight states under scrutiny, unemployment increases and tends to peak after the first three to four quarters. States located in the West and Midwest census appear to display similar magnitudes in their responses. High income states like California, New Jersey, and Massachusetts display much stronger unemployment responses to uncertainty shocks. In general, linking these findings to the responses of income inequality appears to be difficult. States that show strong responses (either positive or negative) tend to feature comparatively weaker responses of unemployment. 
\begin{figure}[!ht]
\fbox{\begin{minipage}[t]{\textwidth}
\centering
\begin{minipage}[t]{.24\textwidth}
\centering
\large \textit{West}\vspace{0.1cm}\hrule\vspace{0.5cm}
\small California\\[0.1cm]
\includegraphics[width=\linewidth]{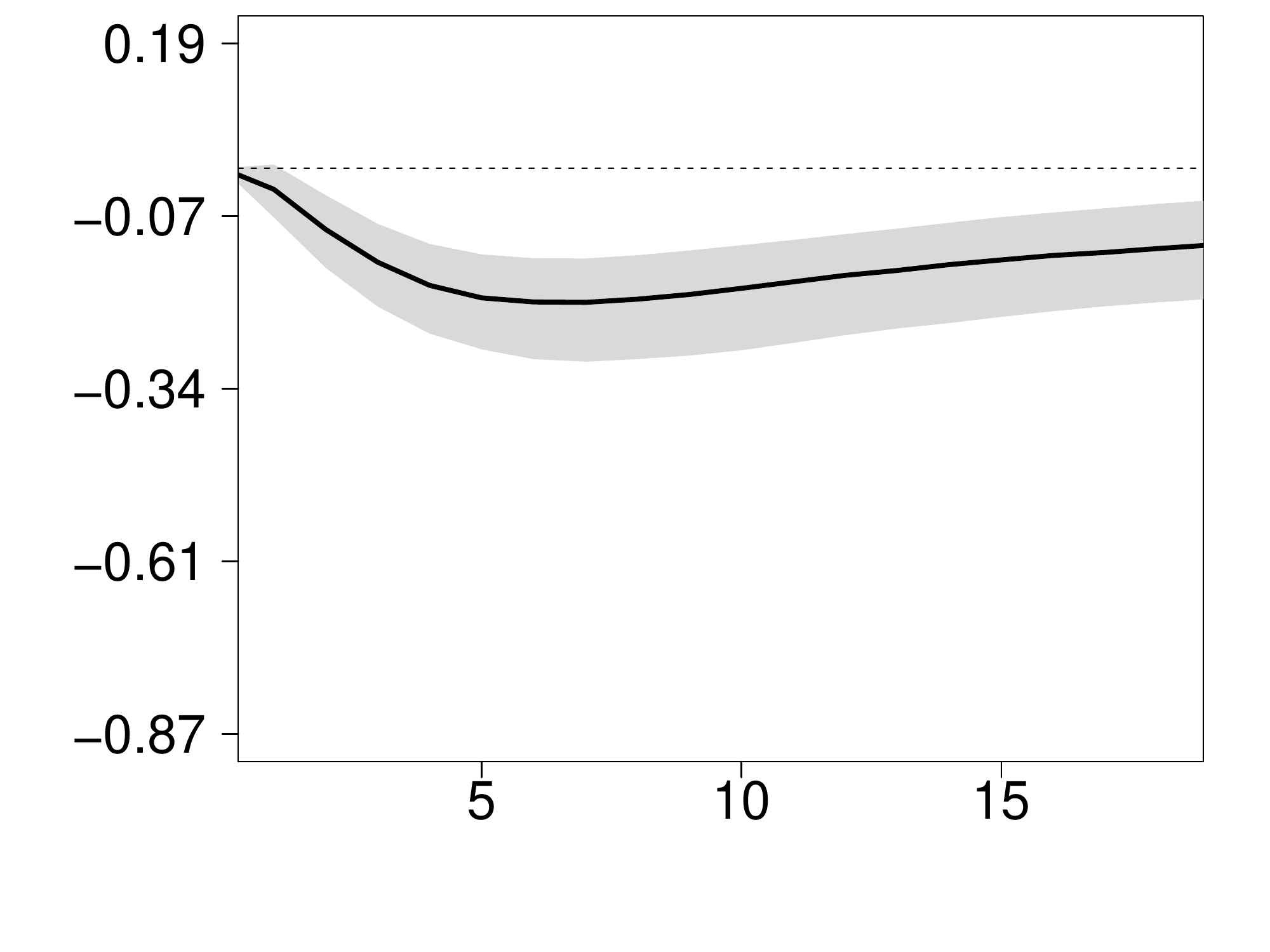}
\small New Mexico\\[0.1cm]
\includegraphics[width=\linewidth]{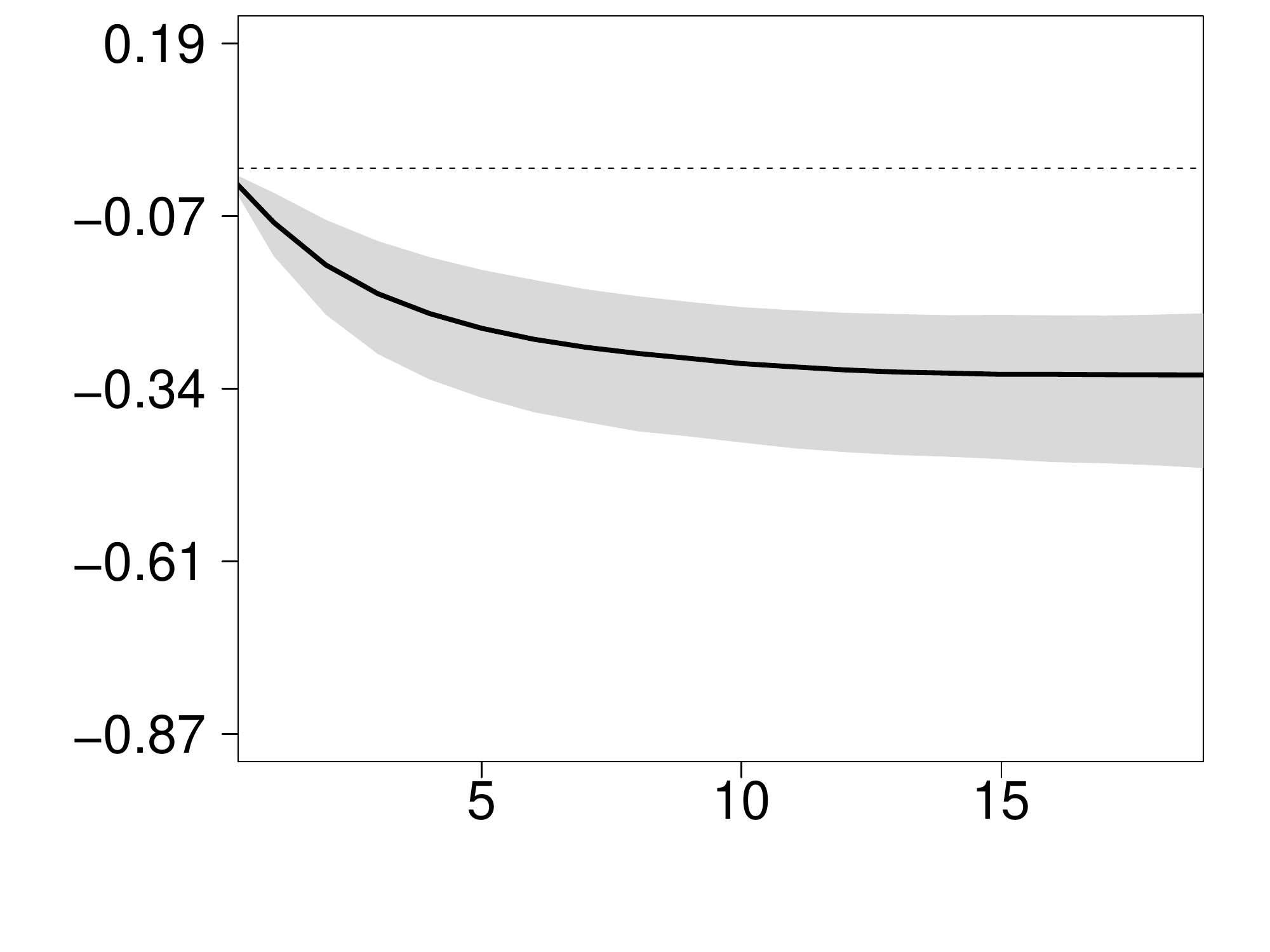}
\end{minipage}
\begin{minipage}[t]{0.24\textwidth}
\centering 
\large \textit{Midwest}\vspace{0.1cm}\hrule\vspace{0.5cm}
\small North Dakota\\[0.1cm]
\includegraphics[width=\linewidth]{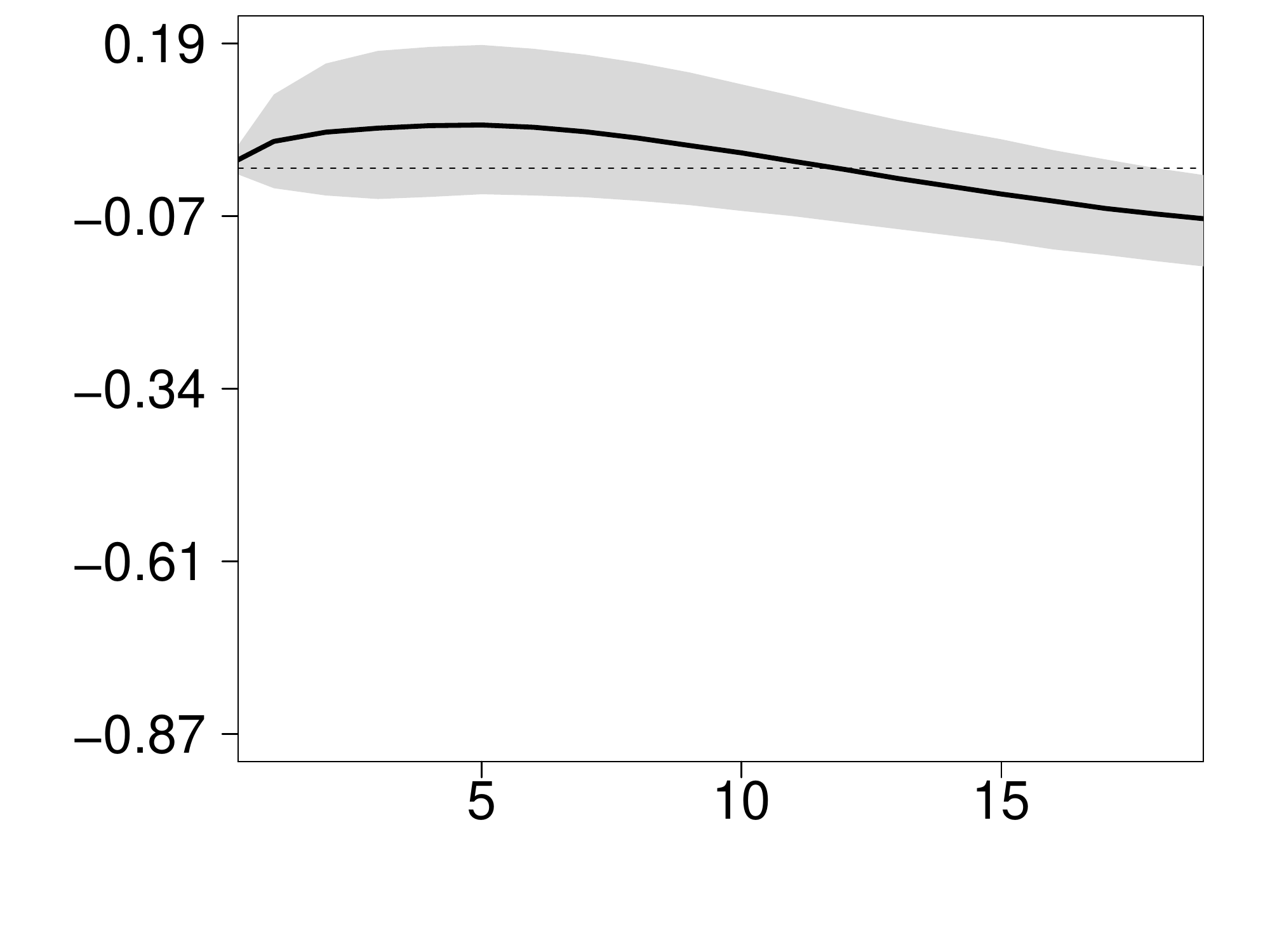}
\small Arkansas\\[0.1cm]
\includegraphics[width=\linewidth]{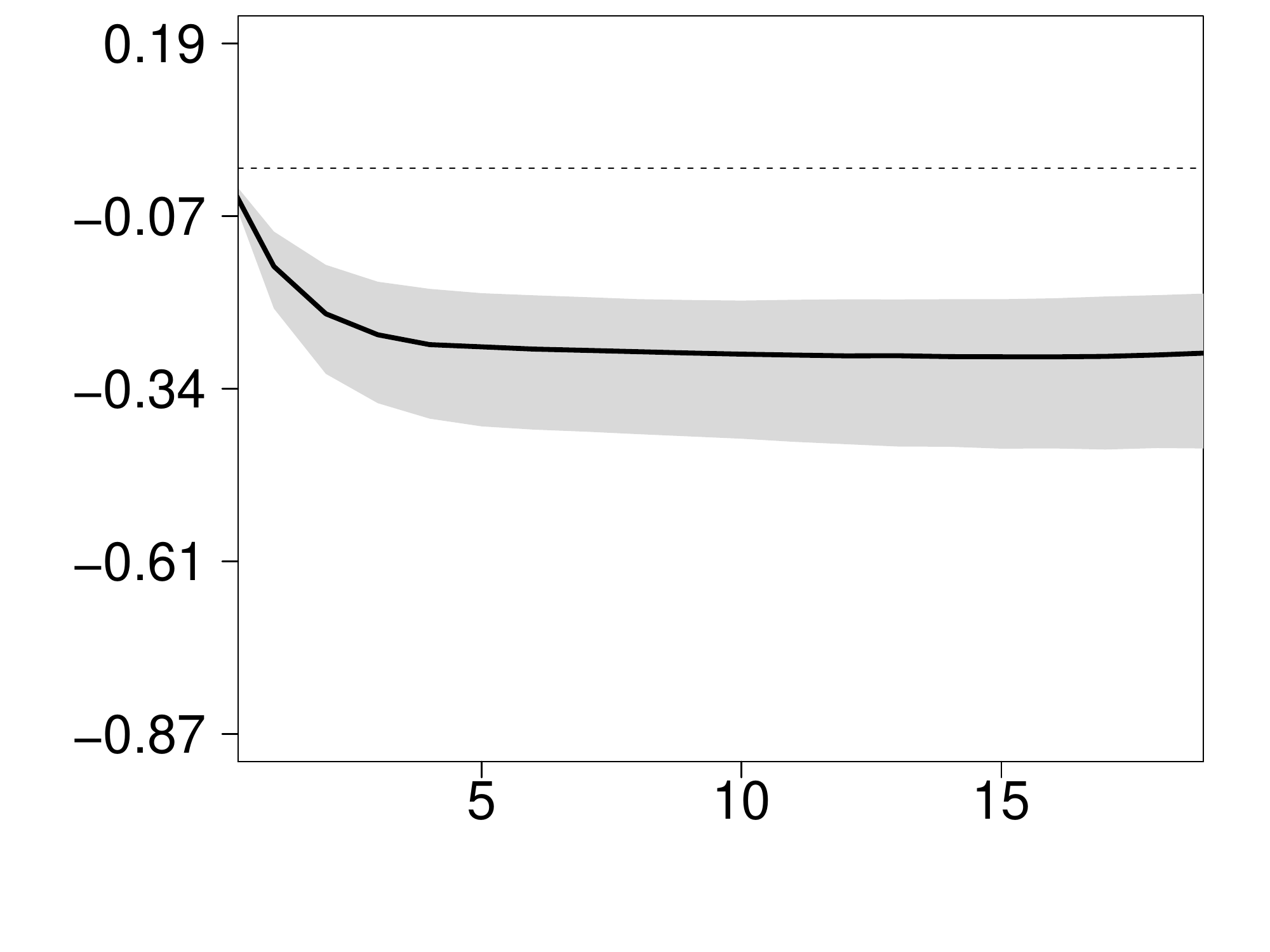}
\end{minipage}
\begin{minipage}[t]{0.24\textwidth}
\centering
\large \textit{South}\vspace{0.1cm}\hrule\vspace{0.5cm}
\small Texas\\[0.1cm]
\includegraphics[width=\linewidth]{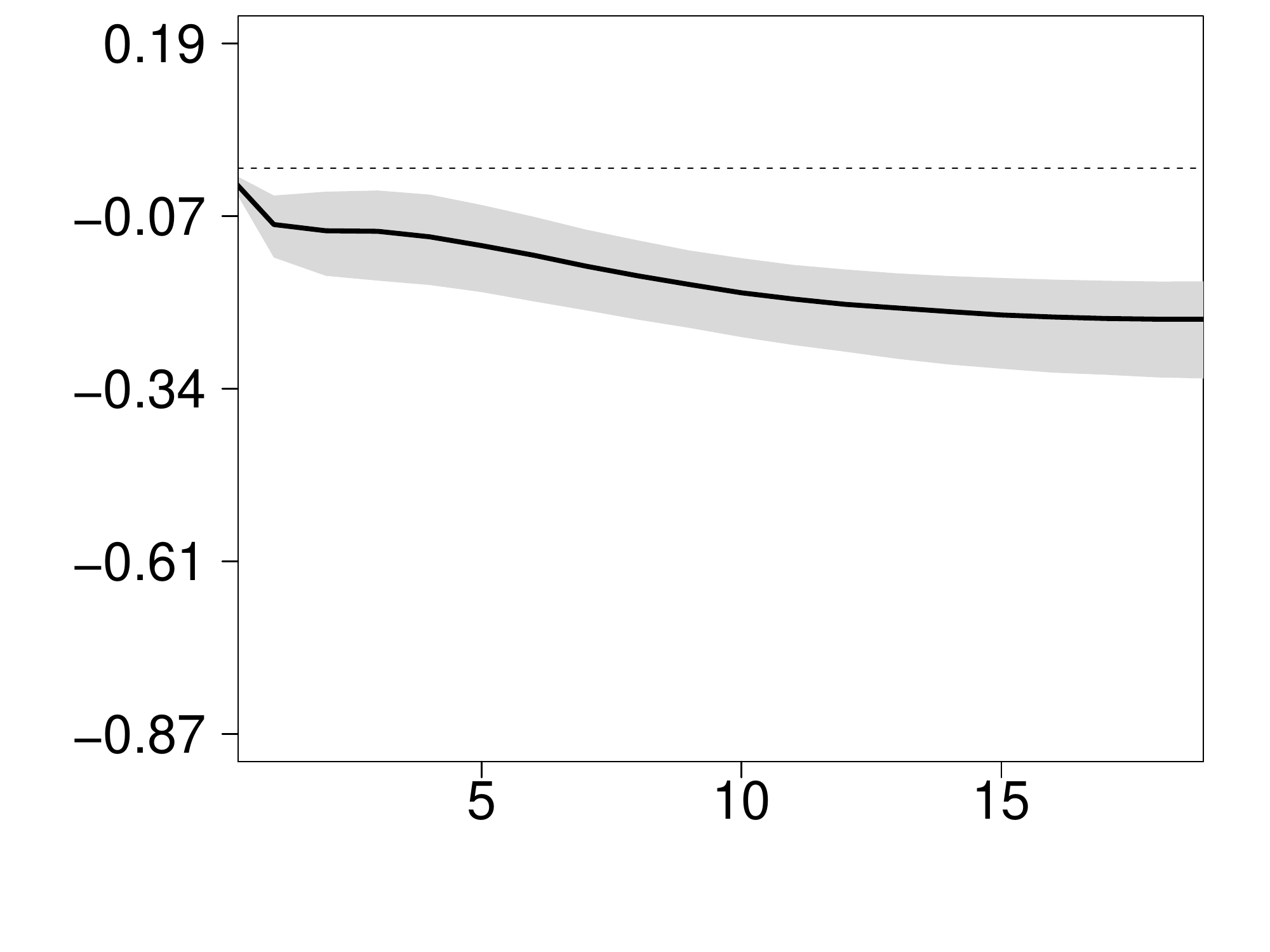}
\small Florida\\[0.1cm]
\includegraphics[width=\linewidth]{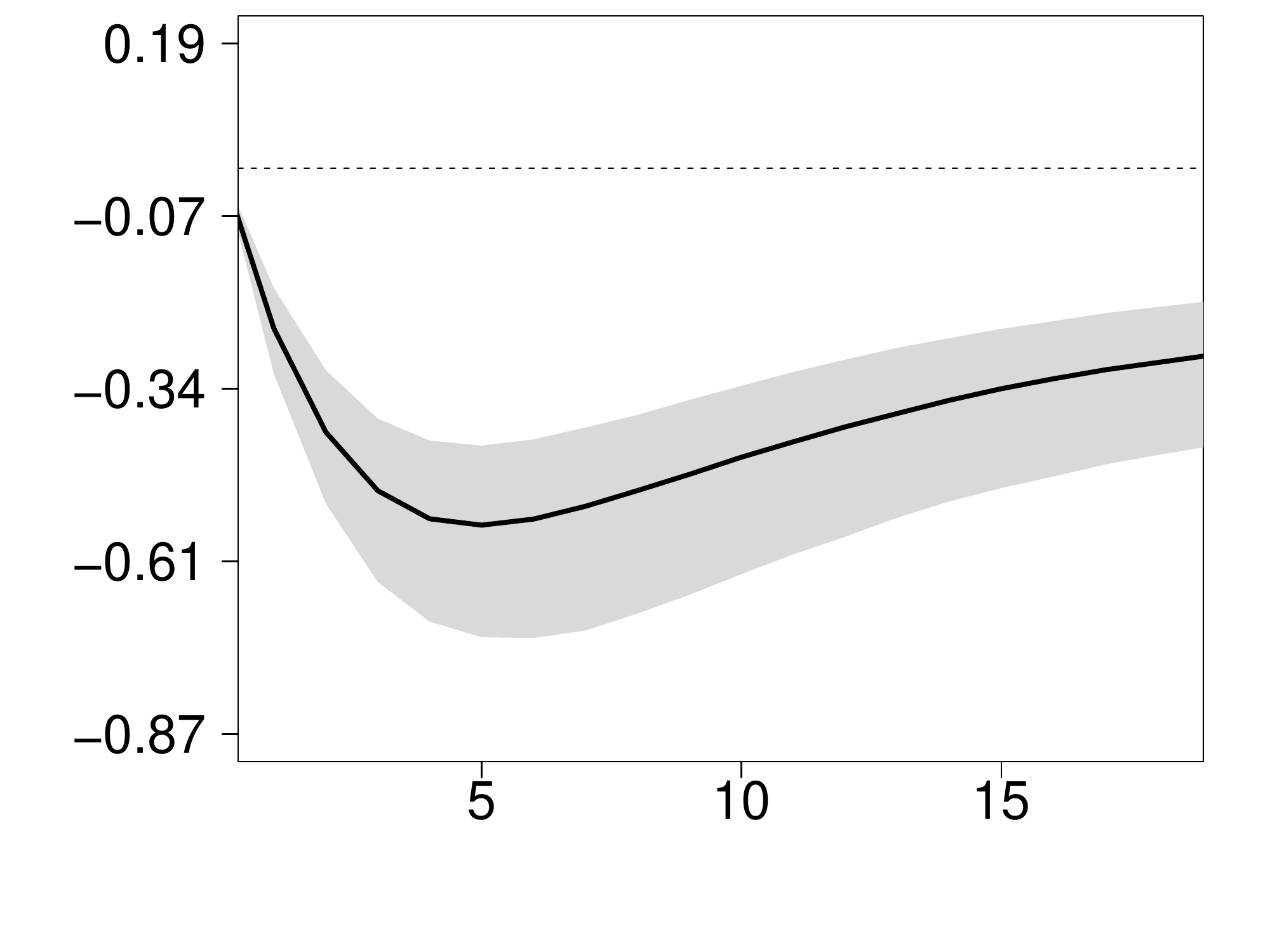}
\end{minipage}
\begin{minipage}[t]{0.24\textwidth}
\centering
\large \textit{Northeast}\vspace{0.1cm}\hrule\vspace{0.5cm}
\small New Jersey\\[0.1cm]
\includegraphics[width=\linewidth]{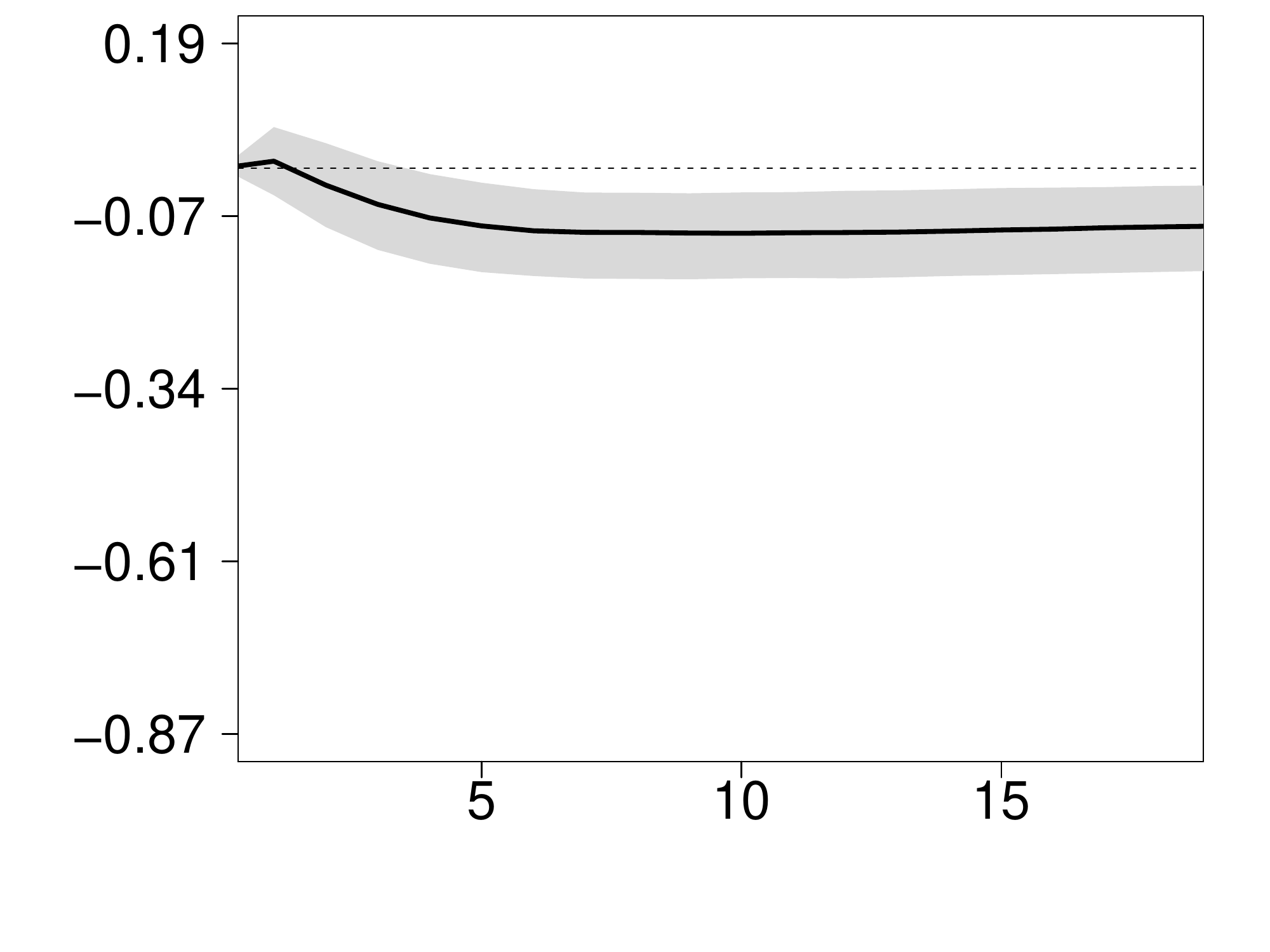}
\small Massachusetts\\[0.1cm]
\includegraphics[width=\linewidth]{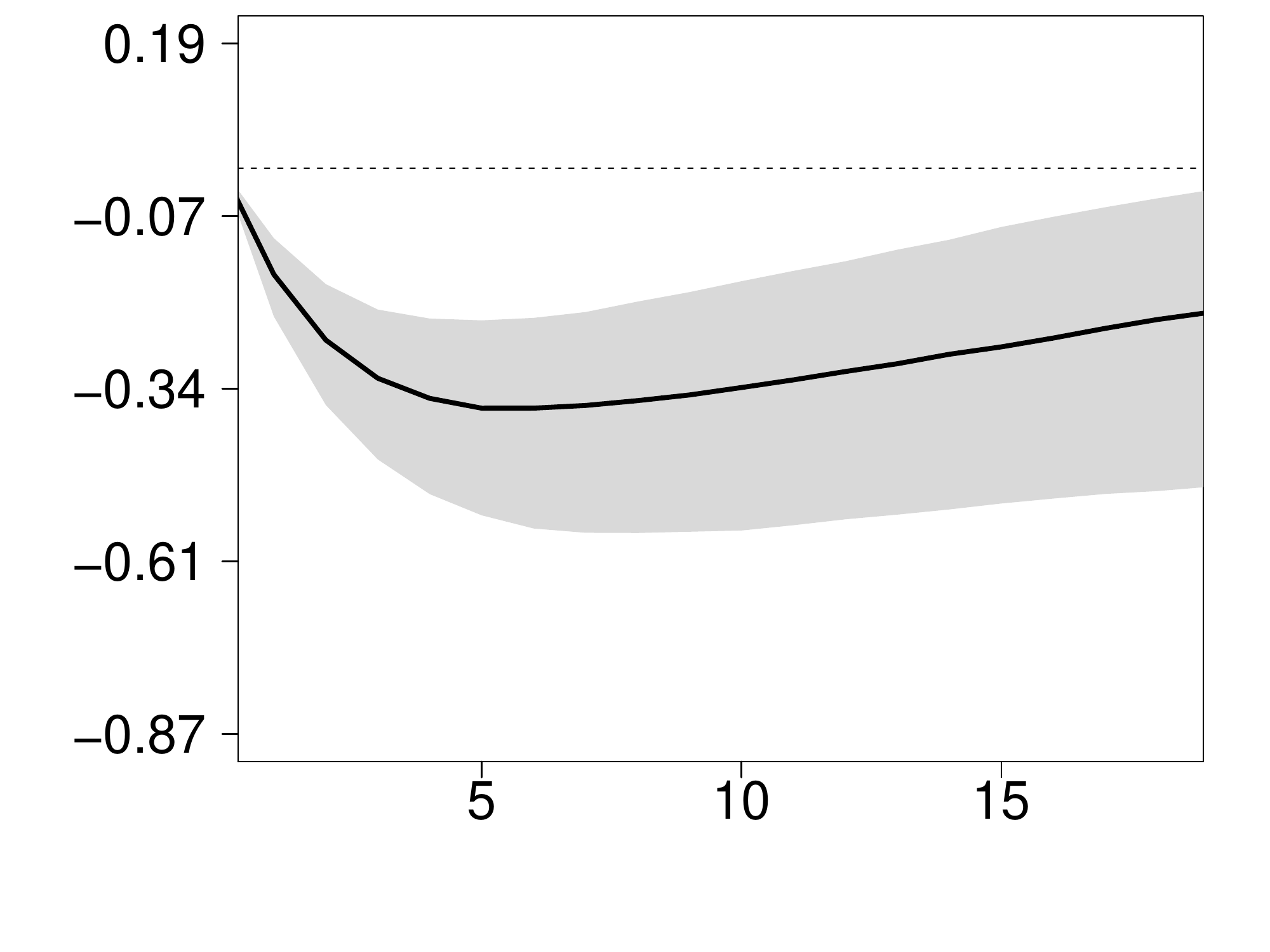}
\end{minipage}
\end{minipage}}
\begin{minipage}{16cm}\vspace{.3cm}
\footnotesize \textit{Notes}: The solid black line is the median response and the (gray) shaded area represents the 16th and 84th percentiles. The dotted line indicates the zero line. Results are based on 5,000 posterior draws. Sample period: 1985:Q1 -- 2017:Q1. Front axis: quarters after impact.
\end{minipage}%
\caption{Impulse response functions for employment in selected US states.}\label{fig:EM_selected}
\end{figure}
The responses of employment (see \autoref{fig:EM_selected}) mirror the responses of unemployment.  All states except North Dakota display decreasing levels of employment that appear to be similar in magnitudes. Compared to the reactions of unemployment we find that employment responses appear to be much more persistent, being statistically significant from zero for up to three years in most states considered. One connection to the  responses of income inequality is that states experiencing persistent declines in employment sometimes also feature persistent reactions of income inequality (see, e.g, Texas, New Jersey and Florida).

Finally, considering the reactions of total personal income points towards a high correlation with income inequality reactions. In general, we find that income declines in most states except North Dakota. The states that display a drop in income share one common feature: a rather slow response that takes a few quarters to react in a statistically significant manner. This holds true for the majority of  states considered except California and North Dakota. Especially for California, we observe immediate declines in total personal income which could partially explain the initial increase in income inequality observed in \autoref{fig:iq_selected}.

\begin{figure}[!ht]
\fbox{\begin{minipage}[t]{\textwidth}
\centering
\begin{minipage}[t]{.24\textwidth}
\centering
\large \textit{West}\vspace{0.1cm}\hrule\vspace{0.5cm}
\small California\\[0.1cm]
\includegraphics[width=\linewidth]{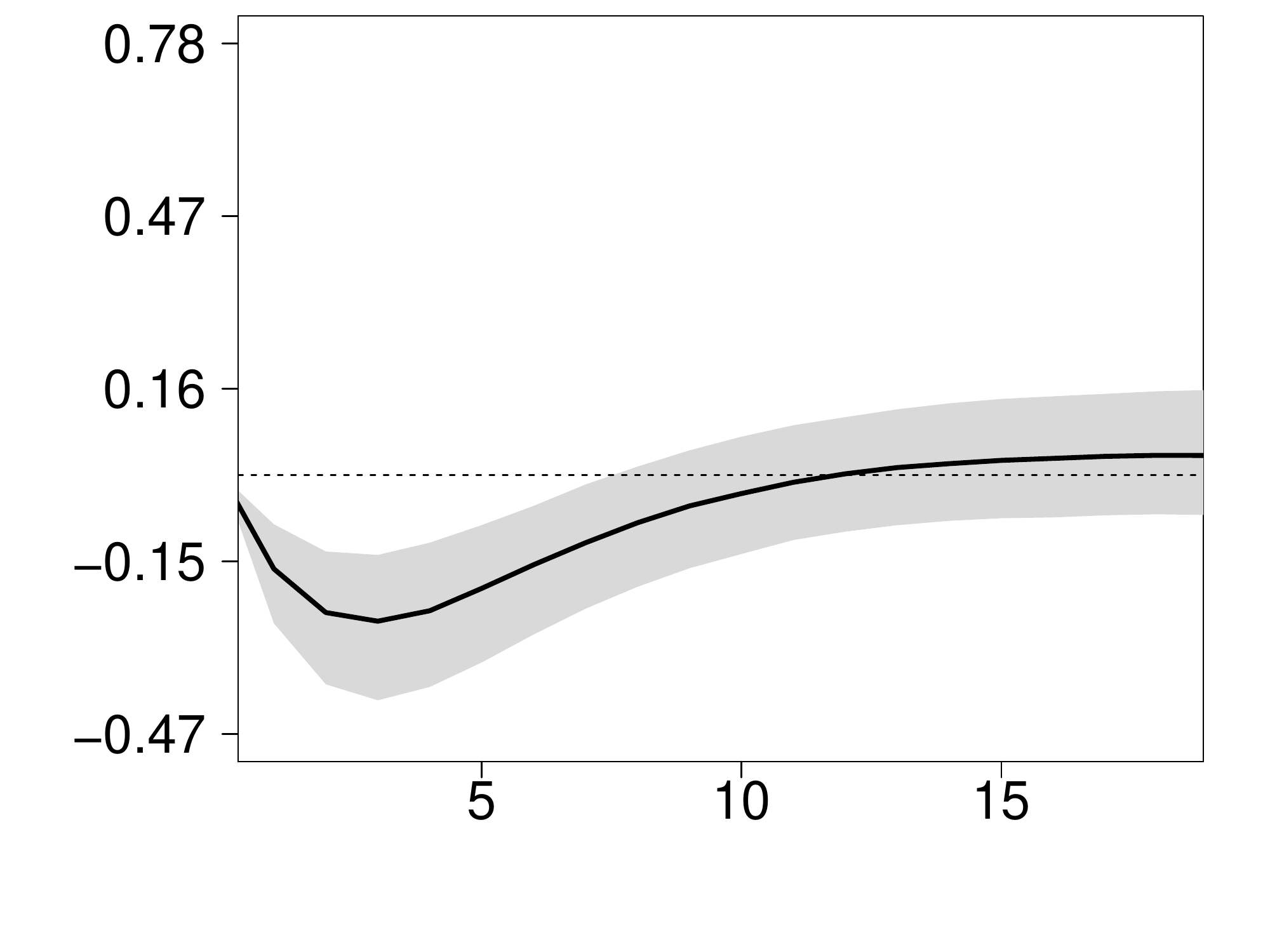}
\small New Mexico\\[0.1cm]
\includegraphics[width=\linewidth]{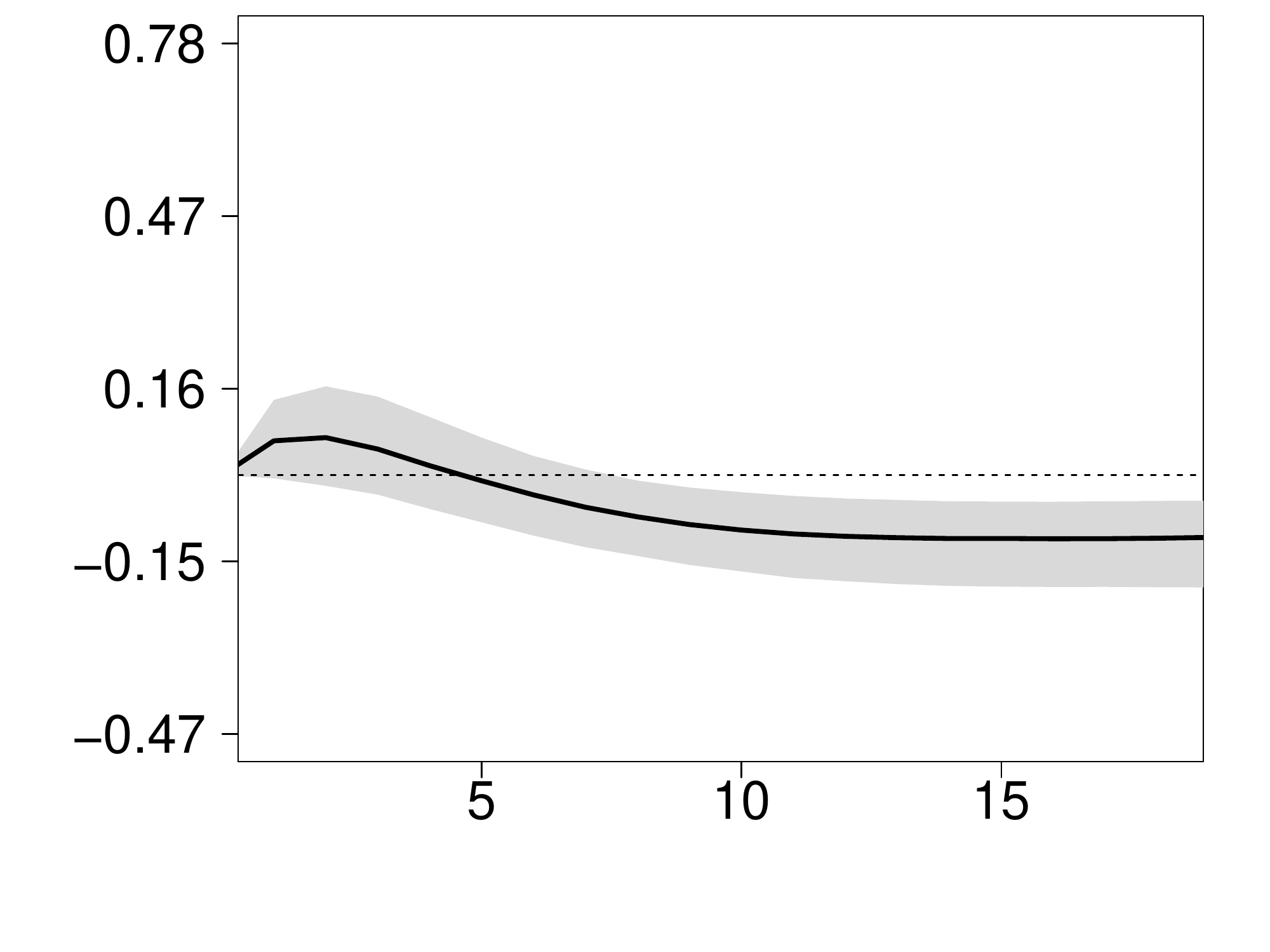}
\end{minipage}
\begin{minipage}[t]{0.24\textwidth}
\centering 
\large \textit{Midwest}\vspace{0.1cm}\hrule\vspace{0.5cm}
\small North Dakota\\[0.1cm]
\includegraphics[width=\linewidth]{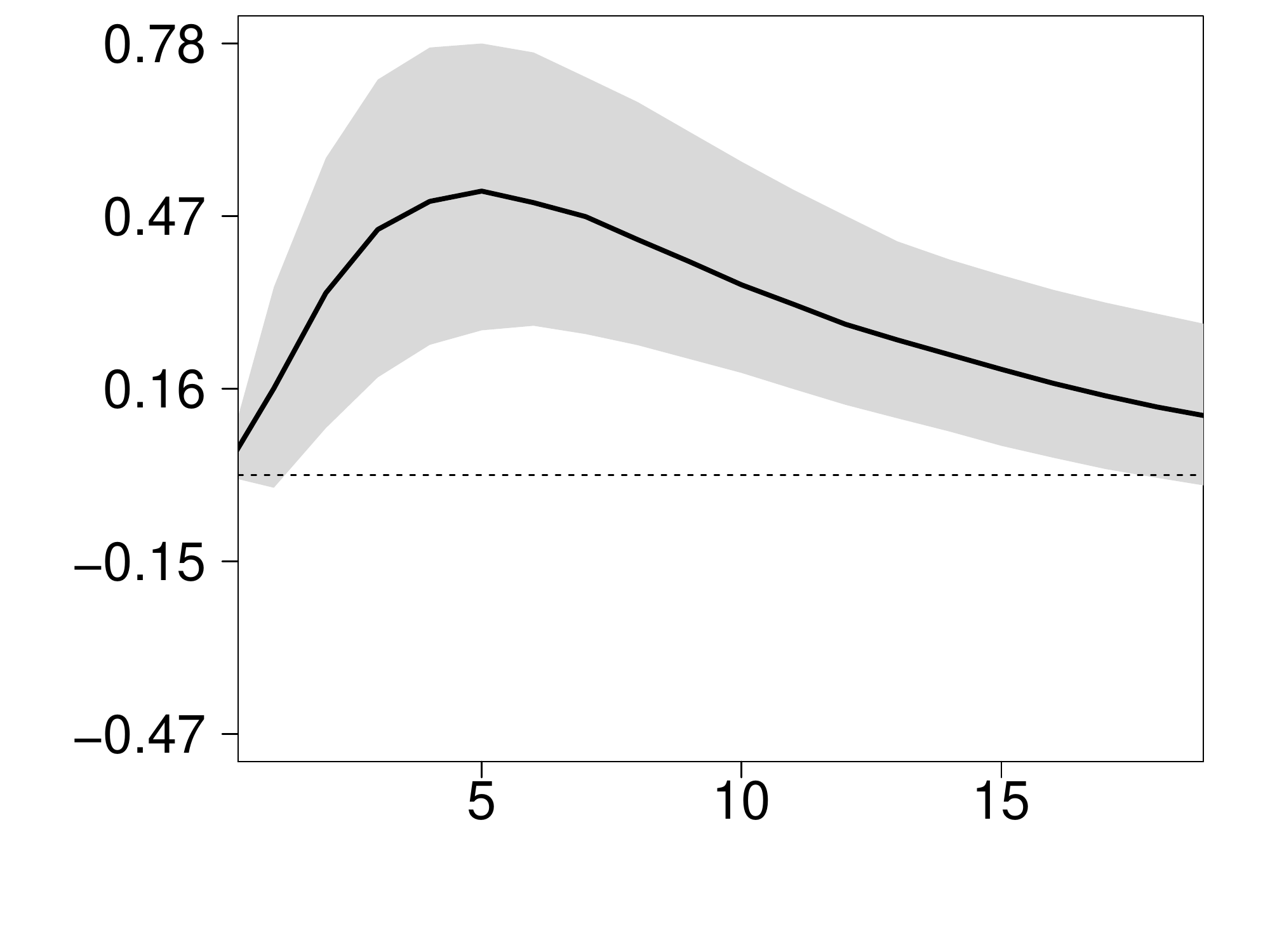}
\small Arkansas\\[0.1cm]
\includegraphics[width=\linewidth]{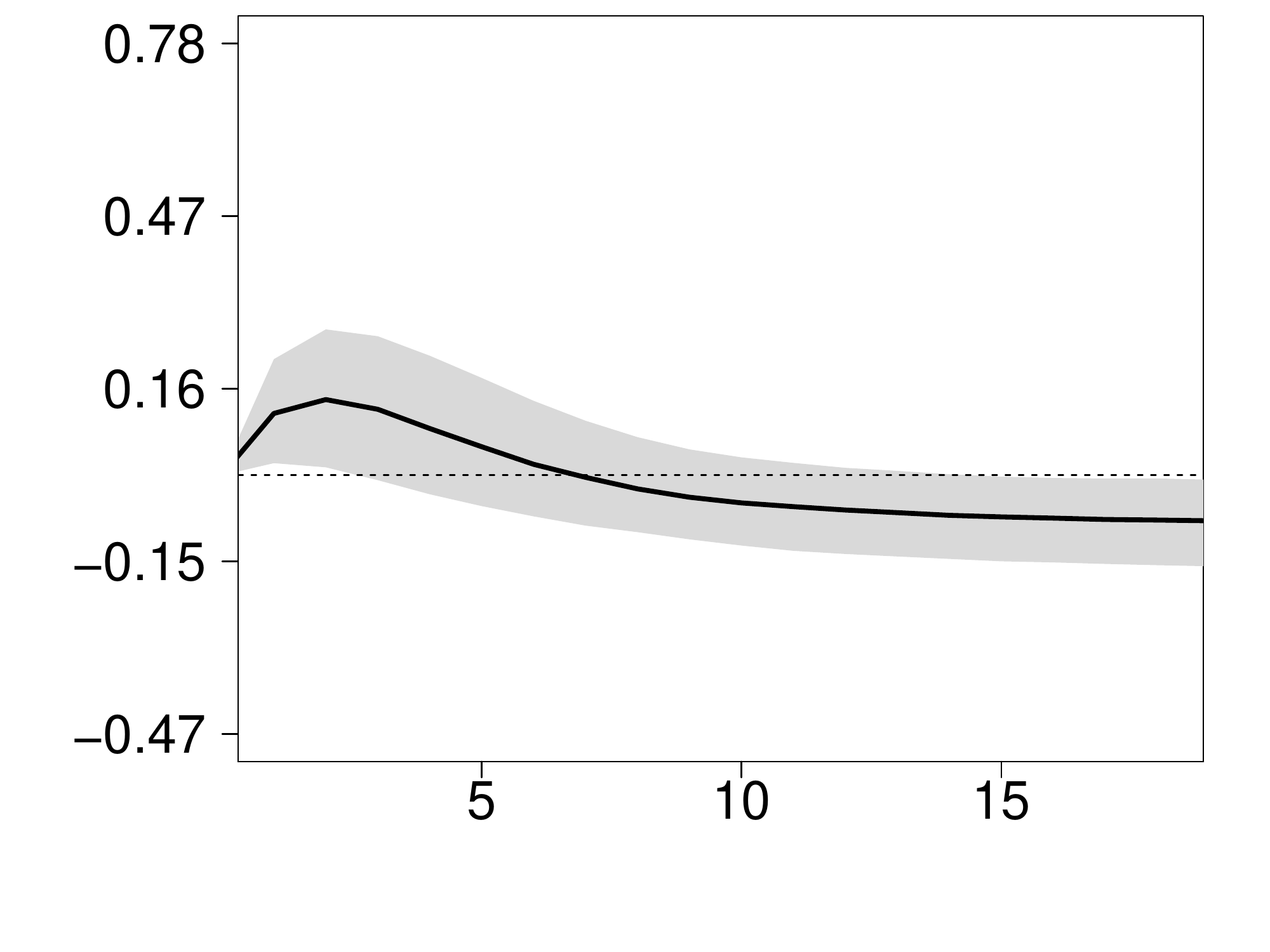}
\end{minipage}
\begin{minipage}[t]{0.24\textwidth}
\centering
\large \textit{South}\vspace{0.1cm}\hrule\vspace{0.5cm}
\small Texas\\[0.1cm]
\includegraphics[width=\linewidth]{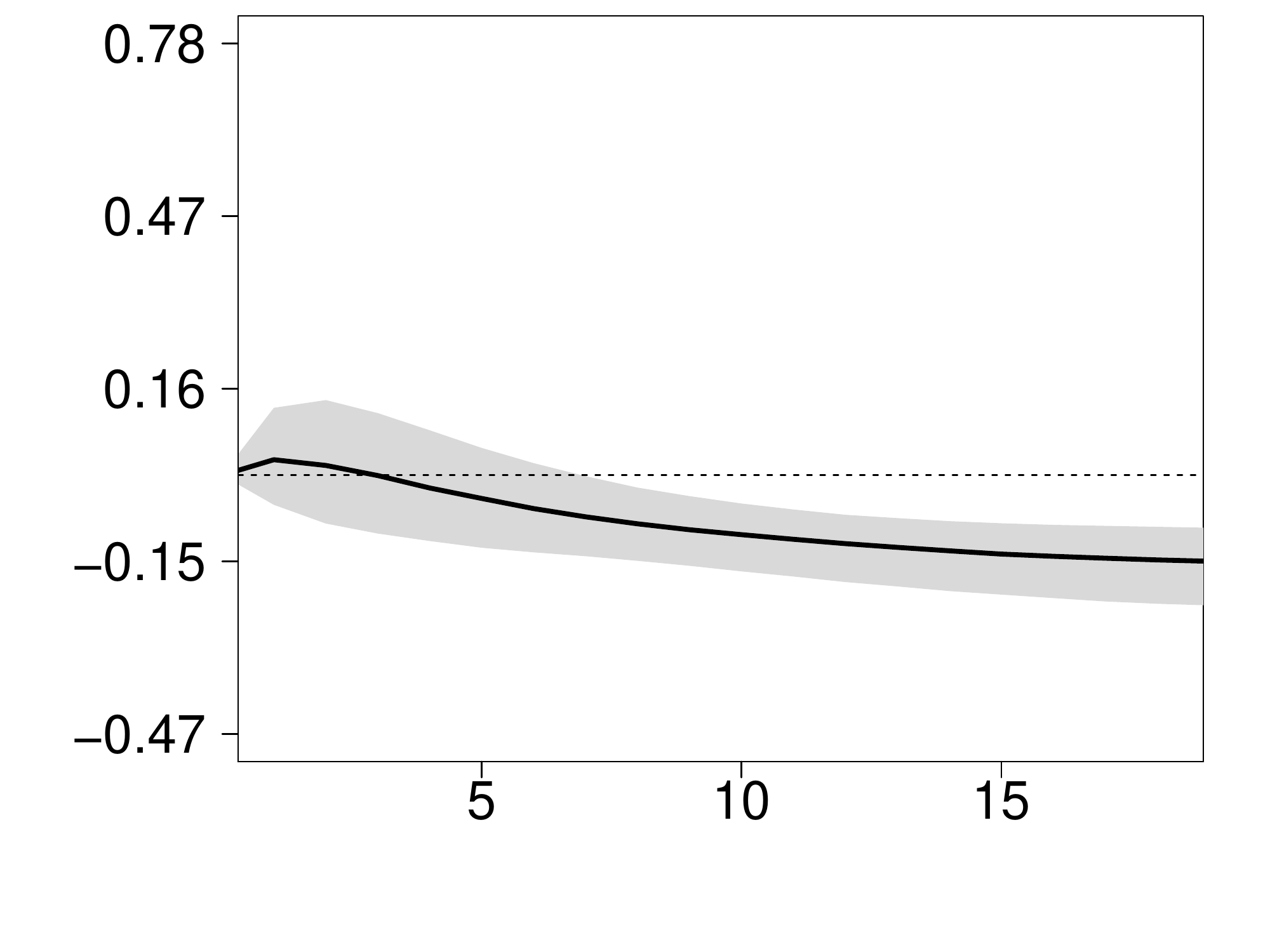}
\small Florida\\[0.1cm]
\includegraphics[width=\linewidth]{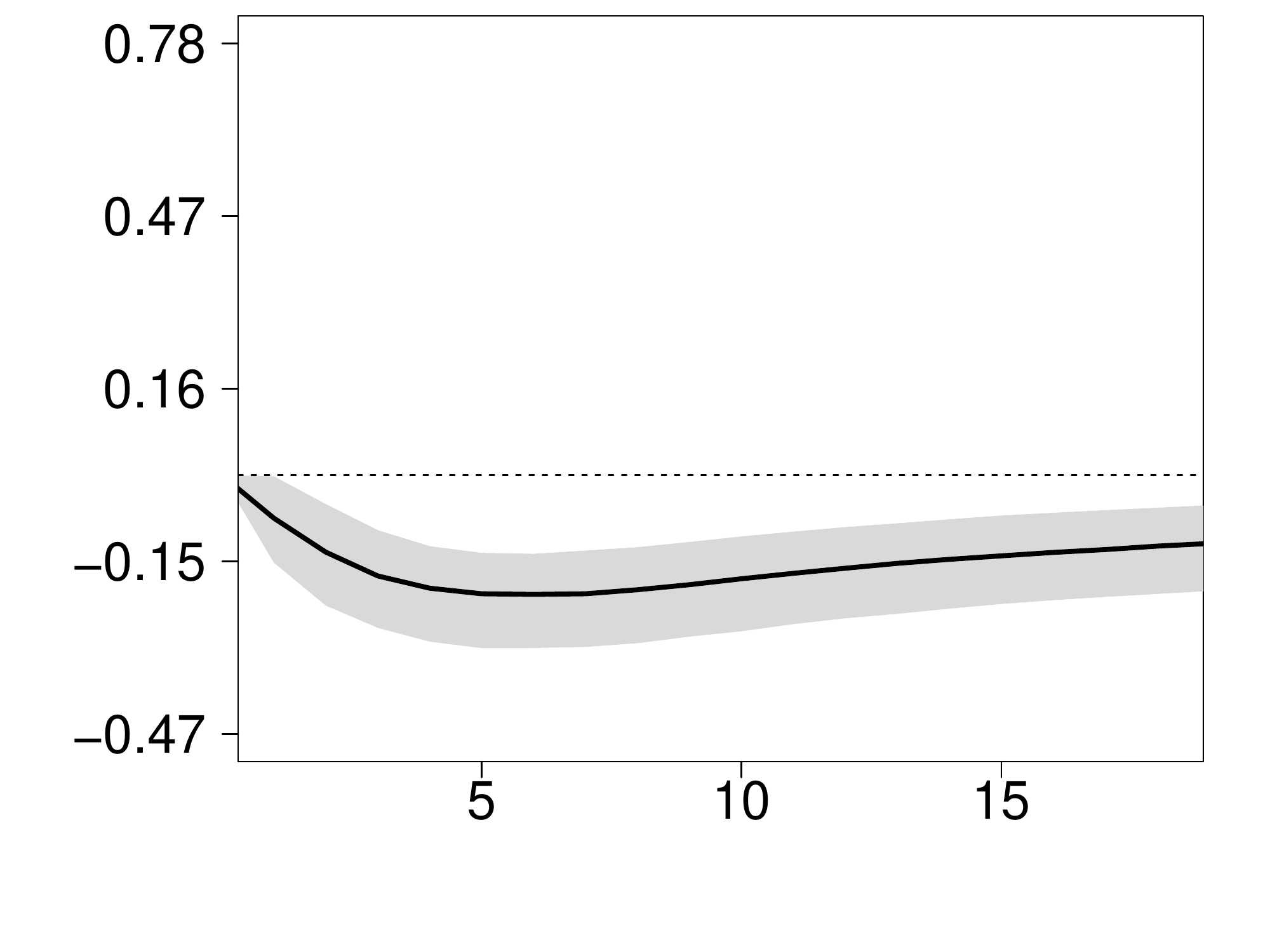}
\end{minipage}
\begin{minipage}[t]{0.24\textwidth}
\centering
\large \textit{Northeast}\vspace{0.1cm}\hrule\vspace{0.5cm}
\small New Jersey\\[0.1cm]
\includegraphics[width=\linewidth]{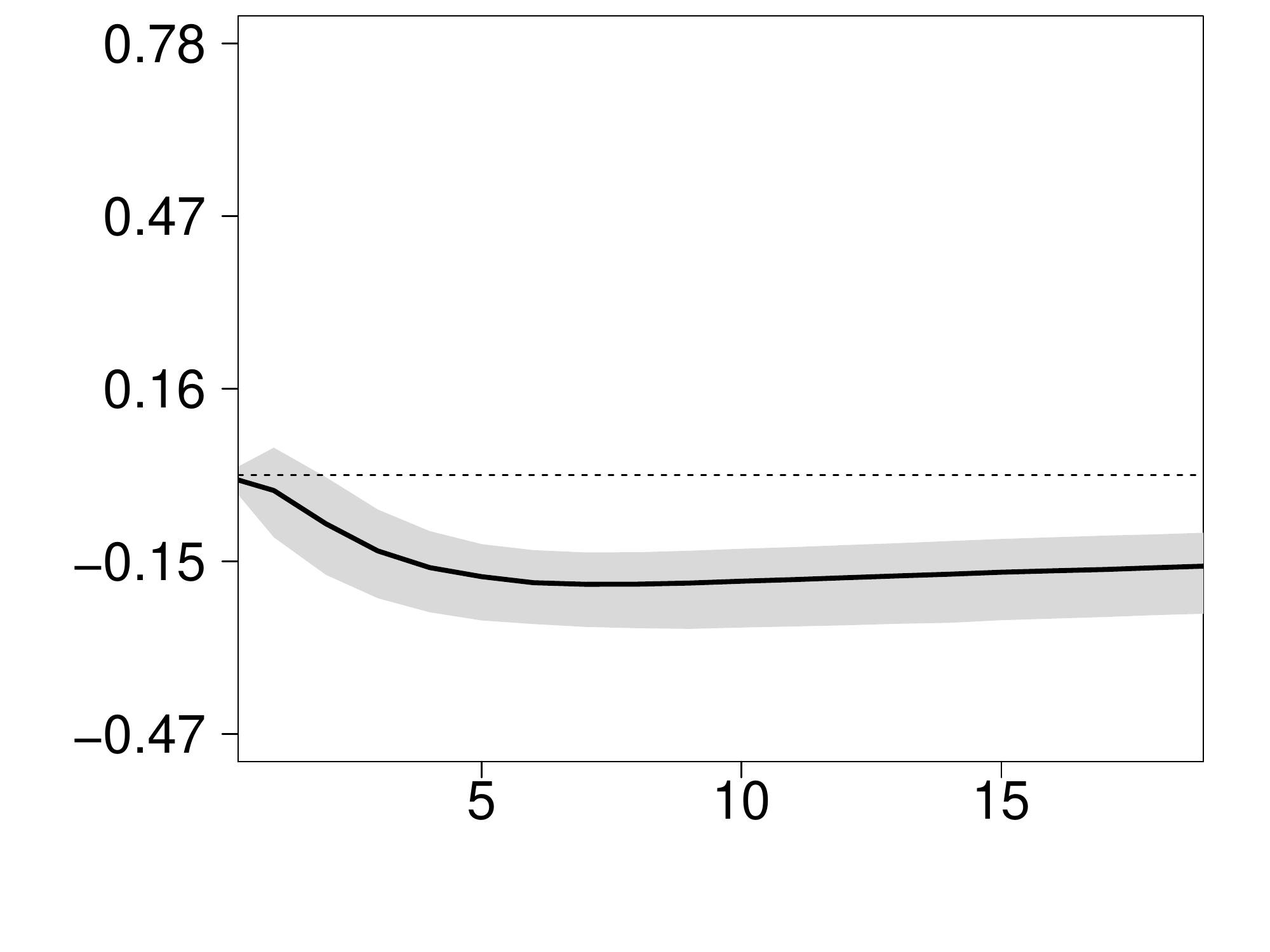}
\small Massachusetts\\[0.1cm]
\includegraphics[width=\linewidth]{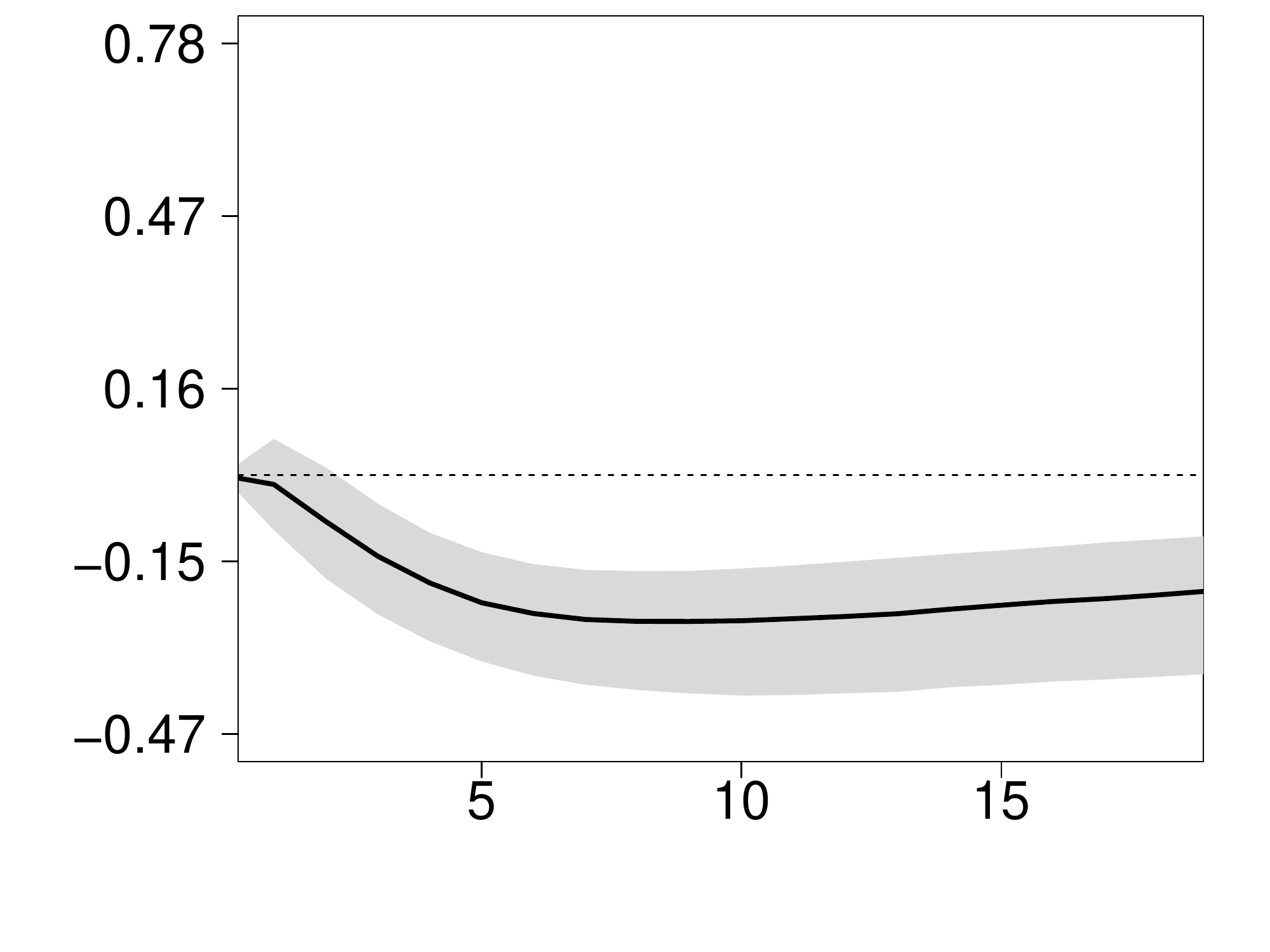}
\end{minipage}
\end{minipage}}
\begin{minipage}{16cm}\vspace{.3cm}
\footnotesize \textit{Notes}: The solid black line is the median response and the (gray) shaded area represents the 16th and 84th percentiles. The dotted line indicates the zero line. Results are based on 5,000 posterior draws. Sample period: 1985:Q1 -- 2017:Q1. Front axis: quarters after impact.
\end{minipage}%
\caption{Impulse response functions for total personal income in selected US states.}\label{fig:YN_selected}
\end{figure}

\subsection{Explaining differences in inequality responses}\label{sec:explanation}
For the purpose of providing additional quantitative information on what determines differences in state-level responses of income inequality, we employ a simple data summary device that involves standard regression analysis. More specifically, we regress the $h$-step ahead responses of income inequality on averages over time of a set of additional state-specific macroeconomic variables for  $h \in \{4, 8, 12\}$ as well as the peak response. While this approach suffers some issues like latent response variables as well as potential model misspecification, it provides a rough gauge on the underlying trends in the data.

In the discussion in Section \ref{sec: IQ_resp} we explained differences in inequality responses by referring to heterogeneity with respect to the composition of income. To provide additional evidence, we employ two measures that aim at representing the income structure at the state level as potential control variables. We consider business income (\textit{bussum}) given in the CPS, calculated as the share of non-farm business and professional practice income for self-employed individuals related to total income. The second measure is the sum of income accruing from dividends, interest and rent (\textit{dir}) from the Federal Reserve Bank of St. Louis data base, which constitutes a part of the total personal income per capita variable. Moreover, to capture differences in the sectoral composition of employment, we calculate averages of the shares of people employed in the agricultural (\textit{agric}), construction (\textit{constr}) and manufacturing (\textit{manu}) sector, again obtained from the FRED data base. The set of explanatory variables is completed by the average level of state-specific unemployment (\textit{unemp}).

Table \ref{tab:reg_results} shows the results of our regression exercise for different horizons of the responses of income inequality and the peak response. A few findings are worth emphasizing. First, notice that across horizons, only one variable appears to be significant. Irrespective of the time horizon considered, we find that dividends, interest rate and rent income per capita feature a strongly negative and significant regression coefficient. This suggests a negative relationship between the magnitude of the responses of income inequality and the  degree of (grossly speaking) non-labor income within a given state. Thus, when considering different time horizons and the peak effect, we find no discernible differences in what variables appear to be important drivers of the inequality responses. Second, the magnitudes between the dividends, interest and rent variable and inequality responses become smaller for higher impulse response horizons. At a first glance, this finding could simply be driven by the fact that impulse responses die out after a few quarters. However, in light of the discussion in Section \ref{sec: IQ_resp} this appears to be the case only in a small number of states. For a moderate number of stlates, we observe persistence medium-run reactions of income inequality. Finally, considering the explanatory power suggests that in the short run, around 14 percent of cross-state variation is captured, rising to as much as 20 percent for explaining the two year ahead responses. 

\begin{table*}[!htbp] \centering
  \caption{Regression of posterior median of inequality responses on state-level macroeconomic quantities} \label{tab:reg_results}
  \label{} 
\begin{tabular}{@{\extracolsep{5pt}}lcccc} 
\\[-1.8ex]\hline 
\hline \\[-1.8ex] 
 & \multicolumn{4}{c}{\textit{Posterior median of response of income inequality}} \\ 
\cline{2-5} 
%\\[-1.8ex] & \multicolumn{4}{c}{} \\ 
\\Horizon (quarters) & $4$ & $8$ & $12$ & $Peak$\\ 
\hline \\[-1.8ex] 
 Agriculture & $-$0.005 & 0.001 & 0.004 & 0.002 \\ 
  & (0.015) & (0.008) & (0.007) & (0.021) \\ 
  & & & & \\ 
 Construction & 0.086 & 0.031 & 0.018 & 0.125 \\ 
  & (0.107) & (0.058) & (0.049) & (0.147) \\ 
  & & & & \\ 
 Manufacturing & $-$0.026 & $-$0.010 & $-$0.008 & $-$0.012 \\ 
  & (0.037) & (0.020) & (0.017) & (0.051) \\ 
  & & & & \\ 
 Dividends, interest and rent & $-$0.243$^{**}$ & $-$0.141$^{**}$ & $-$0.096$^{**}$ & $-$0.292$^{**}$ \\ 
  & (0.096) & (0.052) & (0.044) & (0.132) \\ 
  & & & & \\ 
 Sum of business income & $-$1.354 & $-$0.021 & 0.340 & $-$1.127 \\ 
  & (1.569) & (0.859) & (0.726) & (2.166) \\ 
  & & & & \\ 
 Intercept & 2.165$^{**}$ & 1.272$^{**}$ & 0.842$^{*}$ & 2.177 \\ 
  & (1.067) & (0.584) & (0.494) & (1.473) \\ 
  & & & & \\ 
\hline \\[-1.8ex] 
R$^{2}$ & 0.151 & 0.161 & 0.134 & 0.114 \\ \hline 
\hline \\[-1.8ex] 
\textit{Note:}  & \multicolumn{4}{r}{$^{*}$p$<$0.1; $^{**}$p$<$0.05; $^{***}$p$<$0.01} \\
\end{tabular} 
\end{table*} 

\section{Closing remarks}
In this paper, we investigate the relationship between uncertainty shocks and household income inequality using a novel large-scale econometric framework. Our model enables us to assess how a national uncertainty shock impacts the US economy at the state level, controlling for potential spillovers between states. The results point toward a strong relationship between movements in uncertainty and income inequality for the vast majority of US states under scrutiny. Depending on the income composition within a given state, income inequality increases if the share of labor income is high, while it tends to decrease if the share of capital income is comparatively larger. 

Considering the responses of unemployment, employment and total personal income shed some light on the transmission mechanisms that drive our results. We find that the persistence of the responses of macroeconomic quantities directly translates into persistent reactions of income inequality at the state level. The quantitative contribution of the uncertainty shock in explaining income inequality is assessed by conducting a forecast error variance decomposition. The findings point towards a large degree of heterogeneity across states. For some states, we find that uncertainty shocks play an important role in shaping income inequality dynamics whereas for other states, this role is somewhat smaller but still substantial.
\clearpage

%**********************************************************************************************
\small\singlespacing
\bibliographystyle{./bibtex/cit_econometrica}
\bibliography{./bibtex/lit}
\addcontentsline{toc}{section}{References}
%**********************************************************************************************

\newpage
\onehalfspacing\normalsize

\begin{appendices}\crefalias{section}{appsec}
\setcounter{equation}{0}
\renewcommand\theequation{A.\arabic{equation}}
\section{Prior specification}\label{app:priors}
Estimating the model requires Bayesian methods that involve choosing adequate prior distributions for the model parameters. As described in \cref{sec:econometrics}, we assume that the vectorized VAR coefficients $\bm{\beta}_i$ arise from a common distribution
\begin{equation}
\bm{\beta}_i\sim\mathcal{N}(\bm{\mu},\bm{V}). \label{eq:common}
\end{equation}
Equation (\ref{eq:common}) can be interpreted as a prior distribution on $\bm{\beta}_i$ with mean $\bm{\mu}$ and diagonal variance-covariance matrix $\bm{V}$.  On $\bm{\mu}$, we use a normally distributed prior,
\begin{equation}
\bm{\mu} \sim \mathcal{N}(\bm{0}_M,\bm{V}_0),
\end{equation}
with $\bm{0}_M$ denoting a $M-$dimensional vector of zeros and set $\bm{V}_0 = 10 \times \bm{I}_M$.

For the main diagonal elements of $\bm{V}$, $v_j~(j = 1,\hdots,M)$, we use independent inverted Gamma priors,
\begin{equation}
v_j \sim \mathcal{G}^{-1}(d_0,d_1),
\end{equation}
where the prior hyperparameters $d_0 = d_1 =0.01$ are set to be only weakly informative. 

The coefficient matrices for the national quantities $\bm{D}_p$ and $\bm{S}_q$ are again assigned a Gaussian prior with the mean vector centered on zero with variance $10$. This choice introduces relatively little prior information on the coefficients associated with the national macroeconomic quantities. 

For the factor model in the reduced form errors of the model, we use the following prior setup. The elements $\lambda_{ij}$ of the matrix of factor loadings $\bm{\Lambda}$ for $i = 1,\hdots,L$ and $j = 1,\hdots,F$ are assigned a normally distributed prior, that is, $\lambda_{ij} \sim \mathcal{N}(0,10^2)$. 

In \cref{sec:econometrics}, we mentioned that the log volatilities follow an AR(1) process. To specify priors on the corresponding coefficients, we now introduce the specific law of motion in more detail.  Following \cite{aguilar2000bayesian}, we assume that the logarithms of the main diagonal elements of $\bm{H}_t$ and $\bm{\Omega}_t$  follow independent AR(1) processes,
\begin{align}
h_{jt} &= \phi_{hj} + \rho_{hj}(h_{jt-1} - \phi_{hj}) + \sigma_{hj}\xi_{h,j,t} \quad \text{\hspace{.46cm}for} \quad j = 1, \hdots,F,\\
\omega_{jt}  &= \phi_{\omega j} + \rho_{\omega j}(\omega_{jt-1} - \phi_{\omega j}) + \sigma_{\omega j}\xi_{\omega,j,t} \quad \text{for} \quad j = 1, \hdots,L,
\end{align}
and using $s \in \{h,\omega\}$, we denote the unconditional mean of the log-volatility by $\phi_{sj}$, the autoregressive parameter by $\rho_{sj}$, and  $\sigma^2_{sj}$ is the innovation variance of the processes. The serially uncorrelated white noise shocks $\xi_{s,j,t} \sim \mathcal{N}(0,1)$ are standard normally distributed. The prior specification closely follows \citet{kastner2014ancillarity}, with a normally distributed prior on $\phi_{sj} \sim \mathcal{N}(0, 10^2)$, a Gamma prior on $\sigma^2_{sj} \sim \mathcal{G}(1/2,1/2)$ and a Beta prior on the transformed persistence parameter $(\rho_{sj}+1)/2 \sim \mathcal{B}(25,5)$.

\setcounter{equation}{0}
\renewcommand\theequation{B.\arabic{equation}}
\newpage
\section{Full conditional posterior sampling}\label{app:mcmc}
The prior setup described above translates into a set of full conditional posterior distributions that have a well-known form. In what follows, we only briefly summarize the steps involved in obtaining a valid draw from the joint posterior distribution, and provide additional references that include more details on the exact posterior moments.

Our Gibbs sampler iterates between the following steps:
\begin{enumerate}[label=(\roman*)]
	\item The VAR coefficients in $\bm{\beta}_i$ can be sampled on an equation-by-equation basis, where conditional on $\bm{\Lambda} \bm{f}_t$, the full conditional posterior distribution follows a Gaussian distribution with mean and variance taking a standard form \citep[see, for instance,][]{zellner1996introduction}
	\item  Using the fact that conditional on knowing $\{\bm{\beta}_i\}_{i=1}^N$ the conditional posterior of $\bm{\mu}$ does not depend on the data leads to a Gaussian full conditional posterior distribution that takes a well-known form \citep{koop2003}.

\item The VAR coefficients for the national quantities $\bm{D}$ and $\bm{S}$ are sampled analogously to the state-specific parameters from multivariate Gaussian distributions on an equation-by-equation basis.
	
\item  The free elements in $\bm{\Lambda}$ can, again, be simulated on an equation-by-equation basis. Notice that conditional on the latent factors, $\bm{\Lambda}$ is obtained by estimating a sequence of standard Bayesian regression models with heteroscedastic innovations \citep[see][]{aguilar2000bayesian}

\item For the latent factors $\{\bm{f}_t\}_{t=1}^{T}$, we simulate the full history by drawing from a set of independent Gaussian distributions,
	\begin{align}
	\bm{f}_t|\bullet &\sim \mathcal{N}(\obar{\bm{f}}_t, \bm{P}_t)\\
	\bm{P}_t &= \bm{H}_t - \bm{\Upsilon}_t\bm{\Theta}_t\bm{\Upsilon}_t'\nonumber\\
	\bm{\Upsilon}_t &= \bm{H}_t \bm{\Lambda}'\bm{\Theta}_t^{-1} \nonumber\\
	\obar{\bm{f}}_t &= \bm{\Upsilon}_t \bm{\varepsilon}_t.\nonumber
	\end{align}
	\item The full history of the log-volatilities is sampled using the algorithm outlined in \citet{kastner2014ancillarity}; see also \citet{kastner2016dealing}.
\end{enumerate}
We pick starting values for the parameters of the model and cycle through the algorithm described above for $10,000$ times, discarding the first $5,000$ draws as burn-in. It is worth mentioning that the employed algorithm exhibits excellent mixing and convergence properties.
\newpage

\section{Additional empirical results}\label{app:add_results}
\begin{figure}[!ht]
\centering
\includegraphics[width=0.8\textwidth]{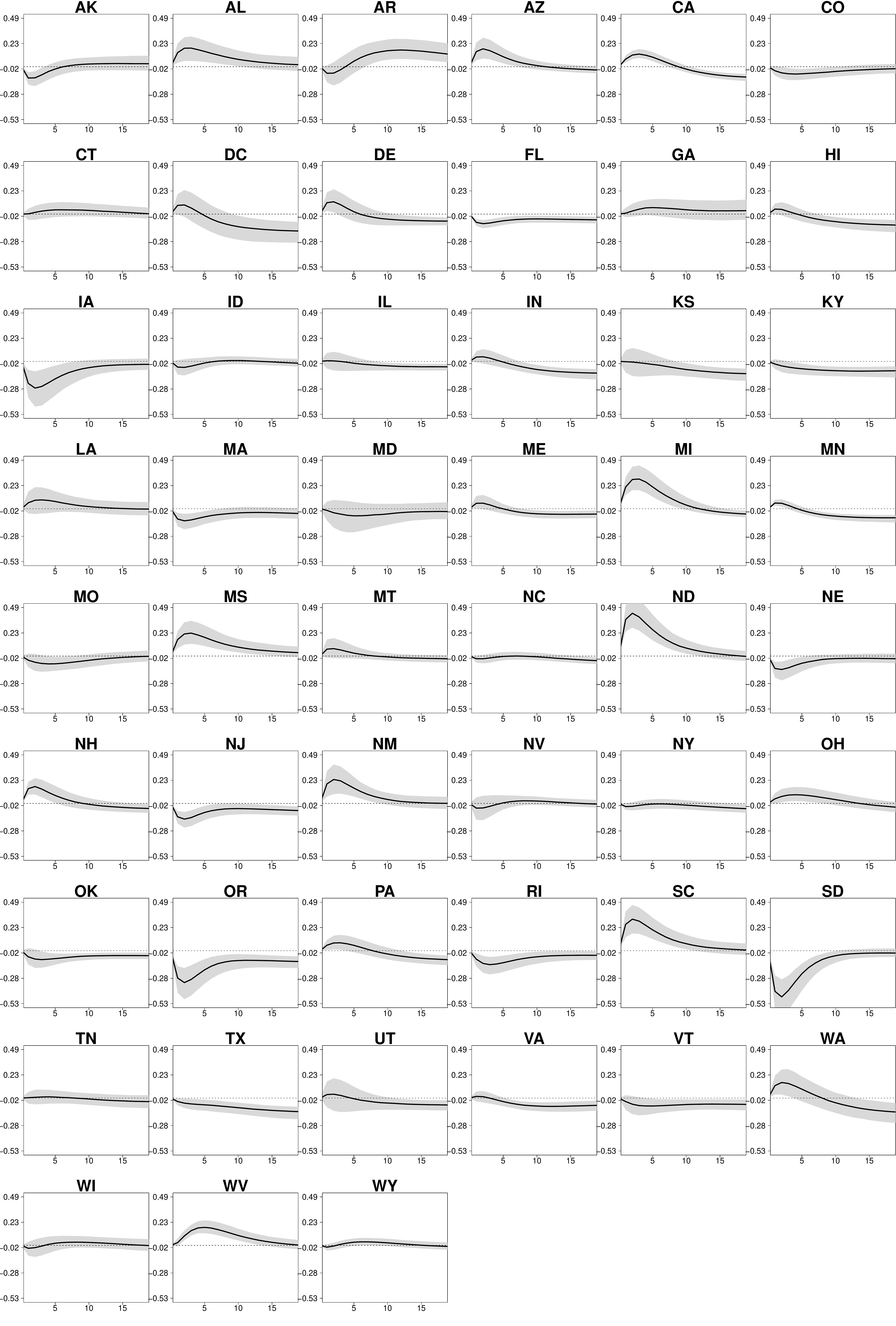}
\caption{Impulse response functions for household income inequality in US states.}\label{fig:irf_iq}
\caption*{\footnotesize{\noindent\textit{Notes}: The solid black line is the median response and the (gray) shaded area represents the 16th and 84th percentiles. The dotted line indicates the zero line. Results are based on 5,000 posterior draws. Sample period: 1985:Q1 -- 2017:Q1. A list of states and abbreviations is given in Appendix \ref{app:statelist}. Front axis: quarters after impact.}}
\end{figure}
\begin{figure}[!ht]
\centering
\includegraphics[width=0.86\textwidth]{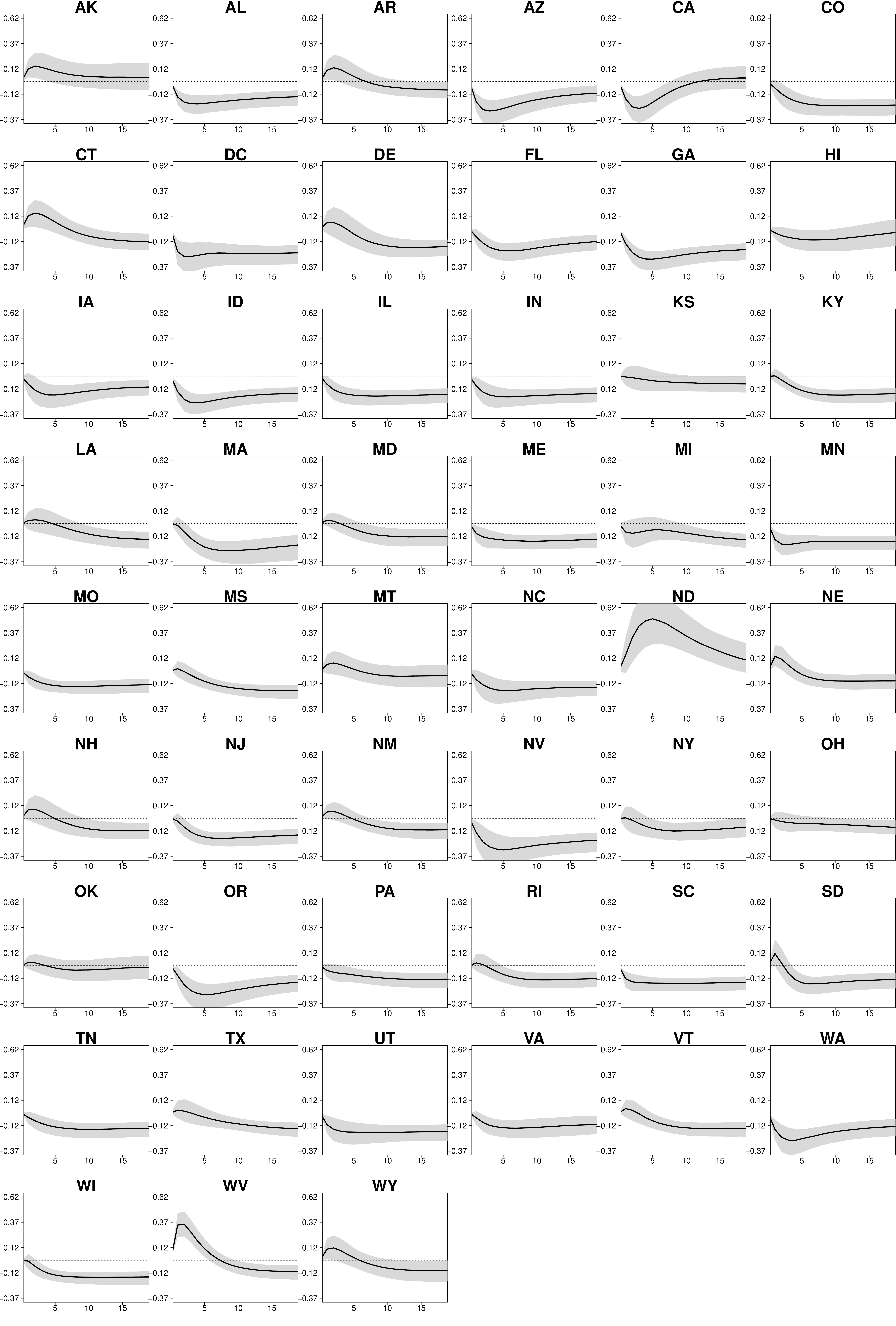}
\caption{Impulse response functions for total personal income in US states.}\label{fig:irf_yn}
\caption*{\footnotesize{\noindent\textit{Notes}: The solid black line is the median response and the (gray) shaded area represents the 16th and 84th percentiles. The dotted line indicates the zero line. Results are based on 5,000 posterior draws. Sample period: 1985:Q1 -- 2017:Q1. A list of states and abbreviations is given in Appendix \ref{app:statelist}. Front axis: quarters after impact.}}
\end{figure}
\begin{figure}[!ht]
\centering
\includegraphics[width=0.86\textwidth]{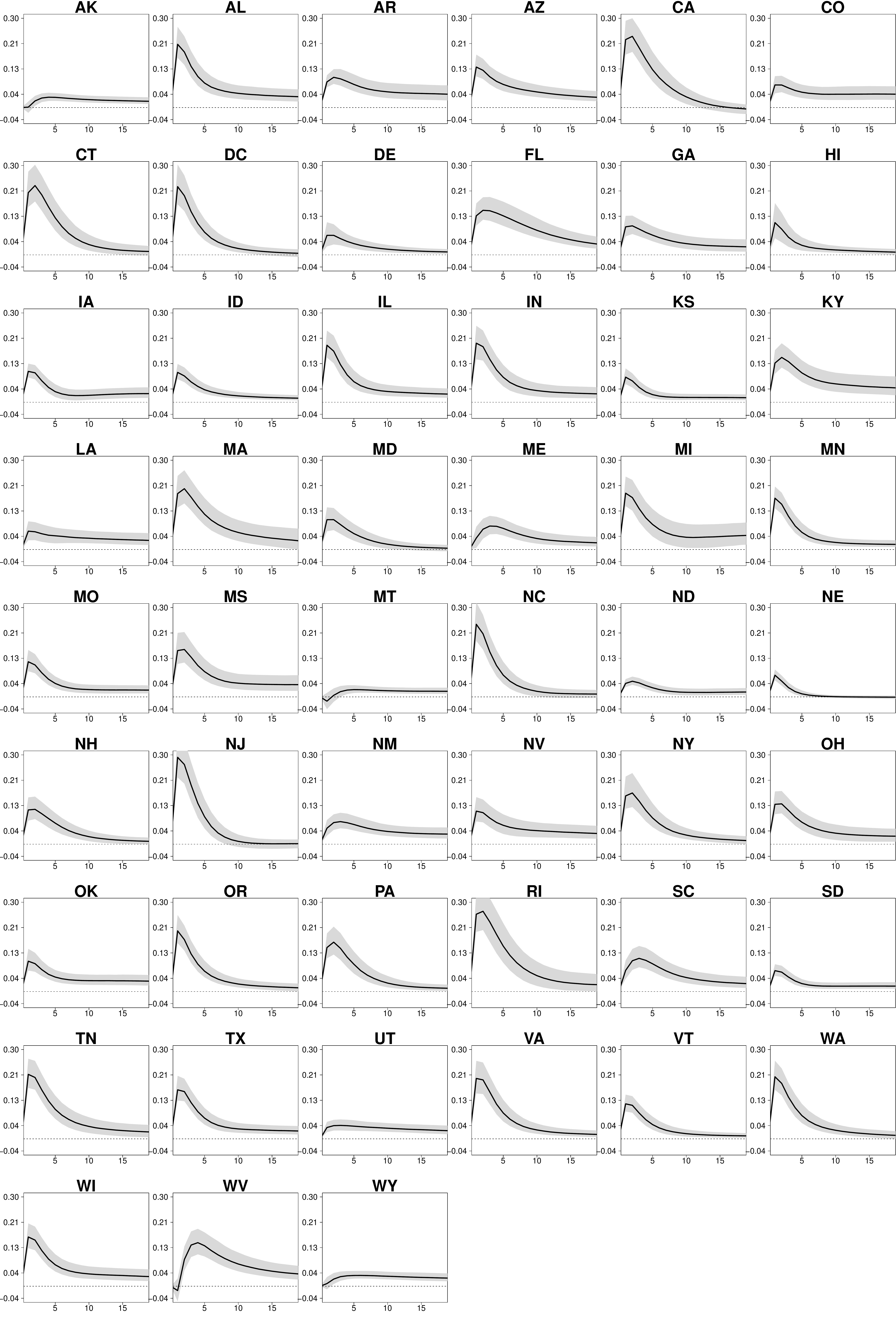}
\caption{Impulse response functions for unemployment in US states.}\label{fig:irf_un}
\caption*{\footnotesize{\noindent\textit{Notes}: The solid black line is the median response and the (gray) shaded area represents the 16th and 84th percentiles. The dotted line indicates the zero line. Results are based on 5,000 posterior draws. Sample period: 1985:Q1 -- 2017:Q1. A list of states and abbreviations is given in Appendix \ref{app:statelist}. Front axis: quarters after impact.}}
\end{figure}
\begin{figure}[!ht]
\centering
\includegraphics[width=0.86\textwidth]{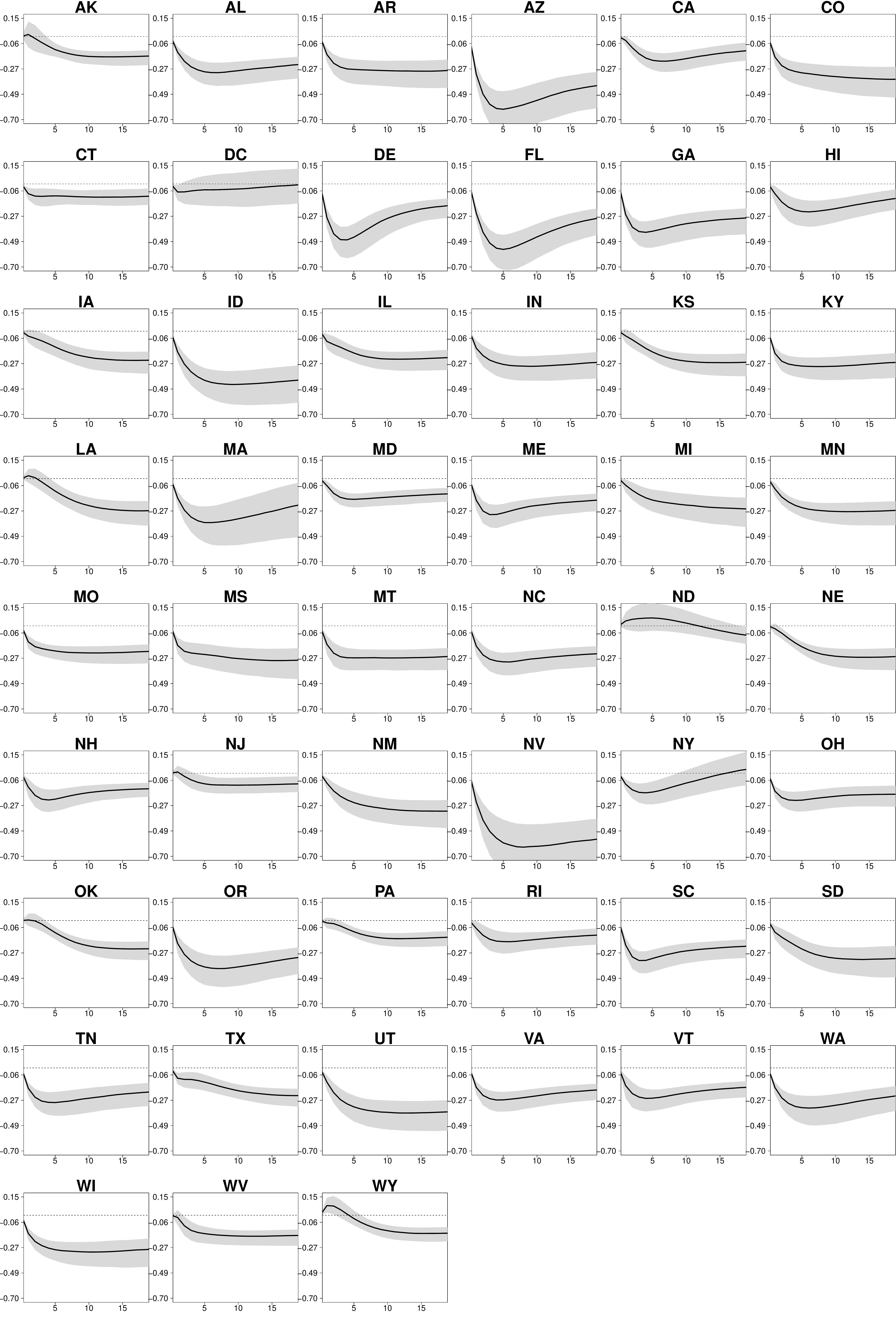}
\caption{Impulse response functions for employment in US states.}\label{fig:irf_em}
\caption*{\footnotesize{\noindent\textit{Notes}: The solid black line is the median response and the (gray) shaded area represents the 16th and 84th percentiles. The dotted line indicates the zero line. Results are based on 5,000 posterior draws. Sample period: 1985:Q1 -- 2017:Q1. A list of states and abbreviations is given in Appendix \ref{app:statelist}. Front axis: quarters after impact.}}
\end{figure}
\begin{figure}[!ht]
\centering
\includegraphics[width=0.86\textwidth]{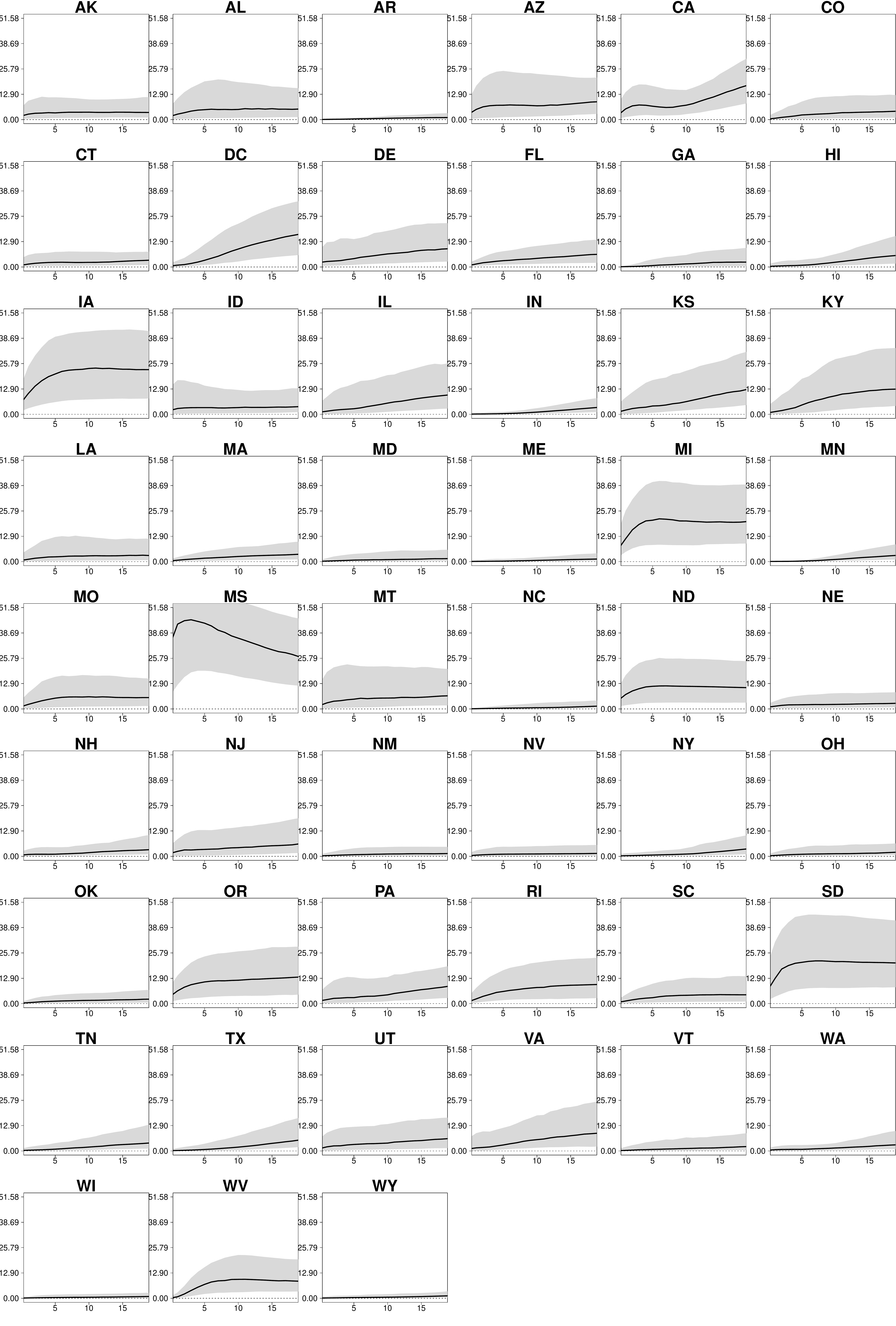}
\caption{Forecast error variance decompositions for household income inequality in US states.}\label{fig:fevd_iq}
\caption*{\footnotesize{\noindent\textit{Notes}: The solid black line is the median and the (gray) shaded area represents the 16th and 84th percentiles. The dotted line indicates the zero line. Results are based on 5,000 posterior draws. Sample period: 1985:Q1 -- 2017:Q1. A list of states and abbreviations is given in Appendix \ref{app:statelist}. Front axis: quarters after impact.}}
\end{figure}
\clearpage

\section{List of states and regional categorization}\label{app:statelist}
\begin{longtable}{p{.1\textwidth} p{.8\textwidth}}
\caption{List of census regions and associated US states.}\\
\toprule
\textbf{Region} & \textbf{States} \\ 
  \midrule\endfirsthead

 & \multicolumn{1}{r}{\textit{ctd.}}\\
 \midrule
\endhead

\midrule
\endfoot

\endlastfoot
\textit{Northeast} & 	Connecticut (CT), Maine (ME), Massachusetts (MA), New Hampshire (NH), New Jersey (NJ), New York (NY), Rhode Island (RI), Vermont (VT) \\
\textit{Midwest}   & 	Illinois (IL), Indiana (IN), Iowa (IA), Kansas (KS), Michigan (MI), Minnesota (MN), Missouri (MO), Nebraska (NE), North Dakota (ND), 
						Ohio (OH), South Dakota (SD), Wisconsin (WI)\\
\textit{South}     & 	Alabama (AL), Arkansas (AR), Delaware (DE), District of Columbia (DC), Florida (FL), Kentucky (KY), Georgia (GA), Louisiana (LA), 
						Maryland (MD), Mississippi (MS), North Carolina (NC), Oklahoma (OK), South Carolina (SC), Tennessee (TN), Texas (TX), Virginia (VA), 
						West Virginia (WV) \\
\textit{West}      & 	Alaska (AK), Arizona (AZ), California (CA), Colorado (CO), Hawaii (HI), Idaho (ID), Montana (MT), Nevada (NV), New Mexico (NM), Oregon (OR), 
						Utah (UT), Washington (WA), Wyoming (WY)\\
\bottomrule
\end{longtable}

\begin{figure}[!ht]
\centering
\includegraphics[width=0.9\textwidth]{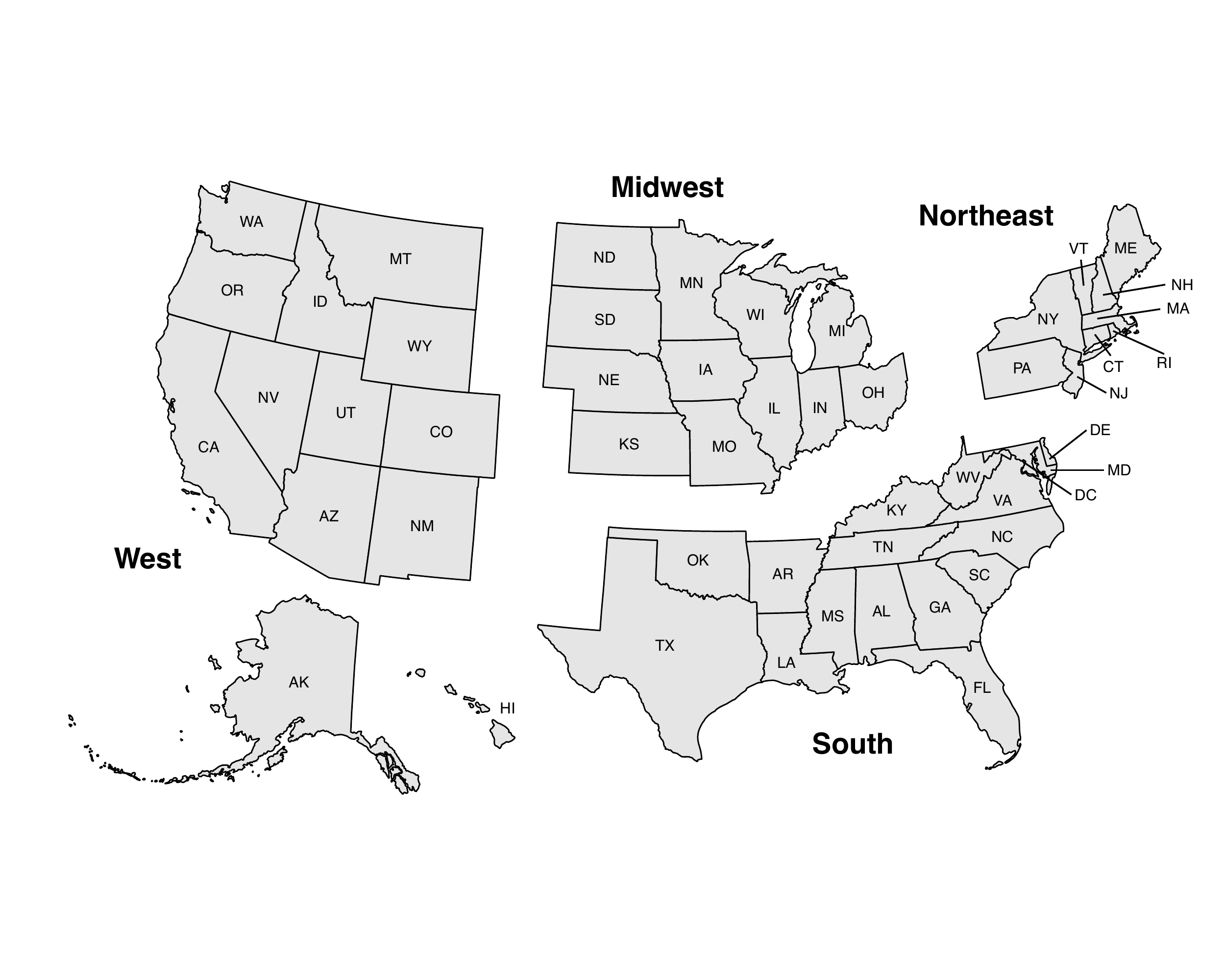}
\caption{Map of the US states and the District of Columbia divided into the four census regions.}\label{fig:map_blank}
\end{figure}
\end{appendices}
\end{document}